\documentclass[11pt,a4paper]{article}
\usepackage[a4paper,margin=2.65cm]{geometry}
\usepackage[T1]{fontenc}
\usepackage[utf8]{inputenc}
\usepackage{lmodern}
\usepackage{amsmath,amssymb,mathtools,amsthm,bm}
\usepackage{microtype}
\usepackage{enumitem}
\usepackage{booktabs}
\usepackage{tabularx}
\usepackage{hyperref}
\usepackage[nameinlink,noabbrev]{cleveref}
\allowdisplaybreaks[2]
\newcommand{\R}{\mathbb{R}}
\newcommand{\dd}{\mathrm{d}}
\newcommand{\Div}{\operatorname{div}}
\newcommand{\tr}{\operatorname{tr}}
\newcommand{\grad}{\nabla}
\newcommand{\half}{\tfrac12}

\newcommand{\Acal}{\mathcal A}
\newcommand{\Dcal}{\mathcal D}

\newcommand{\bv}{\boldsymbol{v}}
\newcommand{\bu}{\boldsymbol{u}}
\newcommand{\bj}{\boldsymbol{j}}
\newcommand{\bq}{\boldsymbol{q}}
\newcommand{\bb}{\boldsymbol{b}}
\newcommand{\bfv}{\boldsymbol{f}}
\newcommand{\bmom}{\boldsymbol{m}}
\newcommand{\bPhi}{\boldsymbol{\Phi}}
\newcommand{\bS}{\boldsymbol{S}}
\newcommand{\bD}{\boldsymbol{D}}
\newcommand{\bI}{\boldsymbol{I}}
\newcommand{\bB}{\boldsymbol{B}}
\newcommand{\bG}{\boldsymbol{G}}
\newcommand{\blambda}{\boldsymbol{\lambda}}
\newcommand{\bP}{\boldsymbol{P}}
\newcommand{\bM}{\boldsymbol{M}}
\newcommand{\bU}{\boldsymbol{U}}
\newcommand{\bV}{\boldsymbol{V}}
\newcommand{\bZ}{\boldsymbol{Z}}
\newcommand{\bQ}{\boldsymbol{Q}}
\newcommand{\bF}{\boldsymbol{F}}
\newcommand{\vp}{\boldsymbol{V}'}
\newcommand{\wprime}{\boldsymbol{W}'}
\newcommand{\Uret}{\boldsymbol{U}'_{\mathrm{ret}}}
\newcommand{\Ufix}{\boldsymbol{U}'_{\mathrm{fix}}}
\newcommand{\Ufixstar}{\boldsymbol{U}'_{\mathrm{fix},*}}

\newtheorem{proposition}{Proposition}[section]
\newtheorem{theorem}[proposition]{Theorem}
\newtheorem{corollary}[proposition]{Corollary}
\theoremstyle{definition}
\newtheorem{definition}[proposition]{Definition}
\newtheorem{example}[proposition]{Example}
\theoremstyle{remark}
\newtheorem{remark}[proposition]{Remark}
\crefname{proposition}{Proposition}{Propositions}
\Crefname{proposition}{Proposition}{Propositions}
\crefname{theorem}{Theorem}{Theorems}
\Crefname{theorem}{Theorem}{Theorems}
\crefname{corollary}{Corollary}{Corollaries}
\Crefname{corollary}{Corollary}{Corollaries}
\crefname{definition}{Definition}{Definitions}
\Crefname{definition}{Definition}{Definitions}
\crefname{example}{Example}{Examples}
\Crefname{example}{Example}{Examples}
\crefname{remark}{Remark}{Remarks}
\Crefname{remark}{Remark}{Remarks}
\hypersetup{hidelinks,
pdftitle={Degenerate symmetric formulation and principal subsystems for chemically reacting multi-temperature fluid mixtures},
pdfauthor={Dieter Bothe; Tommaso Ruggeri}}
\title{Degenerate symmetric formulation and\\[0.20em]principal subsystems for chemically reacting\\[0.20em]multi-temperature fluid mixtures}
\author{Dieter Bothe$^{1}$ \quad Tommaso Ruggeri$^{2,3}$\\[0.45em]
\normalsize $^{1}$Department of Mathematics, Technical University of Darmstadt,\\
\normalsize Peter-Gr\"unberg-Str. 10, 64287 Darmstadt, Germany\\
\normalsize $^{2}$Department of Mathematics, University of Bologna, 40123 Bologna, Italy\\
\normalsize Via Saragozza 8, Bologna, Italy\\
\normalsize $^{3}$Accademia Nazionale dei Lincei, Via della Lungara 10, 00165 Rome, Italy}\date{}

\newcommand{\bibdoi}[1]{\href{https://doi.org/#1}{doi:\nolinkurl{#1}}}
\begin{document}
\maketitle

\begin{abstract}
We construct a degenerate symmetric first-order formulation and principal
subsystems for chemically reacting multi-temperature  fluid mixtures with
independent constituent velocities and temperatures, viscosity, and heat
conduction. Starting from the entropy-admissible Class-III theory of    \mbox{
\cite{BotheRuggeriPaperI2026}}, we characterize when its prescribed constituent Euler fluxes are
generated by scalar potentials in the conservative entropy variables.
Under the stated hypotheses the necessary and sufficient condition is
that the separate Gibbs--Duhem one-forms $\omega_i$ vanish. This implies locally
additive entropy, but it is not required by the general parent theory.
Within this subclass, the complete main field contains multipliers for
both conservative balances and zero-storage constitutive equations. The
spatial Hessians are symmetric, and the temporal Hessian is positive
semidefinite and singular. For the separated transport laws, eliminating
the auxiliary variables gives a factorization of the
positive-semidefinite dissipative principal symbol. We extend the algebraic Boillat--Ruggeri principal-subsystem construction
to this degenerate setting. Commuting restrictions yield a dissipative
common-temperature system and a nested hierarchy of grouped-temperature
Euler subsystems. Classical characteristic interlacing applies within
the nested Euler hierarchy, whose temporal Hessians are positive
definite. Separately, constrained entropy maximization defines an
energy-equivalent common temperature and provides a thermodynamic
Class-III-to-Class-II reduction for additive and nonadditive parent
entropies. For a thermally reconstructed reference state, the full
relative entropy splits exactly into the thermal projection deficit
and the relative entropy of the retained variables. We derive the
Class-II entropy-production identity and distinguish the thermodynamic
projection from principal restrictions and dynamical relaxation limits.
\end{abstract}

\section{Introduction}
\label{sec:introduction}

Chemically reacting fluid mixtures can be described at different levels of
resolution. Classical diffusion theories employ a single barycentric
velocity and a common temperature, with diffusive fluxes prescribed by
constitutive relations such as Fick's law or the Maxwell--Stefan equations.
More detailed models allow different constituent velocities, while the
most resolved descriptions also assign independent temperatures to the
constituents. These levels are commonly called Classes~I, II, and III,
respectively~\cite[Section~7.1]{HutterJohnk2004} and~\cite{BotheDreyer2015}. In   \cite{BotheRuggeriPaperI2026} we developed an entropy-consistent
Class-III theory including chemical reactions, interconstituent exchanges,
viscosity, and heat conduction, allowing nonadditive entropy.

An early general covariant construction of the main field and the
associated symmetric conservative formulation was given by Ruggeri and
Strumia~\cite{RuggeriStrumia1981}. The treatment of fluid mixtures within
Rational Extended Thermodynamics was developed systematically by
M\"uller and Ruggeri~\cite[Chapters~5 and~7]{MuellerRuggeri1998}.
For nonreacting Euler mixtures with constituentwise thermodynamics,
Ruggeri and Simi\'c~\cite{RuggeriSimic2007} constructed the symmetric
hyperbolic multi-temperature  formulation, identified its single-temperature
principal subsystem, and studied the subcharacteristic inequalities,
the Shizuta--Kawashima condition and a binary example.
Subsequent work clarified the energy-equivalent average temperature,
dynamical pressure, and Maxwellian iteration for nonreacting
multi-temperature  mixtures~\cite{GouinRuggeri2008,RuggeriSimic2009}.
See also the systematic account in
\cite[Chapters~28, 29 and~31]{RuggeriSugiyama2021}.
  Related symmetric formulations for single-fluid equations and
hyperbolic--parabolic normal forms appear in
\cite{HughesFrancaMallet1986,KawashimaShizuta1988}.
For multicomponent reactive flows in a different, common-velocity,
common-temperature setting, Giovangigli and Massot
\cite{GiovangigliMassot1998} obtained a symmetric form and a
hyperbolic--parabolic normal form. Their model is distinct from the
present Class-III system with independent constituent velocities and
temperatures and zero-storage constitutive equations.

Here we first determine when the \emph{prescribed} constituent Euler
fluxes of the general reacting Class-III model are generated by scalar
potentials in the conservative entropy variables. This integrability
requires the constituentwise identities $\omega_i=0$,
which are stronger than the mixture-wide thermodynamic compatibility
condition of    \mbox{
\cite{BotheRuggeriPaperI2026}}. As proved in \Cref{subsec:euler-rigidity}, they
imply locally additive entropy,
$\eta(\boldsymbol z,\boldsymbol\rho)=\sum_i\eta_i(z_i,\rho_i)$,
with the partial pressures constrained accordingly. Additivity alone
without compatible pressures is not a converse. This is an additional
condition for the fixed-flux generator, not a consequence of entropy
admissibility of the general parent model. On this precisely identified
subclass we construct the complete degenerate symmetric generator and
its principal subsystems. The    main results  are the necessary-and-sufficient
fixed-flux compatibility    theorem, the complete zero-storage
viscous    and heat-conducting  generator, commuting thermal and constitutive principal
restrictions, and the explicit singular-pencil analysis. The subsequent
entropy-maximizing thermal projection is a distinct result, applicable
also to nonadditive parent entropies.

The mixed source formulation also accommodates locally representable coupled constitutive responses, and the positive-transport case yields an explicit coupling criterion for the Euler and dissipative symbols.

To clarify the purpose of this work, we first recall a general question
concerning quasilinear systems of balance laws endowed with a supplementary
entropy law. For regular balance laws with a compatible supplementary
entropy and a positive-definite entropy Hessian in the conservative
densities, the main field yields a symmetric-hyperbolic formulation.
The issue is whether this construction can also be used for viscous and heat-conducting systems
of hyperbolic--parabolic type. We follow an idea proposed by
Ruggeri~\cite{Ruggeri1983Viscous} in 1983: the Fourier--Navier--Stokes
equations can be rewritten as an enlarged first-order system by adjoining
the viscous stress and the heat flux as constitutive unknowns, together
with their constitutive equations. For a single fluid, the resulting
system consists of five balance equations and nine constitutive equations.
The latter contain no time derivatives of the added unknowns. This lack
of temporal storage is precisely what makes the ensuing symmetric
formulation degenerate. We begin with the general main-field construction
and then return to this extension.

Let $\bU=\bU(\boldsymbol x,t)\in\R^N$ denote the vector of unknowns,
where $\boldsymbol x\in\R^3$ is the spatial variable. We consider a
system of $N$ quasilinear balance laws with a supplementary entropy law,
\begin{equation}
  \partial_\alpha\bF^\alpha(\bU)=\boldsymbol J(\bU),
  \qquad \partial_\alpha h^\alpha(\bU)=g(\bU),
  \label{eq:intro-supplementary-system}
\end{equation}
where $\bF^\alpha(\bU)$ and $\boldsymbol J(\bU)$ are $N$-vectors,
whereas $h^\alpha(\bU)$ and $g(\bU)$ are scalar functions of the state.
We use Einstein summation over $\alpha=0,1,2,3$, with $x^0=t$ and
$\partial_0=\partial_t$. Spatial indices $k,l$ range from 1 to 3.
At each state under consideration, assume that the balance equations
admit at least one first-derivative tuple and that every such tuple
satisfies the supplementary law. The differential coefficient of the
supplementary law then annihilates the kernel of the balance-law
coefficient map and therefore lies in its row space. Evaluating the
supplementary law at a compatible tuple also gives the source identity.
Thus there exists a multiplier vector $\bU'$ satisfying
\begin{equation}
  \dd h^\alpha=\bU'\mathbin{\boldsymbol\cdot}\dd\bF^\alpha,
  \qquad g=\bU'\mathbin{\boldsymbol\cdot}\boldsymbol J,
  \qquad \alpha=0,1,2,3.
  \label{eq:intro-main-field-multipliers}
\end{equation}
For a fixed representation of the balance and constitutive rows, full
row rank of the balance-law coefficient map gives uniqueness of the
multiplier vector.
Changing the auxiliary-row normalization changes its coordinate
representation. This freedom is made precise below. The vector $\bU'$ is the \emph{main field}. Its
definition is the complete system of multiplier identities~\eqref{eq:intro-main-field-multipliers}, not a
representation as the gradient of one selected entropy density.

Introduce the scalar potentials
\begin{equation}
  h'^\alpha=\bU'\mathbin{\boldsymbol\cdot}\bF^\alpha-h^\alpha.
  \label{eq:intro-legendre-potentials}
\end{equation}
\emph{We now make the additional assumption} that the $N$ components
of $\bU'$ form a system of independent local variables. Differentiating
the potentials and using the identities~\eqref{eq:intro-main-field-multipliers}, we obtain
\begin{equation}
  \dd h'^\alpha=\bF^\alpha\mathbin{\boldsymbol\cdot}\dd\bU',
  \qquad
  \bF^\alpha=\frac{\partial h'^\alpha}{\partial\bU'}.
  \label{eq:intro-generator-identity}
\end{equation}
Consequently the field equations take the form
\begin{equation}
  \boxed{\displaystyle
  \frac{\partial^2h'^\alpha}{\partial\bU'\,\partial\bU'}
  \,\partial_\alpha\bU'=\boldsymbol J.}
  \label{eq:intro-symmetric-system}
\end{equation}
Here $h^\alpha$ denotes the supplementary mathematical-entropy
four-vector, while $h'^\alpha$ denotes its Legendre generating
potentials. The prime is not a derivative in the spacetime index
$\alpha$. We use the coefficient-matrix notation
$A^\alpha(\bU'):=D_{\bU'}^2h'^\alpha(\bU')$.
Thus $h'^\alpha$ is a scalar function, whereas $A^\alpha$ is
its symmetric Hessian matrix, and
$\partial_\alpha(\partial h'^\alpha/\partial\bU')=\boldsymbol J$
is the generated balance system. In the enlarged zero-storage model
$A^0$ is only positive semidefinite, not positive definite. Hence
symmetry of $A^\alpha$ must not be conflated with    symmetric hyperbolicity in the sense of Friedrichs. This derivation uses the \emph{full} main field.

For a regular hyperbolic balance-law system whose independent temporal
densities are state coordinates, positive definiteness of the
mathematical-entropy Hessian with respect to those densities guarantees
that the main field forms independent variables. The temporal potential is then the
Legendre transform of the convex entropy density, and its Hessian is
positive definite. Equation~\eqref{eq:intro-symmetric-system} is therefore
symmetric hyperbolic in the sense of Friedrichs. The entropy-gradient
representation used by Boillat~\cite{Boillat1974} applies in this
regular setting. It is, however, only a special representation of the
main field, not its general definition.

The latter was given by Ruggeri and Strumia through the multiplier
identities~\eqref{eq:intro-main-field-multipliers}
\cite{RuggeriStrumia1981}. These identities impose compatibility on
all spacetime flux components. They retain their form when the balance
fluxes, entropy current, and sources are transformed consistently, with
the corresponding dual transformation of the multipliers. They remain
meaningful when the temporal densities form only part of the independent
variables. The Fourier--Navier--Stokes construction recalled above is
an example. See also~\cite[Section~2.3.4]{RuggeriSugiyama2021}.

For an enlarged first-order system with constitutive equations lacking
temporal derivatives, the temporal flux has the block form
\begin{equation}
  \bF^0=(\boldsymbol V,\boldsymbol0),\qquad
  \boldsymbol V\in\mathbb R^M,\quad
  \boldsymbol0\in\mathbb R^{N-M}.
  \label{eq:intro-temporal-splitting}
\end{equation}
Nevertheless, the \emph{complete} main field is still determined by the
multiplier identities~\eqref{eq:intro-main-field-multipliers}, including
the spatial identities. Provided its components furnish independent
variables, the potential relations~\eqref{eq:intro-generator-identity}
continue to hold. Since the last $N-M$ components of $\bF^0$ vanish,
$h'^0$ does not depend on the corresponding main-field coordinates.
Its Hessian in~\eqref{eq:intro-symmetric-system} is therefore singular.
Under the appropriate convexity hypotheses it is positive semidefinite.
The spatial Hessians remain symmetric. We thus obtain a symmetric
first-order representation also for a hyperbolic--parabolic system,
though not a  \emph{} 
  symmetric-hyperbolic  system in the    sense of Friedrichs.
The temporal entropy-gradient construction alone cannot provide the
missing constitutive multipliers.

   The principal-subsystem construction is based on the following observation. The algebraic operation
underlying the Boillat--Ruggeri construction of principal
subsystems~\cite{BoillatRuggeri1997} requires a symmetric generator,
not necessarily a positive-definite temporal Hessian. We may freeze
selected components of the main field at constant values and retain
only the equations conjugate to the remaining components. Equivalently,
we restrict the generating potentials to the corresponding affine
section. Their Hessians remain symmetric even if the parent temporal
Hessian is singular. Positive definiteness, and hence the classical
characteristic-interlacing conclusion, is needed only for the
hyperbolic members of the resulting hierarchy.

Starting from the dissipative constituent equations of the general
Class-III theory, the enlarged first-order system has the temporal flux
structure~\eqref{eq:intro-temporal-splitting}. For the subclass in
which the prescribed constituent Euler fluxes satisfy the separate
integrability conditions characterized in \Cref{subsec:euler-rigidity},
and under the further constitutive and regularity hypotheses stated below,
its complete main-field potentials provide a symmetric generator with a
singular temporal Hessian. No corresponding claim is made here for the
unrestricted parent constitutive class.
Setting the differences of the constituent energy multipliers to
zero imposes equality of the reciprocal temperatures. The resulting
single-temperature, multi-velocity model retains viscous and thermal
dissipation.
Its temporal Hessian remains singular. Fixing at zero the main-field
components conjugate to the zero-storage constitutive equations, and
deleting those equations, yields a principal subsystem with the Euler
fluxes. Its homogeneous principal part is symmetric hyperbolic under
the stated convexity hypotheses. Combining these restrictions yields
nested Euler mixture subsystems with grouped temperatures. These are
algebraic restrictions of the equations and potentials, not assertions
that every solution of the original dissipative system remains on the
restricted manifolds.

In \Cref{sec:thermal-reduction}, entropy-maximizing thermal reduction
is developed for the general parent thermodynamics, including the
nonadditive case. Neither that thermodynamic projection nor a principal
restriction is identified with a dynamical relaxation limit.
The thermodynamic projection also yields an exact
relative-entropy decomposition, including sharp equality cases and
consistent splitting under nested thermal groupings
(\Cref{prop:thermal-relative-entropy-decomposition}).

Section~\ref{sec:parent-preliminaries} recalls the parent model.
Section~\ref{sec:main-field} establishes the flux-potential conditions,
the first-order symmetric generator, and the principal-subsystem
hierarchy. Section~\ref{sec:thermal-reduction} considers the reduction
to a common temperature and further mechanical reductions. The
Appendix relates the conservative Liu multipliers to the corresponding
block of the main field.

\section{Parent Class-III model and notation}
\label{sec:parent-preliminaries}

This section recalls the balances, thermodynamic identities, and constitutive relations of the Class-III model~\cite{BotheRuggeriPaperI2026} needed below.

\subsection{Balance    equations  and objective  exchange  variables}
\label{sec:balances}
Let $n$ be the number of constituents. All local state domains
considered below have strictly positive partial densities $\rho_i>0$.
For constituent $i=1,\ldots,n$, let $\rho_i$, $\bv_i$, and $e_i$ denote partial mass density, velocity, and specific nonkinetic internal energy. The partial Cauchy stress is $\bS_i$, the nonconvective constituent energy flux is $\bq_i$, and $\bb_i$ denotes the body force per unit mass of constituent $i$. The interaction productions of momentum and total energy are $\bfv_i$ and $k_i$, respectively. The constituent mass, momentum, and total-energy balances are those developed in \cite{BotheRuggeriPaperI2026}. Their explicit form is displayed below, together with the nine constitutive rows per constituent, in \eqref{eq:acta-FNS-first-order-form}. We use the first three rows throughout.
The mass production of constituent $i$ is denoted by $r_i$.
Prescribed body forces may depend explicitly on position and time.
This dependence in $\boldsymbol J$ is suppressed in the generator notation.
Their momentum and energy contributions cancel in the entropy pairing.
The interaction productions satisfy the separate conservation constraints
\begin{equation}
 \sum_i r_i=0,\qquad \sum_i\bfv_i=\boldsymbol0,\qquad \sum_i k_i=0.
\label{eq:source-conservation}
\end{equation}
Let $n_R$ denote the number of reaction paths, indexed by
$a=1,\ldots,n_R$. For each path, write
\begin{equation}
 \sum_i\alpha_i^a A_i\rightleftharpoons\sum_i\beta_i^a A_i,
 \qquad \nu_i^a:=\beta_i^a-\alpha_i^a,
 \qquad R_a:=R_a^f-R_a^b.
\label{eq:reaction-paths}
\end{equation}
Here $A_i$ is a chemical-species label (unrelated to the
coefficient matrices $A^\alpha$). $\alpha_i^a$ and $\beta_i^a$
are its nonnegative stoichiometric coefficients on the reactant and
product sides, respectively, and $\nu_i^a$ is the signed net coefficient.
The quantities $R_a^f$ and $R_a^b$ are the forward and backward molar
reaction rates per unit volume and time, so $R_a$ is the net rate.
With $M_i$ the molar mass of species $i$, let
$\boldsymbol r=(r_i)_{i=1}^n$ be the mass-production vector and
$\boldsymbol R=(R_1,\ldots,R_{n_R})^{\sf T}$ the net reaction-rate vector.
Define the $n\times n_R$ \emph{mass-weighted stoichiometric matrix}
$\boldsymbol\Gamma=(\Gamma_{ia})$ by
\begin{equation}
 \Gamma_{ia}:=M_i\nu_i^a,
 \qquad r_i=M_i\sum_a\nu_i^aR_a,
 \qquad \boldsymbol r=\boldsymbol\Gamma\boldsymbol R.
\label{eq:stoichiometric-production-matrix}
\end{equation}
Conservation of mass in each reaction path reads
\[
 \sum_{i=1}^n\Gamma_{ia}=\sum_{i=1}^n M_i\nu_i^a=0,
 \qquad a=1,\ldots,n_R.
\]
Here and below ${\sf T}$ denotes transposition, whereas an italic
superscript $T$ labels thermal or thermo-diffusive quantities.

Define the barycentric velocity, diffusion velocities, diffusion mass fluxes, and mass fractions by
\begin{equation}
 \rho:=\sum_i\rho_i,
 \qquad \rho\bv:=\sum_i\rho_i\bv_i,
 \qquad \bu_i:=\bv_i-\bv,
 \qquad \bj_i:=\rho_i\bu_i,
 \qquad y_i:=\rho_i/\rho,
\label{eq:barycentric-definitions}
\end{equation}
so that $\sum_i\rho_i\bu_i=\sum_i\bj_i=\boldsymbol0$.
Stacked symbols such as $\boldsymbol y=(y_i)_{i=1}^n$,
$\bu=(\bu_i)_{i=1}^n$ and $\boldsymbol B=(\bB_i)_{i=1}^n$
collect constituent entries. Species matrices act on each spatial component. The exact mixture stress and energy decomposition are
\begin{equation}
 \bS=\sum_i(\bS_i-\rho_i\bu_i\otimes\bu_i),
\label{eq:mixture-stress}
\end{equation}
\begin{equation}
 \sum_i\left(\rho_i e_i+\half\rho_i|\bv_i|^2\right)
 =\rho e+\half\rho|\bv|^2+\half\sum_i\rho_i|\bu_i|^2,
 \qquad \rho e:=\sum_i\rho_i e_i.
\label{eq:energy-decomposition}
\end{equation}
The objective momentum and    energy exchange terms relative to the barycentric velocity  are
\begin{equation}
 \bmom_i:=\bfv_i-r_i\bv,
 \qquad
 \varepsilon_i:=k_i-\bv\cdot\bfv_i+\half r_i|\bv|^2.
\label{eq:objective-sources}
\end{equation}
They satisfy the separate conservation constraints
\begin{equation}
 \sum_i\bmom_i=\boldsymbol0,\qquad \sum_i\varepsilon_i=0.
\label{eq:objective-source-conservation}
\end{equation}
Summation of the partial thermal-energy balances gives the consistency identity
\begin{equation}
 \partial_t(\rho e)+\Div(\rho e\bv+\widetilde{\bq})
 =\sum_i\bS_i:\grad\bv_i-\sum_i\bu_i\cdot\bmom_i+\half\sum_i r_i|\bu_i|^2,
 \qquad \widetilde{\bq}:=\sum_i(\bq_i+\rho_i e_i\bu_i).
\label{eq:thermal-energy-balance}
\end{equation}

\subsection{Thermodynamic state and pressure representation}
\label{sec:model-class-specification}
\label{subsec:entropy-flux-pressure}
The Class-III entropy density is a constitutive function of the partial
thermal-energy densities and the partial mass densities,
\begin{equation}
 \rho s=\widehat{\rho s}(\rho_1e_1,\ldots,\rho_ne_n,\rho_1,\ldots,\rho_n).
\label{eq:entropy-state}
\end{equation}
For    brevity, set
\begin{equation}
 z_i:=\rho_i e_i,
 \qquad
 \boldsymbol z:=(z_1,\ldots,z_n)^{\sf T},
 \qquad
 \eta(\boldsymbol z,\boldsymbol\rho):=\widehat{\rho s}(\boldsymbol z,\boldsymbol\rho).
\label{eq:thermal-density-notation}
\end{equation}
Here $\eta$ denotes the physical entropy density $\rho s$ in the variables
$(\boldsymbol z,\boldsymbol\rho)$, where $\boldsymbol\rho=(\rho_1,\ldots,\rho_n)^{\sf T}$.
Throughout the paper, the mathematical entropy is $h^0=-\rho s$.

Throughout, $\eta$ is $C^2$ on an open thermodynamic domain, the partial thermodynamic pressure functions are $C^1$, and all balance and supplementary-law identities are understood for sufficiently smooth fields. Additional derivatives are assumed explicitly where needed.
The thermodynamic stability hypothesis is
\begin{equation}
 D_{(\boldsymbol z,\boldsymbol\rho)}^2\eta\prec0,
\label{eq:negative-definite-entropy-hessian}
\end{equation}
throughout the stable interior domain. We also impose the independent positive-temperature condition
\begin{equation}
 \frac{\partial\eta}{\partial z_i}>0.
\label{eq:positive-temperature-domain}
\end{equation}
The    thermodynamic variables conjugate to $z_i$ and $\rho_i$ in the entropy representation  are
\begin{equation}
 \frac{1}{T_i}:=\frac{\partial\eta}{\partial z_i},
 \qquad
 -\frac{\mu_i}{T_i}:=\frac{\partial\eta}{\partial\rho_i}.
\label{eq:entropy-conjugates}
\end{equation}
We restrict attention to the same local bulk class as in the parent construction: the constituents are nonpolar, the partial Cauchy stresses are symmetric, and no independent material director or other anisotropic structure is present. Define
\begin{equation}
 \bD_i:=\half\bigl(\grad\bv_i+(\grad\bv_i)^{\sf T}\bigr),
 \qquad
 \bD_i^{\circ}:=\bD_i-\frac{1}{3}(\Div\bv_i)\bI.
\label{eq:deformation-tensors}
\end{equation}
Write $\bS_i=-P_i\bI+\bS_i^{\circ}$. In the isolated volumetric
response at fixed thermodynamic state, impose the same divergence
$\Div\bv_i$ on all constituents. The partial thermodynamic pressure
$p_i$ is the value of $P_i$ when this common divergence is zero, as
specified in \cite[Section~3.2]{BotheRuggeriPaperI2026}. Set
$\Pi_i:=P_i-p_i$.

The coupled scalar closure retains the same state functions $p_i$, but may give $P_i\ne p_i$ at zero dilatation when other scalar co-factors are active \cite[Section~3.2 and Appendix~A.5]{BotheRuggeriPaperI2026}. The functions $p_i$ are differentiable state functions and are not determined individually by the entropy potential. The selected entropy-flux representation yields the weighted Euler relation
\begin{equation}
 \rho s=\sum_i\left[\frac{\rho_i e_i+p_i}{T_i}-\frac{\rho_i\mu_i}{T_i}\right].
\label{eq:weighted-euler}
\end{equation}
It constrains only the weighted pressure combination and does not determine the individual $p_i$. With
\begin{equation}
 \phi_i:=\frac{\rho_i e_i+p_i-\rho_i\mu_i}{T_i},
\label{eq:entropy-transport-density}
\end{equation}
one has
\begin{equation}
 \rho s=\sum_i\phi_i,
 \qquad
 \rho s\bv+\bPhi=\sum_i\left(\phi_i\bv_i+\frac{\bq_i}{T_i}\right),
\label{eq:constituent-entropy-transport-form}
\end{equation}
where $\bPhi$ is the nonconvective entropy flux.

The sum in \eqref{eq:constituent-entropy-transport-form} is an
algebraic decomposition, not an assumption of constituent thermodynamic
additivity: each $\phi_i$ can still depend on the full mixture state.
The star will be reserved for the entropy-maximizing reconstruction in
Section~\ref{sec:thermal-reduction}.
   The entropy-flux selection and weighted Euler relation used in this decomposition are derived in
\cite{BotheRuggeriPaperI2026}. The fixed-flux compatibility
theorem below gives constituentwise entropy separability together with
compatible partial pressures.

 The noninteraction part of the thermodynamic diffusion co-factor is
\begin{equation}
 \bB_i:=\rho_i\grad\frac{\mu_i}{T_i}-\frac{1}{T_i}\grad p_i-(\rho_i e_i+p_i)\grad\frac{1}{T_i},
 \qquad \sum_i\bB_i=0.
\label{eq:raw-diffusion-cofactor}
\end{equation}

\subsection{Entropy production and constitutive relations}
\label{sec:entropy-principle}
The selected entropy-production representation is
\begin{align}
\zeta={}&\sum_i\frac{1}{T_i}\left(\bS_i^{\circ}:\bD_i^{\circ}-\Pi_i\Div\bv_i\right)
+\sum_i\bq_i\cdot\grad\frac{1}{T_i}
\label{eq:entropy-production}\\
&-\sum_i\bu_i\cdot\left[\bB_i+\frac{1}{T_i}\left(\bmom_i-\half r_i\bu_i\right)\right]
+\sum_i\frac{\varepsilon_i}{T_i}-\sum_aR_a\Acal_a,
\nonumber
\end{align}
where the quantities $\Acal_a$ are the temperature-weighted chemical reaction affinities, defined by
\begin{equation}
 \Acal_a:=\sum_iM_i\nu_i^a\frac{\mu_i}{T_i}.
\label{eq:classIII-affinity}
\end{equation}
Equivalently, the source pairing is
\begin{equation}
 \sum_i\frac{1}{T_i}\left[-\left(\mu_i-\half|\bu_i|^2\right)r_i-\bu_i\cdot\bmom_i+\varepsilon_i\right].
\label{eq:ruggeri-source-pairing}
\end{equation}
For the separated baseline laws below, the coefficients
$\eta_i^{\mathrm v},\lambda_i^{\mathrm v},\kappa_i,d_i^T,\bM,b_{ij}$
and the positive one-way reaction rates $R_a^f,R_a^b$ are prescribed
Galilean-invariant $C^1$ functions of the conservative state $\bV$
defined in \eqref{eq:conservative-variables}, with no dependence on
its derivatives or on independent stress and heat-flux variables.
In the isotropic class considered here, any velocity dependence is
through rotation-invariant combinations of the relative velocities
$\bu_i$. In the baseline of \cite[Section~6.2]{BotheRuggeriPaperI2026}, the $d_i^T$ are functions of the thermodynamic state alone. The one-way rates
satisfy \eqref{eq:reaction-closure}. The displayed gradient terms enter
through the constitutive laws themselves. More general local responses
are treated separately in \Cref{rem:mixed-degenerate-lift}.
The constituent viscous and heat-flux laws are
\begin{equation}
 \bS_i^{\circ}=2\eta_i^{\mathrm v}\bD_i^{\circ},
 \qquad \Pi_i=-\lambda_i^{\mathrm v}\Div\bv_i,
 \qquad \eta_i^{\mathrm v},\lambda_i^{\mathrm v}\ge0,
\label{eq:viscous-closure}
\end{equation}
\begin{equation}
 \bq_i-d_i^T\bu_i=\kappa_i\grad\frac{1}{T_i},
 \qquad \kappa_i\ge0,
 \qquad \sum_i d_i^T=0.
\label{eq:heat-closure}
\end{equation}
Here $d_i^T$ is the scalar thermo-diffusive coupling coefficient
(the superscript $T$ is a label, not a transpose), while $\kappa_i$
is the coefficient multiplying the reciprocal-temperature gradient.
The  condition $\sum_i d_i^T=0$ is a simplifying constitutive assumption of the present Class-III model, not a consequence of the entropy principle. At unequal temperatures, the replacement $d_i^T\mapsto d_i^T-y_i\sum_jd_j^T$ generally changes the individual heat fluxes. It is therefore not a normalization without loss of generality. See \mbox{
\cite[Section~6.2]{BotheRuggeriPaperI2026}}.
The  parameters $\eta_i^{\mathrm v}$ and $\lambda_i^{\mathrm v}$
denote, respectively, shear and bulk viscosity. The superscript
$\mathrm v$ distinguishes $\eta_i^{\mathrm v}$ from the constituent
entropy functions $\eta_i$ used later. $\lambda_i^{\mathrm v}$
is a scalar coefficient and is distinct from the bold vector
$\blambda$ used as a diffusion constraint multiplier.

At each point the relative velocities lie in $\Dcal_{\boldsymbol y}:=\{\boldsymbol a:\sum_i y_i\boldsymbol a_i=0\}$. The dual force space is the zero-sum space. Define the species-space projector $\bP_{\boldsymbol y}$ by $(\bP_{\boldsymbol y})_{ij}:=\delta_{ij}-y_j$, where $\delta_{ij}$ is the Kronecker symbol. Then
\begin{equation}
 \sum_i\boldsymbol a_i\cdot\boldsymbol c_i
 =\sum_i\boldsymbol a_i\cdot(\bP_{\boldsymbol y}^{\sf T}\boldsymbol c)_i
 \qquad(\boldsymbol a\in\Dcal_{\boldsymbol y}).
\label{eq:projection-properties2}
\end{equation}
Define
\begin{equation}
 \widetilde{\bB}_i:=\bB_i-d_i^T\grad\frac{1}{T_i}+\frac{1}{T_i}\left(\bmom_i-\half r_i\bu_i\right),
 \qquad \boldsymbol{\mathfrak b}:=\bP_{\boldsymbol y}^{\sf T}\widetilde{\boldsymbol B}.
\end{equation}
The projected Maxwell--Stefan closure is
\begin{equation}
 \boldsymbol{\mathfrak b}=-\bM\bu,
\label{eq:resistance-closure}
\end{equation}
where $\bM=\bM^{\sf T}\succeq0$ is the constituent resistance
matrix, acting componentwise on spatial vectors. Its row sums vanish.
For strict positivity of the resistance quadratic form on
$\Dcal_{\boldsymbol y}$, we additionally require that the kernel consist
only of vectors constant across constituents:
\begin{equation}
 \sum_{j=1}^n(\bM)_{ij}=0\quad (i=1,\ldots,n),
 \qquad
 \ker\bM=\{(c,\ldots,c)^{\sf T}:c\in\R\}.
\label{eq:resistance-properties}
\end{equation}
   Vanishing column sums ensure that $\bM$ maps into the zero-sum force space. For the symmetric constitutive matrix used here, this condition is equivalent to the vanishing row sums in \eqref{eq:resistance-properties}. Thus  $\bM\bu$ lies in the zero-sum force space    independently of  strict coercivity.
For the constitutive model with pairwise resistance,
\begin{equation}
 (\bM\bu)_i=\sum_{j\ne i}f_{ij}\rho_i\rho_j(\bu_i-\bu_j),
 \qquad f_{ij}=f_{ji}\ge0.
\label{eq:pairwise-resistance}
\end{equation}
Writing
\begin{equation}
 \bG_i:=\bB_i-d_i^T\grad\frac{1}{T_i}-\frac{r_i}{2T_i}\bu_i,
\end{equation}
the projected force relation and the corresponding momentum production are
\begin{align}
 \widetilde{\bB}_i&=-(\bM\bu)_i+y_i\blambda,
\label{eq:gauge-representation}\\
 \bmom_i&=T_i[-(\bM\bu)_i+y_i\blambda-\bG_i],
\label{eq:momentum-reconstruction}
\end{align}
where conservation of total momentum determines
\begin{equation}
 \blambda=\frac{\sum_jT_j[(\bM\bu)_j+\bG_j]}{\sum_jy_jT_j}.
\label{eq:momentum-multiplier}
\end{equation}

The pairwise thermal-relaxation law is
\begin{equation}
 \varepsilon_i^T=\sum_j b_{ij}\left(\frac{1}{T_i}-\frac{1}{T_j}\right),
 \qquad b_{ij}=b_{ji}\ge0,
 \qquad b_{ii}=0.
\label{eq:thermal-relaxation}
\end{equation}
It satisfies $\sum_i\varepsilon_i^T=0$, and its entropy production is
\begin{equation}
 \zeta_{\mathrm{relax}}=\sum_i\frac{\varepsilon_i^T}{T_i}
 =\half\sum_{i,j}b_{ij}\left(\frac{1}{T_i}-\frac{1}{T_j}\right)^2\ge0.
\label{eq:thermal-relaxation-dissipation}
\end{equation}
For strictly positive one-way rates, the logarithmic reaction law is
\begin{equation}
 \log\frac{R_a^f}{R_a^b}=-\frac{\gamma_a}{\mathcal R}\Acal_a,
 \qquad \gamma_a>0,
\label{eq:reaction-closure}
\end{equation}
where $\mathcal R$ denotes the universal gas constant (distinguished from the reaction-rate vector $\boldsymbol R$). The  choice $\gamma_a=1$ gives the local-detailed-balance normalization used in \mbox{
\cite{BotheRuggeriPaperI2026}}. Arbitrary $\gamma_a>0$ retains the entropy-admissibility property below. The  associated reaction contribution satisfies
\begin{equation}
 -R_a\Acal_a=\frac{\mathcal R}{\gamma_a}(R_a^f-R_a^b)\log\frac{R_a^f}{R_a^b}\ge0.
\label{eq:reaction-dissipation}
\end{equation}
For the specified constitutive model, the full objective energy exchange is $\varepsilon_i=\varepsilon_i^T$. An additional energy redistribution $\varepsilon_i^{\mathrm{add}}$, with $\sum_i\varepsilon_i^{\mathrm{add}}=0$, would contribute $\sum_i\varepsilon_i^{\mathrm{add}}/T_i$ to entropy production. Conservation of energy alone does not make that contribution vanish at unequal temperatures.
The resulting bulk entropy production is
\begin{align}
\zeta={}&\sum_i\frac{2\eta_i^{\mathrm v}}{T_i}|\bD_i^{\circ}|^2
+\sum_i\frac{\lambda_i^{\mathrm v}}{T_i}(\Div\bv_i)^2
+\sum_i\kappa_i\left|\grad\frac{1}{T_i}\right|^2
\label{eq:closed-zeta}\\
&+\sum_i\bu_i\cdot(\bM\bu)_i
+\half\sum_{i,j}b_{ij}\left(\frac{1}{T_i}-\frac{1}{T_j}\right)^2
-\sum_aR_a\Acal_a\ge0.
\nonumber
\end{align}

\par\begingroup
In the process-resolved alternative of \cite[Section~6.6]{BotheRuggeriPaperI2026},
a molar process rate $W_\ell=W_\ell^f-W_\ell^b$ has prescribed reaction
increments $\omega_a^\ell$ and objective-energy increments $h_i^\ell$,
with opposite increments in the forward and reverse directions
and $\sum_i h_i^\ell=0$. The corresponding sources and rate ratio are
\[
 R_a=\sum_\ell\omega_a^\ell W_\ell,\qquad
 \varepsilon_i^{\mathrm{proc}}=\sum_\ell h_i^\ell W_\ell,\qquad
 \log\frac{W_\ell^f}{W_\ell^b}
 =\frac{1}{\mathcal R}\left(\sum_i\frac{h_i^\ell}{T_i}
                         -\sum_a\omega_a^\ell\Acal_a\right).
\]
This replaces the corresponding separated reaction and energy-exchange
mechanisms. The assigned energy increments must be included in the
entropy change even when their constituent sum vanishes. If reactive
momentum is also resolved, admissibility requires the complete joint
source pairing described in Section~7 of that paper.
\par\endgroup

On a common-temperature manifold define
\begin{equation}
 \bq^{\mathrm h}:=\sum_i[\bq_i+(\rho_i e_i+p_i)\bu_i],
 \qquad \boldsymbol\tau_i:=\bS_i+p_i\bI.
\label{eq:thermal-flux-and-irreversible-stress}
\end{equation}
The exact physical nonconvective total-energy flux satisfies
\begin{equation}
 \bq^{\mathrm{tot}}=\bq^{\mathrm h}-\sum_i\boldsymbol\tau_i\bu_i+\half\sum_i\rho_i|\bu_i|^2\bu_i.
\label{eq:physical-versus-thermal-energy-flux}
\end{equation}
If $\widehat{\bq}:=\bq^{\mathrm{tot}}-\sum_i\mu_i\bj_i$, then
\begin{equation}
 \bPhi=\frac{\widehat{\bq}}T+\frac{1}T\sum_i\boldsymbol\tau_i\bu_i-\frac{1}{2T}\sum_i\rho_i|\bu_i|^2\bu_i.
\label{eq:physical-shifted-entropy-flux}
\end{equation}

\subsection{Chemically inherited momentum}
For later use in the mechanical reduction, the chemically inherited momentum law of the Class-II construction~\cite{BotheDreyer2015} is recorded explicitly:
\begin{equation}
 \bfv_i^C-r_i\bv_i=-\sum_jC^{\mathrm{chem}}_{ij}(\bv_i-\bv_j),
\label{eq:BD-chemical-force}
\end{equation}
For positive molar masses $M_i$ and reaction paths with
$\sum_k\alpha_k^a>0$, mass conservation gives
$\sum_k\alpha_k^aM_k=\sum_k\beta_k^aM_k>0$.
The chemical transfer matrix is
\begin{equation}
 C^{\mathrm{chem}}_{ij}=\sum_a\frac{M_iM_j}{\sum_k\alpha_k^aM_k}
 \left(R_a^f\beta_i^a\alpha_j^a+R_a^b\alpha_i^a\beta_j^a\right).
\label{eq:chemical-matrix}
\end{equation}
Stoichiometric mass conservation gives $r_i=\sum_j(C^{\mathrm{chem}}_{ij}-C^{\mathrm{chem}}_{ji})$. In barycentric variables,
\[
 \bmom_i^C=-\sum_jC^{\mathrm{chem}}_{ij}(\bu_i-\bu_j)+r_i\bu_i.
\]

 This resolved chemical-momentum law is an alternative to the aggregate source closure in Section~\ref{sec:entropy-principle}, not an additional term to be appended to it unchanged. Its associated energy exchange and complete reactive entropy production must be specified consistently, as in \mbox{
\cite{BotheRuggeriPaperI2026}}.

 The complete derivations, event-resolved alternatives, coupled constitutive relations, and internal-state ionization application are given in \cite{BotheRuggeriPaperI2026}. The present paper uses only the identities and constitutive relations summarized above.

\section{Complete main field, symmetric generators, and principal subsystems}
\label{sec:main-field}

The general Class-III model recalled in \Cref{sec:parent-preliminaries}
obeys the entropy principle and the mixture-wide compatibility
\eqref{eq:raw-diffusion-cofactor}. It does not require the individual
thermodynamic diffusion co-factors to vanish. Before writing the enlarged
first-order system, we specify and characterize the additional
\emph{fixed-flux integrability} restriction used for the symmetric
generator in this paper. The constitutive relations and balance laws
remain those of    \mbox{
\cite{BotheRuggeriPaperI2026}}, but the main-field generator and its standard
Euler principal subsystems will be asserted only within this subclass.
Thermodynamic projections in \Cref{sec:thermal-reduction} are treated
separately and need not satisfy this restriction.

\paragraph{Entropy-sign and generator convention.}
Recall the convention $h^0=-\rho s$. On the conservative convexity domain,
the entropy balance is written
\begin{equation}
  \partial_\alpha h^\alpha=g=-\zeta\le0.
\label{eq:entropy-sign-convention}
\end{equation}
Here $g$ is the source of the mathematical entropy law and
$\zeta=-g$ is the physical entropy production.
Constituent and reaction indices are summed only when a sum is displayed.
The four-vector $(h^0,h^1,h^2,h^3)$ collects the mathematical entropy
density and its three flux components. Its balance is called
supplementary because it follows from the field equations rather than
providing an additional independent equation. The generating potentials are denoted throughout by $h'^{\alpha}$,
as in the Introduction. The symbols $h_E'^{\alpha}$ denote
only the Euler restriction of these potentials. A bar denotes
restriction of a generator to a constant main-field section.

\paragraph{Complete state and temporal flux.}
For each constituent define the conservative block
\begin{equation}
 \bV_i=(\rho_i,\rho_i\bv_i,E_i),\qquad
 E_i:=\rho_i e_i+\half\rho_i|\bv_i|^2,\qquad
 \bV=(\bV_i)_{i=1}^n\in\R^{5n}.
\label{eq:conservative-variables}
\end{equation}
The constitutive block consists of the symmetric irreversible stresses
$\boldsymbol\tau_i=\bS_i+p_i\bI$ and the physical constituent heat
fluxes $\bq_i$:
\begin{equation}
 \bZ=(\boldsymbol\tau_i,\bq_i)_{i=1}^n\in\R^{9n},\qquad
 \bU=(\bV,\bZ)\in\R^{14n},\qquad
 \bF^0(\bU)=\begin{pmatrix}\bV\\\boldsymbol0_{9n}\end{pmatrix}.
\label{eq:complete-mixture-state}
\end{equation}
Symmetric tensors have six independent components. Thus $M=5n$ and
$N=14n$ in the notation of the Introduction. No constitutive unknown is
set to zero in this definition. Since $h^0=-\eta(\boldsymbol z(\bV),
\boldsymbol\rho)$ is independent of $\bZ$, the temporal multiplier
identity determines only the conservative block of the main field.
The spatial identities must determine the complementary block.

\paragraph{Conservative entropy multipliers.}
For the reversible-flux integrability test, we first determine the $5n$
conservative entropy multipliers:
\begin{equation}
  \vp := \frac{\partial h^0}{\partial\bV}
  =\left(\frac{\mu_i-\half|\bv_i|^2}{T_i},
           \frac{\bv_i}{T_i},-\frac{1}{T_i}\right)_{i=1}^n
  =\bigl(\Lambda_i^\rho,\boldsymbol\Lambda_i^{\bv},\Lambda_i^E\bigr)_{i=1}^n.
\label{eq:U-prime-main-field}
\end{equation}
They follow from the common thermodynamic entropy of    \mbox{
\cite{BotheRuggeriPaperI2026}}, without
any prior assumption that the complete dissipative main field exists.
Their explicit constituent components follow from
\eqref{eq:entropy-conjugates} and the definition of $E_i$.

\subsection{Constitutive subclass: conservative convexity and fixed-flux integrability}
\label{subsec:euler-rigidity}

We first test the \emph{prescribed reversible constituent fluxes},
which will later appear upon restriction of the full dissipative generator.
This is a test of the constitutive subclass, not an assumption that the
starting physical model is an Euler system. Put
$\bS_i=-p_i\bI$ and $\bq_i=\boldsymbol0$ for this flux test.
Let $\boldsymbol F_E^k$ collect the constituent mass, momentum, and
total-energy fluxes in the $k$th spatial direction:
\[
 \boldsymbol F_{E,i}^k
 =\left(\rho_i v_{i,k},\,
        \rho_i\bv_i v_{i,k}+p_i\boldsymbol e_k,\,
        (E_i+p_i)v_{i,k}\right),
 \qquad \boldsymbol F_E^k=(\boldsymbol F_{E,i}^k)_{i=1}^n.
\]
Here $\boldsymbol e_k$ is the $k$th Cartesian basis vector.
The subscript $E$ specifies these prescribed Euler fluxes. The full
$14n$-component spatial flux of the enlarged system, introduced later,
is denoted by $\boldsymbol F^k$. Its balance and auxiliary blocks
are displayed in \eqref{eq:full-stored-and-spatial-flux-blocks}.
To recover these \emph{same} Euler fluxes as a principal subsystem of a
symmetric generator, we require scalar spatial
potentials $h_E'^k$ satisfying
\begin{equation}
  \boldsymbol F_E^k(\vp)
  =\frac{\partial h_E'^{k}}{\partial \vp},
  \qquad k=1,2,3.
\label{eq:main-field-flux-potential-condition}
\end{equation}
The temporal Euler potential is the Legendre transform
$h_E'^0(\vp)=\vp\cdot\bV-h^0(\bV)$, so that
$\partial h_E'^0/\partial \vp=\bV$.
These four scalar functions are the generators: their first derivatives
recover the densities and fluxes, and their Hessians are the symmetric
coefficient matrices. With a positive-definite conservative
entropy Hessian,
\eqref{eq:main-field-flux-potential-condition} gives the local
symmetric-hyperbolic Euler formulation.

The advantage of this formulation is that the coefficient of the
time derivative is positive definite and the spatial coefficient
matrices are symmetric. At each admissible state this gives real Euler
characteristic speeds and a positive quadratic form for energy
estimates. It is the structure used in local smooth-solution theory,
subject to the further regularity and source hypotheses discussed in
\Cref{rem:degenerate-qualitative-scope} below. It does not prevent shock
formation or establish global smoothness. The same positive quadratic form    restricts to the retained subspaces of  the principal subsystems.

The two requirements are established separately below.
The first result proves conservative convexity from the thermodynamic
hypotheses in \eqref{eq:negative-definite-entropy-hessian} and~\eqref{eq:positive-temperature-domain},
while the rigidity theorem characterizes spatial flux integrability for
the prescribed constituent Euler fluxes.

\begin{proposition}[Positive-definite conservative entropy Hessian]
\label[proposition]{prop:general-conservative-convexity}
Assume \eqref{eq:negative-definite-entropy-hessian} and~\eqref{eq:positive-temperature-domain} at an interior state with $\rho_i>0$. Then the mathematical entropy
\[
  h^0(\bV)=-\rho s(\bV)=-\eta(\boldsymbol z(\bV),\boldsymbol\rho)
\]
has
\begin{equation}
  D_{\bV}^2h^0\succ0
\label{eq:positive-conservative-entropy-hessian}
\end{equation}
at that state. Consequently $h^0$ is locally strictly convex in the conservative variables from \eqref{eq:conservative-variables}. More generally it is strictly convex on every convex conservative-coordinate subset on which the stated hypotheses hold. No global convexity of a possibly nonconvex admissible state domain is asserted.
\end{proposition}

\begin{proof}
For an affine conservative variation $\delta\bV_i=(\delta\rho_i,\delta(\rho_i\bv_i),\delta E_i)$,
\begin{equation}
  \delta z_i
  =\delta E_i-\bv_i\cdot\delta(\rho_i\bv_i)
   +\half|\bv_i|^2\delta\rho_i,
  \qquad
  \delta^2z_i
  =-\frac{1}{\rho_i}\left|\delta(\rho_i\bv_i)-\bv_i\delta\rho_i\right|^2.
\label{eq:thermal-energy-second-variation}
\end{equation}
Hence
\begin{equation}
  D_{\bV}^2h^0[\delta\bV,\delta\bV]
  =-D^2\eta[(\delta\boldsymbol z,\delta\boldsymbol\rho),(\delta\boldsymbol z,\delta\boldsymbol\rho)]
  +\sum_i\frac{1}{\rho_iT_i}
  \left|\delta(\rho_i\bv_i)-\bv_i\delta\rho_i\right|^2.
\label{eq:general-conservative-convexity}
\end{equation}
The first term is positive definite in $(\delta\boldsymbol z,\delta\boldsymbol\rho)$ and the second is nonnegative. If both vanish, then $\delta\boldsymbol z=\delta\boldsymbol\rho=0$ and $\delta(\rho_i\bv_i)=0$ for every $i$, which also gives $\delta E_i=0$. Thus the quadratic form is positive for every nonzero conservative variation.
\end{proof}

\par\begingroup
The criterion is also necessary at positive densities. Indeed, if
$D_{\bV}^2h^0\succ0$, choose
$\delta(\rho_i\bv_i)=\bv_i\delta\rho_i$ and arbitrary
$(\delta\boldsymbol z,\delta\boldsymbol\rho)$ in
\eqref{eq:general-conservative-convexity} to obtain $-D^2\eta\succ0$.
Choosing instead $\delta\boldsymbol z=\delta\boldsymbol\rho=0$
and a nonzero momentum increment for a single constituent gives
$\partial_{z_i}\eta=1/T_i>0$. Thus, for $\rho_i>0$,
positive definiteness of the conservative entropy Hessian is equivalent to the two pointwise
thermodynamic conditions
\eqref{eq:negative-definite-entropy-hessian} and
\eqref{eq:positive-temperature-domain}.
\par\endgroup

\begin{theorem}[Fixed-flux RET rigidity and characterization of the Euler flux potentials]
\label[theorem]{prop:euler-flux-rigidity}
Assume \eqref{eq:negative-definite-entropy-hessian} and~\eqref{eq:positive-temperature-domain} to hold, $\rho_i>0$, and $C^1$ partial thermodynamic pressures $p_i$. Work locally on a thermodynamic product neighbourhood in the constituent variables, shrinking the chart when necessary, times an open product neighbourhood of independently variable constituent velocities. The additivity conclusion below is local and does not assert global factorization on an arbitrary connected state domain. Put
\begin{equation}
  \vartheta_i:=\frac{1}{T_i},
  \qquad
  \chi_i:=-\frac{\mu_i}{T_i},
  \qquad
  \omega_i:=\vartheta_i\,\dd p_i+(z_i+p_i)\,\dd\vartheta_i+\rho_i\,\dd\chi_i.
\label{eq:individual-gibbs-duhem-one-form}
\end{equation}
Here $\bF_E^k$ is the $5n$-component flux of the prescribed
constituent Euler balance equations and $\vp$ is the corresponding
$5n$-component multiplier block defined in \eqref{eq:U-prime-main-field}.
The test concerns differentiation with respect to those conservative
multipliers, before introducing the auxiliary coordinates of the
complete main field.
Then the spatial Euler flux-potential condition \eqref{eq:main-field-flux-potential-condition} holds if and only if
\begin{equation}
  \omega_i=0,
  \qquad i=1,\ldots,n.
\label{eq:individual-gibbs-duhem}
\end{equation}
When these identities hold, the spatial potentials may be chosen as
\begin{equation}
  h_E'^{k}=\sum_i\frac{p_i}{T_i}v_{i,k},
  \qquad k=1,2,3.
\label{eq:general-euler-spatial-generator}
\end{equation}
Moreover, on the stated local neighbourhood, \eqref{eq:individual-gibbs-duhem} implies
\begin{equation}
  \eta(\boldsymbol z,\boldsymbol\rho)
  =\sum_i\eta_i(z_i,\rho_i)+C,
  \qquad
  p_i=p_i(z_i,\rho_i),
\label{eq:ret-additive-entropy}
\end{equation}
where the constituent functions are normalized by
$\eta_i=\vartheta_i(z_i+p_i)+\rho_i\chi_i$, and $C$ is a constant.
With this normalization, the weighted Euler relation
\eqref{eq:weighted-euler} implies $C=0$.
\end{theorem}

\begin{proof}
Recall the transport densities in \eqref{eq:entropy-transport-density},
and write
\begin{equation}
  \phi_i=\vartheta_i(z_i+p_i)+\rho_i\chi_i,
  \qquad \widetilde h^k:=-\sum_i\phi_i v_{i,k}.
\label{eq:constituent-legendre-density}
\end{equation}
A direct calculation using the Euler fluxes $\boldsymbol F_E^k$
and the conservative multipliers \eqref{eq:U-prime-main-field},
together with the entropy conjugates \eqref{eq:entropy-conjugates}, gives
\begin{equation}
  \vp\mathbin{\boldsymbol\cdot}\dd\boldsymbol F_E^k
  -\dd\widetilde h^k
  =\sum_i v_{i,k}\,\omega_i.
\label{eq:euler-integrability-residual}
\end{equation}
Suppose first that a flux potential $h_E'^{k}$ exists. Define the associated supplementary mathematical-entropy flux by
\[
  h^k:=\vp\mathbin{\boldsymbol\cdot}\boldsymbol F_E^k-h_E'^{k}.
\]
Because $\dd h_E'^{k}=\boldsymbol F_E^k\mathbin{\boldsymbol\cdot}\dd \vp$, the Legendre identity gives
$\dd h^k=\vp\mathbin{\boldsymbol\cdot}\dd\boldsymbol F_E^k$. Hence
\[
  \dd(h^k-\widetilde h^k)=\sum_i v_{i,k}\omega_i.
\]
The right-hand side contains only thermodynamic differentials, with no terms proportional to $\dd v_{i,l}$. Hence all velocity derivatives of $h^k-\widetilde h^k$ vanish, so this difference depends only on the thermodynamic variables. Its differential must therefore be unchanged when the velocities vary at fixed thermodynamic state. Varying one $v_{i,k}$ at a time in the right-hand side changes it by $\delta v_{i,k}\,\omega_i$, which forces $\omega_i=0$ for every $i$.

Conversely, if \eqref{eq:individual-gibbs-duhem} holds, direct differentiation gives
\[
  \dd\!\left(\sum_i\vartheta_i p_i v_{i,k}\right)
  =\boldsymbol F_E^k\mathbin{\boldsymbol\cdot}\dd \vp,
\]
which proves \eqref{eq:general-euler-spatial-generator}.
The prescribed Euler fluxes are $C^1$ in the local
main-field coordinates. Thus this differential identity makes
$h_E'^{k}$ a $C^2$ potential under the stated $C^1$ pressure
assumption.
Finally, \eqref{eq:individual-gibbs-duhem} and~\eqref{eq:constituent-legendre-density} yield
\begin{equation}
  \dd\phi_i=\vartheta_i\,\dd z_i+\chi_i\,\dd\rho_i.
\label{eq:constituent-additive-differential}
\end{equation}
Thus $\phi_i$ depends locally only on $(z_i,\rho_i)$. Consequently
$\vartheta_i=\partial_{z_i}\phi_i$ and $\chi_i=\partial_{\rho_i}\phi_i$ also depend locally only on $(z_i,\rho_i)$. The Gibbs relation gives
$\dd\eta=\sum_i(\vartheta_i\dd z_i+\chi_i\dd\rho_i)=\dd\sum_i\phi_i$. Taking $\eta_i:=\phi_i$, connectedness of the local coordinate neighbourhood gives \eqref{eq:ret-additive-entropy}. Since the weighted Euler relation is precisely $\eta=\sum_i\phi_i$, it fixes $C=0$. The pressure statement then follows from $\vartheta_i p_i=\phi_i-z_i\vartheta_i-\rho_i\chi_i$ and $\vartheta_i>0$.
\end{proof}

\begin{corollary}[Diffusion co-factors and temporal Euler generator]
\label[corollary]{cor:ret-compatible-classIII}
Under the hypotheses of \Cref{prop:euler-flux-rigidity} and the
weighted Euler relation \eqref{eq:weighted-euler}, the subclass
\eqref{eq:individual-gibbs-duhem} has vanishing thermodynamic diffusion
co-factors,
\begin{equation}
  \bB_i\equiv\boldsymbol0,
  \qquad i=1,\ldots,n,
\label{eq:ret-B-vanishes}
\end{equation}
and its temporal Euler generator is
\begin{equation}
  h_E'^{0}=\sum_i\frac{p_i}{T_i}.
\label{eq:general-euler-temporal-generator}
\end{equation}
These are consequences of the fixed-flux compatibility theorem, not
additional hypotheses. They concern the conservative Euler block.
The dissipative generator also uses the auxiliary constitutive equations.
In the ideal-gas example below, entropy of mixing is accounted for by
the terms proportional to $-\rho_i\log\rho_i$ in the constituent
entropy densities.
\end{corollary}

\begin{remark}[Relation to simple mixtures in the sense of Ingo M\"uller]
\label[remark]{rem:mueller-simple-mixtures}
The converse requires compatible partial pressures: if
$\eta=\sum_i\eta_i(z_i,\rho_i)$ and
$\vartheta_i p_i=\eta_i-z_i\vartheta_i-\rho_i\chi_i$, then
$\phi_i=\eta_i$ and \eqref{eq:individual-gibbs-duhem} holds. Thus
additive entropy alone, with otherwise unspecified $p_i$, is insufficient.

\par\begingroup
For example, starting from compatible additive binary data, replace
$p_1$ by $p_1+T_1a$ and $p_2$ by $p_2-T_2a$, where $a$ is a smooth
nonconstant thermodynamic function. The weighted Euler relation is
unchanged, whereas the new one-forms are $\omega_1=\dd a$ and
$\omega_2=-\dd a$. Hence even closed constituent one-forms need not
satisfy the fixed-flux compatibility condition.
\par\endgroup

The local separability in \eqref{eq:ret-additive-entropy} is the
multi-temperature  counterpart of the thermodynamic structure of simple
fluid mixtures in the sense of M\"uller~\cite{Mueller1973SimpleMixtures}.
See also the explicit common-temperature characterization in
\cite[Section~8.4]{MuellerMueller2023}. Indeed, with
$s_i:=\eta_i/\rho_i$, local inversion of the constituent caloric relation
gives the specific Helmholtz free energy
$\psi_i=e_i-T_i s_i$ as a function of $(T_i,\rho_i)$ alone. Consequently,
\[
 \psi_i=\psi_i(T_i,\rho_i),\qquad
 p_i=p_i(T_i,\rho_i),\qquad
 \mu_i=\mu_i(T_i,\rho_i).
\]
At common temperature the mixture Helmholtz free-energy density is
$\rho\psi=\sum_i\rho_i\psi_i(T,\rho_i)$, recovering that
simple-mixture thermodynamic structure.

Here simplicity is a conclusion of fixed-flux integrability,
not an additional assumption in the parent model of \cite{BotheRuggeriPaperI2026}. The
comparison concerns local thermodynamic potentials and pressures, not
an identification of all gradient-dependent constitutive assumptions
in the two theories. Nor does the theorem alone impose ideal-gas
equations of state or suppress reactions, momentum exchange, thermal
relaxation, or distinct constituent velocities and temperatures.
These mechanisms remain possible through the balances and their
entropy-admissible constitutive terms. The restrictive point is the
absence of cross-constituent dependence in the local thermodynamic
potentials. Independently variable constituent velocities prevent
cancellation of the Gibbs--Duhem residuals between    constituent equations, so each
$\omega_i$ must vanish separately. Failure of this test rules out the
specified flux-potential representation. It is not, by itself, a
failure of the second law or a proof of ill-posedness.

\par\begingroup
The integrability test is also independent of constant constituent
energy references. Under $z_i^\sharp=z_i+\rho_i\Delta e_i$ and
$\chi_i^\sharp=\chi_i-\vartheta_i\Delta e_i$, with $p_i$ and
$\vartheta_i$ unchanged, direct substitution gives $\omega_i^\sharp=\omega_i$.
\par\endgroup
\par\begingroup
For the reacting constitutive model, the same change of reference also
requires $\varepsilon_i^\sharp=\varepsilon_i+r_i\Delta e_i$.
Preservation of zero total energy production for the full independently
admissible reaction network requires
$\boldsymbol\Gamma^{\sf T}\Delta\boldsymbol e=0$.
For resolved processes the corresponding increments are
$h_i^{\ell,\sharp}=h_i^\ell+\Delta e_i M_i\sum_a\nu_i^a\omega_a^\ell$.
Keeping the assigned reaction energy equal to zero after such a shift
generally changes the separated Class-III closure at unequal temperatures
\cite[Remark~2.1 and Section~6.6]{BotheRuggeriPaperI2026}.
\par\endgroup
\end{remark}

\begin{example}[Calorically perfect ideal-gas mixture]
\label[example]{ex:ideal-hyperbolicity}
Let
\begin{equation}
  e_i=e_i^0+c_{V,i}T_i,
  \qquad
  p_i=\rho_iR_iT_i,
  \qquad
  s_i=s_i^0+c_{V,i}\log\frac{T_i}{T_i^0}
        -R_i\log\frac{\rho_i}{\rho_i^0},
\label{eq:ideal-gas-realization}
\end{equation}
with $c_{V,i}>0$,    $R_i=\mathcal R/M_i>0$ the specific gas constant of constituent $i$, and $\rho s=\sum_i\rho_i s_i$. Each constituent satisfies its own Gibbs--Duhem identity and hence \eqref{eq:individual-gibbs-duhem}. Therefore the general spatial generator
\eqref{eq:general-euler-spatial-generator} reduces to
$h_E'^k=\sum_i\rho_iR_i v_{i,k}$ for $k=1,2,3$. Proposition~\ref{prop:general-conservative-convexity} supplies
a positive-definite conservative entropy Hessian, so the Class-III Euler principal part is symmetric hyperbolic on the positive-density, positive-temperature state domain.
\end{example}

\begin{corollary}[Adding auxiliary variables does not remove the fixed-flux obstruction]
\label[corollary]{cor:no-go-nonadditive-generator}
Let $\widetilde h'^k(\vp,\wprime)$, $k=1,2,3$, be $C^2$ spatial generator
potentials depending on additional main-field coordinates $\wprime$.
The argument applies to any admissible auxiliary block. In the
specific $14n$-field mixture studied below, this block consists of
$(\boldsymbol\Sigma,\bQ)$. Suppose that, at a fixed value
$\wprime=\wprime_0$, their derivatives with respect to $\vp$ reproduce
the prescribed Class-III Euler mass, momentum, and energy fluxes.
Then $h_E'^k(\vp):=\widetilde h'^k(\vp,\wprime_0)$ is itself an Euler flux
potential. Under the hypotheses of \Cref{prop:euler-flux-rigidity},
the constituentwise identities \eqref{eq:individual-gibbs-duhem} and
local additivity \eqref{eq:ret-additive-entropy} follow.
In particular, auxiliary variables with no time derivatives do not
remove this obstruction when their fixed-value restriction leaves the
Euler fluxes unchanged.
\end{corollary}

\begin{remark}[Rewriting the equations versus changing their fluxes]
\label[remark]{rem:representation-versus-modification}
An invertible reparametrization of the physical state or a constant invertible
linear combination of the balance equations preserves the entropy-pair
identity $\dd h^k=\vp\cdot\dd\boldsymbol F_E^k$ when all quantities
are transformed consistently. \Cref{cor:no-go-nonadditive-generator}
likewise applies to auxiliary extensions whose restriction to a
constant auxiliary section reproduces the prescribed Euler fluxes.
These statements concern the density and flux identities, not merely
whether two systems have the same solutions.
\end{remark}

\paragraph{From the constitutive subclass to the complete dissipative system.}
We now return to the \emph{full} viscous, heat-conducting, chemically
reacting mixture, with its original balance and constitutive fluxes.
The following first-order equations display the physical model of
   \mbox{
\cite{BotheRuggeriPaperI2026}}. Their fluxes and constitutive equations  can be written for
the general parent class, but the    source representation without spatial derivatives  and complete
symmetric generator established below use precisely the fixed-flux
subclass characterized above. In particular, no such generator is
claimed here for arbitrary nonzero diffusion co-factors.

\paragraph{The first-order Fourier--Navier--Stokes form.}
Displaying the balance and constitutive equations together makes the
first-order structure explicit. In \cite[Section~1, equations~(1.1)
and~(1.5)]{Ruggeri1983Viscous}, the single-fluid system has 14 unknowns:
five fluid variables, six independent components of the symmetric
viscous stress, and three components of the heat flux. The present construction uses the \emph{classical} NSF first-order
formulation in Section~1 of that paper, not the distinct hyperbolic
relaxation theory developed subsequently. The system below reproduces
that first-order NSF structure for a single constituent without
mixture exchange. It retains separate
deviatoric, spherical, and heat-conduction relations. With the variables just defined, we have
\begin{equation}
\begin{aligned}
 \partial_t\rho_i+\Div(\rho_i\bv_i)
  &=r_i,\\
 \partial_t(\rho_i\bv_i)
  +\Div(\rho_i\bv_i\otimes\bv_i+p_i\bI-\boldsymbol\tau_i)
  &=\rho_i\bb_i+\bfv_i,\\
 \partial_t E_i
  +\Div\big[(E_i+p_i)\bv_i-\boldsymbol\tau_i\bv_i+\bq_i\big]
  &=\rho_i\bb_i\cdot\bv_i+k_i,\\
 -\frac{1}{2}\operatorname{Div}\mathbb A_i
  &=-\frac{1}{2\eta_i^{\mathrm v}}\bS_i^{\circ},\\
 -\Div\bv_i
  &=\frac{1}{\lambda_i^{\mathrm v}}\Pi_i,\\
 -\grad(1/T_i)
  &=-\frac{1}{\kappa_i}(\bq_i-d_i^T\bu_i).
\end{aligned}
\label{eq:acta-FNS-first-order-form}
\end{equation}
For the third-order tensor $\mathbb A_i$ in the deviatoric row,
\begin{equation}
 (\mathbb A_i)_{abk}
 :=\delta_{ak}(v_i)_b+\delta_{bk}(v_i)_a
   -\frac{2}{3}\delta_{ab}(v_i)_k,
 \qquad (\operatorname{Div}\mathbb A_i)_{ab}
 :=\partial_k(\mathbb A_i)_{abk}=2(D_i^\circ)_{ab}.
\label{eq:acta-third-order-velocity-tensor}
\end{equation}
Here $\operatorname{Div}$ contracts the last tensor index, and repeated
spatial indices are summed. The normalized constitutive rows in
\eqref{eq:acta-FNS-first-order-form} are exactly the Newton deviatoric,
bulk-viscous, and thermo-Fourier equations, with five, one, and three
scalar components, respectively. The constant row scalings specify
the normalization. The corresponding multiplier identities
\eqref{eq:intro-main-field-multipliers} are verified below.
The state-dependent transport coefficients enter the
constitutive source terms. Since the constitutive rows contain no time derivatives,
this is a $14n$-variable first-order differential--algebraic system.
The normalization assumes strictly positive viscosity and heat
conductivity. A mixed extension for vanishing coefficients is discussed
in \Cref{rem:mixed-degenerate-lift}.
The sources $r_i$, $\bfv_i$ and $k_i$ are those recalled in Section~\ref{sec:parent-preliminaries}.
For the general parent the closure for $\bfv_i$ can still involve gradients.
On the subclass \eqref{eq:individual-gibbs-duhem}, the diffusion
co-factors vanish by \eqref{eq:ret-B-vanishes}, and the remaining
sources acquire the required    dependence on the enlarged state without spatial derivatives, as verified in
\Cref{prop:degenerate-acta-generator}. The term $d_i^T\bu_i$ is the
specified thermo-diffusive contribution to the heat flux. The source closures and restrictions on reactive momentum transfer are
those of \cite{BotheRuggeriPaperI2026}. No new exchange laws are imposed here.
Eliminating the nine constitutive unknowns per constituent recovers the
second-order viscous    and  heat-conducting system. The additional fixed-flux integrability conditions were established
in \Cref{subsec:euler-rigidity}. We now verify the complete
main-field identities and the dissipative symbol on that subclass.

\subsection{Main-field components, generators, and the entropy identity}
\label{subsec:complete-field-and-generator}
We now follow the construction based on the supplementary law rather
than introducing only the conservative entropy gradient. For the $14n$-equation first-order system just displayed,
we use $\Lambda_i^\rho$, $\boldsymbol\Lambda_i^{\bv}$ and
$\Lambda_i^E$ for the conservative multipliers, and
$\boldsymbol\Sigma_i$, $\bQ_i$ for the stress and heat-flux multipliers.
Together these are the components of the \emph{complete} main field
after the constant recombination of the six stress rows of
\eqref{eq:acta-FNS-first-order-form} into one symmetric-tensor equation,
as detailed in the proof below.
For the specific normalized constitutive rows written above, the
identities~\eqref{eq:intro-main-field-multipliers} determine the
multipliers when the indicated rank conditions hold. An alternative
choice of rows requires a consistent dual transformation of the
auxiliary multipliers and the source pairing. The explicit expressions are given below, followed
immediately by the generating functions and their verification.

\paragraph{Complete main field: conservative and constitutive multipliers.}
For the constituent total-energy density $E_i$ introduced above,
\begin{equation}
  \dd(\rho_i e_i)
  =\dd E_i-\bv_i\cdot\dd(\rho_i\bv_i)
  +\half|\bv_i|^2\dd\rho_i.
\label{eq:thermal-total-differential}
\end{equation}
From \eqref{eq:entropy-conjugates} and~\eqref{eq:thermal-total-differential}, the conservative multipliers are the first three components of \eqref{eq:complete-main-field-components} below. They are not, by themselves, the complete main field.

The conservative multiplier block $\vp$, already fixed by the entropy
density in \eqref{eq:U-prime-main-field}, supplies only the first $5n$
components of the complete main field. The remaining components must
satisfy the spatial identities. They are not determined by $h^0$ alone.
Following the zero-storage construction of \cite{Ruggeri1983Viscous}, we introduce
\begin{equation}
  \boldsymbol\tau_i=\bS_i+p_i\bI=\bS_i^{\circ}-\Pi_i\bI,
  \qquad \boldsymbol\Sigma_i:=\frac{\boldsymbol\tau_i}{T_i},
\label{eq:extended-stress-main-field}
\end{equation}
and    obtain  the \emph{complete} main field:
\begin{equation}
  \bU'=(\vp,\wprime),\qquad
  \wprime=(\boldsymbol\Sigma,\bQ).
\label{eq:acta-enlarged-main-field}
\end{equation}
Here $\boldsymbol\Sigma=(\boldsymbol\Sigma_i)_{i=1}^n$ and
$\bQ=(\bQ_i)_{i=1}^n$. In the chosen normalization
$\bQ_i=\bq_i$. The conservative multipliers and the
auxiliary components are explicitly
\begin{equation}
 \Lambda_i^\rho=\frac{\mu_i-\frac{1}2|\bv_i|^2}{T_i},\quad
 \boldsymbol\Lambda_i^{\bv}=\frac{\bv_i}{T_i},\quad
 \Lambda_i^E=-\frac{1}{T_i},\quad
 \boldsymbol\Sigma_i=\frac{\bS_i^\circ-\Pi_i\bI}{T_i},\quad
 \bQ_i=\bq_i.
 \label{eq:complete-main-field-components}
\end{equation}
\paragraph{Notation.}
The physical state is $\bU=(\bV,\bZ)$, where $\bV$ comprises
conservative densities and $\bZ$ comprises stress and heat flux.
Its complete dual main field is \emph{always}
$\bU'=(\vp,\wprime)$, with $\vp$ conjugate to $\bV$
and $\wprime=(\boldsymbol\Sigma,\bQ)$ the auxiliary multiplier block.
The primes label dual coordinates. $\bZ$ and $\wprime$
are not the same variables. In particular, the lowercase
$\boldsymbol z=(\rho_i e_i)_{i=1}^n$ is the vector of partial
thermal-energy densities, whereas uppercase $\bZ$ is the block
of physical constitutive unknowns. The four scalar generating potentials
are $h'^\alpha(\bU')$. $h_E'^\alpha(\vp)$ are their
Euler restrictions, not another definition of the main field.
Later, $\Uret$ and $\Ufix$ will label \emph{arbitrary}
retained and frozen linear coordinate blocks, and will not replace
$\vp$ and $\wprime$ as physical definitions.

The six independent entries of $\boldsymbol\Sigma_i$ jointly
represent the five deviatoric and the one spherical stress directions:
$\boldsymbol\Sigma_i^\circ=\bS_i^\circ/T_i$ and
$\frac{1}{3}\operatorname{tr}\boldsymbol\Sigma_i=-\Pi_i/T_i$.
Thus no separate stress-multiplier symbols are necessary.
The choice $\bQ_i=\bq_i$ is the normalization of the
present zero-storage generator. A change to the $\bq_i/T_i^2$
normalization used in the 1983 formulation requires a corresponding
change of the auxiliary rows and generating functions, not a simple
replacement of symbols. The present Fourier row uses $\grad(1/T_i)$.
The temperature-gradient normalization of \cite[Section~4]{Ruggeri1983Viscous}
retains the physical balance rows and changes the auxiliary heat-flux
multiplier and the spatial potential, as checked below.

These are all $14n$ multiplier components. Their spatial multiplier
identities are verified in \Cref{prop:degenerate-acta-generator}, under
the fixed-flux integrability hypothesis already characterized in
\Cref{prop:euler-flux-rigidity,cor:ret-compatible-classIII}.
The complementary multipliers cannot be deduced from the temporal entropy
density alone. Their validity is checked against the spatial fluxes and
the source pairing.

 Let $\boldsymbol e_k$ be the $k$th Cartesian basis vector. Tensor derivatives with respect to symmetric $\boldsymbol\Sigma_i$ use
the Frobenius pairing $\boldsymbol A:\boldsymbol B=\sum_{a,b}A_{ab}B_{ab}$.
All matrix symmetry assertions use an orthonormal basis for this pairing
(equivalently, off-diagonal tensor components scaled by $\sqrt2$),
together with the Euclidean pairing for vectors.

 \begin{proposition}[Local rank and normalization of the auxiliary main field]
\label[proposition]{prop:auxiliary-normalization-rank}
For the fixed normalized rows of \eqref{eq:acta-FNS-first-order-form},
consider the physical coordinates $(\bV,\boldsymbol\tau,\bq)$ and the
main-field coordinates $(\vp,\boldsymbol\Sigma,\bQ)$ of
\eqref{eq:complete-main-field-components}. Under the assumptions of
\Cref{prop:general-conservative-convexity}, this change of variables
has full rank $14n$ and is locally invertible.
For the prescribed entropy flux, the spatial multiplier
identities determine the auxiliary block uniquely whenever they are
solvable.

More generally, an invertible \emph{constant} recombination of the
constitutive equations by a matrix $S\in GL(9n)$ preserves the
supplementary-law identities if their auxiliary multipliers transform
by $S^{-\mathsf T}$. It preserves the rank and the scalar generating
potentials. State-dependent changes of the auxiliary fluxes
require a separate check of the multiplier identities.
\end{proposition}
\begin{proof}
At fixed conservative variables, the derivatives of
$\boldsymbol\Sigma_i=\boldsymbol\tau_i/T_i$ and $\bQ_i=\bq_i$
with respect to $(\boldsymbol\tau_i,\bq_i)$ form the block diagonal
matrix $(T_i^{-1}I_6)\oplus I_3$. The full Jacobian is block
triangular, with upper-left block $D^2_{\bV}h^0\succ0$.
Its determinant is therefore nonzero (in orthonormal tensor
coordinates, it equals $\det(D^2_{\bV}h^0)\prod_iT_i^{-6}$).
To check uniqueness relative to these rows and the selected
entropy flux, subtract two spatial multiplier identities with the same
conservative multipliers. For every $k$, the auxiliary differences satisfy
\[
 \sum_i\left[-(\Delta\boldsymbol\Sigma_i\boldsymbol e_k)\cdot\dd\bv_i
              -\Delta Q_{i,k}\,\dd\vartheta_i\right]=0.
\]
The thermodynamic dual variables and constituent velocities are
independent local coordinates. Independent variations of $\bv_i$ and
$\vartheta_i$ therefore give $\Delta\boldsymbol\Sigma_i=0$ and
$\Delta\bQ_i=0$ for every $i$.
For a constant change of constitutive rows, put
$M=\operatorname{diag}(I_{5n},S)$,
$\widetilde{\bF}^{\alpha}=M\bF^{\alpha}$,
$\widetilde{\boldsymbol J}=M\boldsymbol J$ and
$\widetilde{\bU}'=M^{-\mathsf T}\bU'$.
Then $\widetilde{\bU}'\cdot\dd\widetilde{\bF}^{\alpha}
=\bU'\cdot\dd\bF^{\alpha}$ and
$\widetilde{\bU}'\cdot\widetilde{\boldsymbol J}
=\bU'\cdot\boldsymbol J$.
The Legendre potentials are unchanged because
$\widetilde{\bU}'\cdot\widetilde{\bF}^{\alpha}
=\bU'\cdot\bF^{\alpha}$. Invertibility preserves the rank.

For the heat-row example, put $\widetilde{\bQ}_i=\bq_i/T_i^2$.
Replacing its auxiliary spatial flux $-T_i^{-1}\boldsymbol e_k$ by
$T_i\boldsymbol e_k$ and multiplying its source by $T_i^2$ gives the
equivalent temperature-gradient equation. Both the differential pairing
and the source pairing are preserved, since
\[
 \widetilde Q_{i,k}\,\dd T_i
 =Q_{i,k}\,\dd(-T_i^{-1}),\qquad
 \widetilde{\bQ}_i\cdot(T_i^2\boldsymbol J_{\bQ_i})
 =\bQ_i\cdot\boldsymbol J_{\bQ_i}.
\]
The balance fluxes and the entropy four-vector are the same physical
functions of the state. The spatial potential in the new dual variables
has heat term $T_i\widetilde Q_{i,k}$ in place of $Q_{i,k}\Lambda_i^E$.
\end{proof}

With the mathematical entropy sign
$h^0=-\rho s$, the complete source pairing is
\begin{equation}
 g=\bU'\mathbin{\boldsymbol\cdot}\boldsymbol J=-\zeta,
 \qquad\zeta\ge0.
\label{eq:full-main-field-production-convention}
\end{equation}
The viscous and heat contributions come from the auxiliary source rows, and the full identity is verified in the    proof of \mbox{
\Cref{prop:degenerate-acta-generator}}.

\paragraph{Generating functions and proof.}
We now construct the generating potentials $h'^\alpha$ of the complete system (whose derivatives yield the temporal and spatial fluxes),
using precisely the notation of \eqref{eq:intro-legendre-potentials}--
\eqref{eq:intro-generator-identity}. To distinguish them from the
conservative Euler potentials we write the latter as $h_E'^\alpha$.
Both the multiplier-coordinate and physical forms of the generators
are given together in \Cref{prop:degenerate-acta-generator}. They are
valid only on the fixed-flux compatible subclass characterized in
\Cref{prop:euler-flux-rigidity,cor:ret-compatible-classIII}.

For the complete first-order system
\eqref{eq:complete-mixture-state}--\eqref{eq:acta-FNS-first-order-form},
the following result constructs the generator in main-field coordinates
$\bU'=(\vp,\wprime)$, with the constitutive multiplier block specified above. In quasilinear form,
\begin{equation}
 \begin{pmatrix}P&0\\0&0\end{pmatrix}\partial_t
 \begin{pmatrix}\vp\\\wprime\end{pmatrix}
 +\sum_{k=1}^3
 \begin{pmatrix}a^k&b^k\\(b^k)^{\sf T}&0\end{pmatrix}
 \partial_k\begin{pmatrix}\vp\\\wprime\end{pmatrix}
 =\boldsymbol{J}(\bU'),\qquad P\succ0.
 \label{eq:acta-FNS-symmetric-block-form}
\end{equation}
Here $\boldsymbol{J}(\bU')$ collects the physical balance sources and the algebraic
constitutive terms. Explicitly, the coefficient matrices are
$A^\alpha=\partial^2h'^\alpha/
(\partial\bU'\,\partial\bU')$ for the potentials defined below.
$P$ and $a^k$ are $5n\times5n$ blocks and $b^k$ is $5n\times9n$.
The block $a^k$ may depend on $\wprime$. The zero block in the temporal matrix expresses precisely
the    absence of time derivatives in  the Newton and Fourier    constitutive equations.
The singular temporal matrix distinguishes this formulation from a  symmetric-hyperbolic relaxation model.

\begin{proposition}[Degenerate first-order generator with positive transport coefficients]
\label[proposition]{prop:degenerate-acta-generator}
Assume the RET-compatible subclass of \Cref{cor:ret-compatible-classIII} and the constitutive relations of Section~\ref{sec:entropy-principle}, with $\eta_i^{\mathrm v}>0$, $\lambda_i^{\mathrm v}>0$, and $\kappa_i>0$. Use the complete multiplier field already displayed in
\eqref{eq:extended-stress-main-field}--\eqref{eq:acta-enlarged-main-field}.
These coordinates must satisfy the spatial compatibility identities
of the supplementary entropy law, which are verified below. The full-rank coordinate property is established in
\Cref{prop:auxiliary-normalization-rank}. Hence the $14n$
main-field components are independent local coordinates despite the
zero temporal flux block.
  With $h_E'^{\alpha}$ from \eqref{eq:general-euler-spatial-generator} and~\eqref{eq:general-euler-temporal-generator}, set
\begin{align}
  h'^{0}(\bU')&:=h_E'^{0}(\vp)
  =\sum_i\frac{p_i}{T_i},
\label{eq:extended-time-generator}
\\
  h'^{k}(\bU')&:=h_E'^{k}(\vp)
  +\sum_i\left[
    \frac{\boldsymbol\Lambda_i^{\bv}\cdot(\boldsymbol\Sigma_i\boldsymbol e_k)}{\Lambda_i^E}
    +Q_{i,k}\Lambda_i^E
  \right]
  =\sum_i\frac{p_iv_{i,k}-(\boldsymbol\tau_i\bv_i)_k-q_{i,k}}{T_i},
  \qquad k=1,2,3.
\label{eq:extended-spatial-generator}
\end{align}
Then the physical constituent mass, momentum, and total-energy fluxes are generated by the $\vp$ derivatives of \eqref{eq:extended-time-generator} and~\eqref{eq:extended-spatial-generator}. The $\boldsymbol\Sigma$ and $\bQ$ equations have zero temporal storage and reproduce the Newton and thermo-Fourier laws with the source choice \eqref{eq:acta-zero-storage-sources} below. The resulting first-order system has symmetric coefficient matrices and a positive-semidefinite temporal Hessian whose kernel is exactly the $\boldsymbol\Sigma$--$\bQ$ auxiliary subspace. Its Legendre four-vector is minus the selected physical entropy four-vector, and, on solutions of the auxiliary constitutive equations, the negative of its source contraction gives exactly the closed entropy production \eqref{eq:closed-zeta}.
\end{proposition}

\begin{proof}
The temporal part of the fundamental multiplier identity gives the
conservative components in \eqref{eq:complete-main-field-components}: differentiating
$h^0=-\eta$ with respect to the total-energy variables and using
\eqref{eq:thermal-total-differential} yields precisely the first
three entries in \eqref{eq:complete-main-field-components}. Because
the last $9n$ temporal densities vanish, their multipliers are obtained
instead by the \emph{spatial} identities. We verify them, including
the source pairing, by constructing the potentials explicitly.

Since $\boldsymbol\Lambda_i^{\bv}/\Lambda_i^E=-\bv_i$, differentiation of \eqref{eq:extended-spatial-generator} gives
\begin{align}
  \frac{\partial h'^{k}}{\partial\boldsymbol\Lambda_i^{\bv}}
  &=\rho_i\bv_i v_{i,k}-\bS_i\boldsymbol e_k,
\label{eq:extended-generated-momentum-flux}
\\
  \frac{\partial h'^{k}}{\partial\Lambda_i^E}
  &=\left(\rho_i e_i+\half\rho_i|\bv_i|^2\right)v_{i,k}
    -(\bS_i\bv_i)_k+Q_{i,k},
\label{eq:extended-generated-energy-flux}
\end{align}
while the mass flux is unchanged from the Euler generator. The auxiliary differential rows are
\begin{equation}
  \sum_k\partial_k
  \frac{\partial h'^{k}}{\partial\boldsymbol\Sigma_i}
  =-\bD_i,
  \qquad
  \sum_k\partial_k
  \frac{\partial h'^{k}}{\partial\bQ_i}
  =-\grad\frac{1}{T_i}.
\label{eq:extended-zero-storage-left}
\end{equation}
Introduce the positive-definite viscosity operator on symmetric tensors
\begin{equation}
  \mathbb V_i\boldsymbol D
  :=\frac{2\eta_i^{\mathrm v}}{T_i}\boldsymbol D^{\circ}
   +\frac{\lambda_i^{\mathrm v}}{T_i}(\tr\boldsymbol D)\bI.
\label{eq:temperature-weighted-viscosity-operator}
\end{equation}
The five deviatoric and one scalar viscous rows in
\eqref{eq:acta-FNS-first-order-form} are equivalent to the tensorial
row used in the generator. Indeed, with the residuals of those rows
written with all terms on the left,
\[
 \boldsymbol R_i^\circ=-\bD_i^\circ+
       \frac{\bS_i^\circ}{2\eta_i^{\mathrm v}},\qquad
 R_i^b=-\Div\bv_i-\frac{\Pi_i}{\lambda_i^{\mathrm v}},
\]
one has, using \eqref{eq:extended-stress-main-field},
\[
 \boldsymbol R_i^\Sigma:=\boldsymbol R_i^\circ+
          \frac{1}{3}R_i^b\bI
 =-\bD_i+\mathbb V_i^{-1}\boldsymbol\Sigma_i,
 \qquad
 \boldsymbol\Sigma_i:\boldsymbol R_i^\Sigma
 =\frac{\bS_i^\circ}{T_i}:\boldsymbol R_i^\circ
  -\frac{\Pi_i}{T_i}R_i^b.
\]
Thus the factor $1/3$ belongs to the recombination of the rows,
not to the spherical part of the stress main-field component,
$\frac{1}{3}\operatorname{tr}\boldsymbol\Sigma_i=-\Pi_i/T_i$.
Write $\boldsymbol{J}=(\boldsymbol J_{\vp},\boldsymbol J_{\boldsymbol\Sigma},\boldsymbol J_{\bQ})$ for the complete source vector and its blocks. $\boldsymbol J_{\cdot}$ denotes a source block, not a derivative. The auxiliary source components
\begin{equation}
  \boldsymbol J_{\boldsymbol\Sigma_i}=-\mathbb V_i^{-1}\boldsymbol\Sigma_i,
  \qquad
  \boldsymbol J_{\bQ_i}=-\frac{\bQ_i-d_i^T\bu_i}{\kappa_i}
\label{eq:acta-zero-storage-sources}
\end{equation}
turn \eqref{eq:extended-zero-storage-left} into \eqref{eq:viscous-closure} and~\eqref{eq:heat-closure}. Define for this proof
\[
  \boldsymbol\xi_i:=\frac{\bQ_i-d_i^T\bu_i}{\kappa_i}.
\]
Along solutions of the heat-flux constitutive equation, $\boldsymbol\xi_i=\grad(1/T_i)$. Because \eqref{eq:ret-B-vanishes} holds, the resistance co-factor is
\[
  \widetilde{\bB}_i=-d_i^T\boldsymbol\xi_i
   +\frac{1}{T_i}\left(\bmom_i-\half r_i\bu_i\right),
  \qquad
  \bP_{\boldsymbol y}^{\sf T}\widetilde{\boldsymbol B}=-\bM\bu,
\]
so \eqref{eq:gauge-representation}, \eqref{eq:momentum-reconstruction}, and~\eqref{eq:momentum-multiplier}, with the reciprocal-temperature gradients replaced by $\boldsymbol\xi$, express $\bmom_i$ as a function of the enlarged state without spatial derivatives.

The system is therefore
\begin{equation}
  \partial_t\frac{\partial h'^{0}}{\partial \bU'}
  +\sum_{k=1}^3\partial_k\frac{\partial h'^{k}}{\partial \bU'}
  =\boldsymbol{J}(\bU').
\label{eq:degenerate-generator-system}
\end{equation}
For clarity, the block labels $\vp$ and $\wprime$
are fixed throughout the physical $14n$-field formulation.
The complete generated densities and fluxes have the block structure
\begin{equation}
 \bF^0(\bU)=\frac{\partial h'^0}{\partial\bU'}
 =\begin{pmatrix}\bV\\\boldsymbol0_{9n}\end{pmatrix},
 \qquad
 \bF^k(\bU)=\frac{\partial h'^k}{\partial\bU'}
 =\begin{pmatrix}\bF_{\mathrm{bal}}^k(\vp,\wprime)\\
                  \bF_{\mathrm{aux}}^k(\vp)\end{pmatrix}.
\label{eq:full-stored-and-spatial-flux-blocks}
\end{equation}
The first block has $5n$ components and the second has $9n$.
In the generator normalization the constitutive spatial block is
\[
 \bF_{\mathrm{aux}}^k
 =\bigl(-\operatorname{sym}(\bv_i\otimes\boldsymbol e_k),
           \Lambda_i^E\boldsymbol e_k\bigr)_{i=1}^n,
 \qquad \operatorname{sym}\boldsymbol A
 :=(\boldsymbol A+\boldsymbol A^{\sf T})/2.
\]
It does not vanish even at $\wprime=0$. Only the auxiliary
\emph{temporal densities} vanish: the auxiliary
\emph{spatial fluxes} generally do not. At $\wprime=0$ the retained
balance-flux block is precisely $\bF_E^k(\vp)$. The Euler principal
subsystem is obtained by restricting the \emph{potential} and deleting
its auxiliary rows, not by declaring the full auxiliary spatial-flux
block zero.
All coefficient matrices are Hessians and hence symmetric. By \Cref{prop:general-conservative-convexity} and local Legendre duality,
\begin{equation}
  D_{\bU'}^2h'^{0}
  =\begin{pmatrix}
      D_{\vp}^2h_E'^{0} & 0\\
      0 & 0
    \end{pmatrix}
  \succeq0,
  \qquad
  D_{\vp}^2h_E'^{0}
  =(D_{\bV}^2h^0)^{-1}\succ0,
\label{eq:degenerate-temporal-hessian}
\end{equation}
which proves the stated kernel characterization.

Using the Legendre relation \eqref{eq:intro-legendre-potentials} and
the generator identity \eqref{eq:intro-generator-identity}, we recover
the entropy four-vector from the explicit potentials, without introducing
another definition. The stress term in
\eqref{eq:extended-spatial-generator} is homogeneous of degree one in
the complete main field and hence cancels in the Legendre expression.
The term $Q_{i,k}\Lambda_i^E$ contributes $-Q_{i,k}/T_i$.
Consequently $h^0=-\rho s$, and the spatial components coincide with
minus the selected physical entropy transport
\eqref{eq:constituent-entropy-transport-form}.
Differentiating the inverse Legendre relation with the generated
fluxes already verified above gives the full multiplier identities
\eqref{eq:intro-main-field-multipliers}, including the auxiliary rows,
without repeating the general calculation of the Introduction.
Contraction with the complete main field yields the source relation
$\zeta=-\bU'\mathbin{\boldsymbol\cdot}\boldsymbol J(\bU')$,
as already recorded in \eqref{eq:full-main-field-production-convention}.
The auxiliary stress contribution is
\[
  -\boldsymbol\Sigma_i:\boldsymbol J_{\boldsymbol\Sigma_i}
  =\boldsymbol\Sigma_i:\mathbb V_i^{-1}\boldsymbol\Sigma_i
  =\frac{2\eta_i^{\mathrm v}}{T_i}|\bD_i^{\circ}|^2
   +\frac{\lambda_i^{\mathrm v}}{T_i}(\Div\bv_i)^2.
\]
For the heat--resistance block, \eqref{eq:projection-properties2} and~\eqref{eq:resistance-closure} give
$\sum_i\bu_i\cdot\widetilde{\bB}_i=-\sum_i\bu_i\cdot(\bM\bu)_i$. Therefore
\begin{align*}
 \sum_i\left[\bQ_i\cdot\boldsymbol\xi_i
 -\frac{\bu_i}{T_i}\cdot(\bmom_i-\half r_i\bu_i)\right]
 &=\sum_i\left[(\bQ_i-d_i^T\bu_i)\cdot\boldsymbol\xi_i
              -\bu_i\cdot\widetilde{\bB}_i\right]\\
 &=\sum_i\left[\kappa_i|\boldsymbol\xi_i|^2
              +\bu_i\cdot(\bM\bu)_i\right].
\end{align*}
The remaining conservative source contraction is the thermal-relaxation and chemical part already evaluated in \eqref{eq:thermal-relaxation-dissipation} and~\eqref{eq:reaction-dissipation}. This recovers \eqref{eq:closed-zeta} with the same signs and completes the proof.
\end{proof}

The algebraic production of the enlarged state, before using any auxiliary equation, is
\begin{align}
 \zeta_{\mathrm{lift}}(\bU'):=-\bU'\mathbin{\boldsymbol\cdot}\boldsymbol{J}(\bU')
 ={}&\sum_i\boldsymbol\Sigma_i:\mathbb V_i^{-1}\boldsymbol\Sigma_i
 +\sum_i\frac{|\bQ_i-d_i^T\bu_i|^2}{\kappa_i}
 +\sum_i\bu_i\cdot(\bM\bu)_i\nonumber\\
 &+\zeta_{\mathrm{relax}}-\sum_aR_a\Acal_a\ge0.
\label{eq:lifted-state-production}
\end{align}
It equals the gradient expression \eqref{eq:closed-zeta} when the stress and heat-flux rows hold. Generator restrictions below act on \eqref{eq:lifted-state-production}. Deleting an auxiliary row does not retain that row as a gradient constraint.

\subsection{Mixed zero-storage formulation and dissipative symbol}
\label{subsec:degenerate-generator}
We now record the mixed variant needed when transport coefficients vanish.

\begin{remark}[Semidefinite mixed formulation without constitutive inversion]
\label[remark]{rem:mixed-degenerate-lift}
The inverses in \eqref{eq:acta-zero-storage-sources} are used only in the reduced $14n$-variable formulation. To include vanishing viscosity or heat-conductivity modes without constitutive inversion, introduce auxiliary algebraic variables consisting of independent symmetric tensors $\boldsymbol{\mathsf D}_i$ and vectors $\boldsymbol\xi_i$, and leave the generators \eqref{eq:extended-time-generator} and~\eqref{eq:extended-spatial-generator} independent of them. The resulting generator coordinates are
$\widehat{\bU}'=(\bU',\boldsymbol{\mathsf D},\boldsymbol\xi)$,
of dimension $23n$. The new rows have zero fluxes in every spacetime
direction and enforce algebraic constraints. Thus the multiplier
extension is selected as part of this representation. Uniqueness in
these purely algebraic directions is not inferred from the
differential rank condition in the Introduction. Choose
\begin{align}
  \boldsymbol J_{\boldsymbol\Sigma_i}&=-\boldsymbol{\mathsf D}_i,
  &\boldsymbol J_{\boldsymbol{\mathsf D}_i}&=\boldsymbol\Sigma_i-\mathbb V_i\boldsymbol{\mathsf D}_i,
\label{eq:mixed-stress-sources}
\\
  \boldsymbol J_{\bQ_i}&=-\boldsymbol\xi_i,
  &\boldsymbol J_{\boldsymbol\xi_i}&=\bQ_i-d_i^T\bu_i-\kappa_i\boldsymbol\xi_i.
\label{eq:mixed-heat-sources}
\end{align}

In this mixed formulation the retained momentum source is also defined using the independent variable $\boldsymbol\xi_i$: set
\[
 \bG_i=-d_i^T\boldsymbol\xi_i-\frac{r_i}{2T_i}\bu_i
\]
This is the specialization of the earlier $\bG_i$ (defined immediately before \eqref{eq:gauge-representation}), using $\bB_i=0$ and replacing the thermal gradient by $\boldsymbol\xi_i$. We use \eqref{eq:momentum-reconstruction} and~\eqref{eq:momentum-multiplier} with this value. These source formulas contain no division by $\kappa_i$ and therefore remain defined when $\kappa_i=0$.

The mixed differential and algebraic rows impose $\boldsymbol{\mathsf D}_i=\bD_i$, $\boldsymbol\xi_i=\grad(1/T_i)$ and the original constitutive laws without division by a dissipative coefficient. Their source contraction is
\begin{align}
  -\boldsymbol\Sigma_i:\boldsymbol J_{\boldsymbol\Sigma_i}
  -\boldsymbol{\mathsf D}_i:\boldsymbol J_{\boldsymbol{\mathsf D}_i}
  &=\boldsymbol{\mathsf D}_i:\mathbb V_i\boldsymbol{\mathsf D}_i\ge0,
\label{eq:mixed-stress-production}
\\
  -\bQ_i\cdot \boldsymbol J_{\bQ_i}
  -\boldsymbol\xi_i\cdot \boldsymbol J_{\boldsymbol\xi_i}
  &=d_i^T\bu_i\cdot\boldsymbol\xi_i+\kappa_i|\boldsymbol\xi_i|^2.
\label{eq:mixed-heat-production}
\end{align}
The first term on the right of \eqref{eq:mixed-heat-production} cancels with the corresponding term in the momentum-source contraction, leaving the heat and resistance contributions of \eqref{eq:closed-zeta}. Thus the    generating potentials and entropy identities remain valid  for positive-semidefinite viscosity and conductivity blocks. A singular constitutive block precludes the inverse-based
representation on its null directions. Coercivity and regularity of
the eliminated equations depend on the resulting differential
operator. The mixed differential--algebraic representation remains
defined at singular constitutive blocks. The mixed first-order formulation, its   $14n$-variable    formulation for positive transport coefficients, and the second-order formulation obtained by constitutive substitution are representations of the same constitutive model, not thermodynamic reductions between model classes.

   Equivalent unreduced formulations need not yield the same principal restrictions. The Euler-flux restriction below uses
the   $14n$-variable generator  with positive transport coefficients. On its section $\bQ=0$, the
eliminated algebraic variable is
$\boldsymbol\xi_i=-d_i^T\bu_i/\kappa_i$. Setting all auxiliary
coordinates of the $23n$-variable mixed generator to zero would instead
set $\boldsymbol\xi_i=0$, and can change the retained momentum source.
Thus constitutive elimination and principal restriction need not
commute. A zero-coefficient Euler restriction requires a specified
mixed section and    a separate verification of the restricted source. It is not obtained by extending
formulas containing $1/\kappa_i$ to $\kappa_i=0$.

\par\begingroup
\paragraph{General constitutive source realization.}
The same mixed construction applies beyond diagonal transport laws.
On the fixed-flux compatible subclass, put
$\boldsymbol Y=(\boldsymbol{\mathsf D},\boldsymbol\xi)$ and
$\wprime=(\boldsymbol\Sigma,\bQ)$.
Suppose that, after constitutive substitution, a selected local response
$\widehat{\wprime}(\vp,\boldsymbol Y)$ and conservative source
$\boldsymbol J_{\vp}(\vp,\boldsymbol Y)$ satisfy
\[
 \widehat\zeta:=-\vp\cdot\boldsymbol J_{\vp}
                  +\boldsymbol Y\cdot\widehat{\wprime}\ge0
\]
for all admissible arguments. Here tensor components are paired by the
Frobenius product. Leave the potentials independent of $\boldsymbol Y$
and choose
\[
 \boldsymbol J_{\wprime}=-\boldsymbol Y,\qquad
 \boldsymbol J_{\boldsymbol Y}
       =\wprime-\widehat{\wprime}(\vp,\boldsymbol Y).
\]
The auxiliary differential rows impose
$\boldsymbol Y=((\bD_i)_i,(\grad(1/T_i))_i)$, and the zero-flux rows impose
$\wprime=\widehat{\wprime}$. On the entire enlarged state space,
\[
 -(\vp,\wprime,\boldsymbol Y)\cdot
   (\boldsymbol J_{\vp},-\boldsymbol Y,\wprime-\widehat{\wprime})
 =\widehat\zeta .
\]
Thus this source realization requires no constitutive inverse, including for coupled or
nonlinear responses that admit this local representation. The temporal
Hessian and the physical fluxes remain those of the mixed formulation.
\par\endgroup
\par\begingroup
This construction preserves an already selected constitutive model.
It does not derive reciprocity or replace the balance anchoring,
dimensional-parity, conservation and mechanism-selection requirements
of \cite[Section~3.1 and Appendix~A]{BotheRuggeriPaperI2026}.
For a coupled scalar mobility law, representation in terms of the
prescribed dilatations must first be justified. It is not implied by
positive semidefiniteness alone.
\par\endgroup

\end{remark}

\begin{remark}[Principal dissipative symbol after constitutive elimination]
\label[remark]{rem:factorized-dissipative-symbol}
For the separated Newton--Fourier laws
\eqref{eq:viscous-closure}--\eqref{eq:heat-closure}, with the
conservative-state dependence specified in \Cref{sec:entropy-principle},
fix an interior state and eliminate the stress and intrinsic heat-flux variables. Let $\mathcal L(\boldsymbol n)$ denote the real first-derivative symbol: replace each spatial derivative $\partial_k$ by $n_k$ in the linearized strain-rate and reciprocal-temperature-gradient expressions. For an increment
$\delta\vp=(\delta\Lambda_i^\rho,\delta\boldsymbol\Lambda_i^{\bv},
\delta\Lambda_i^E)_{i=1}^n$, this map is explicitly
\[
 \mathcal L(\boldsymbol n)\delta\vp
 =\bigl((\operatorname{sym}(\delta\bv_i\otimes\boldsymbol n))_{i=1}^n,
          (\boldsymbol n\,\delta\vartheta_i)_{i=1}^n\bigr),
 \quad
 \delta\bv_i=T_i(\delta\boldsymbol\Lambda_i^{\bv}
                         +\bv_i\delta\Lambda_i^E),
 \quad \delta\vartheta_i=-\delta\Lambda_i^E.
\]
The stress and heat blocks are ordered as in the constitutive block.
With
\begin{equation}
  \mathbb R
  :=\operatorname{diag}\left(
      (\mathbb V_i)_{i=1}^n,
      (\kappa_i\bI)_{i=1}^n
    \right)\succeq0,
\label{eq:dissipative-constitutive-block}
\end{equation}
the    positive-semidefinite  symbol associated with the highest-order    viscous and heat-conduction  contribution factorizes as
\begin{equation}
  \mathcal B(\boldsymbol n)
  =\mathcal L(\boldsymbol n)^{\sf T}\mathbb R\,\mathcal L(\boldsymbol n)
  \succeq0.
\label{eq:factorized-dissipative-symbol}
\end{equation}
Indeed, the auxiliary spatial-flux derivative is
$-\mathcal L(\boldsymbol n)$, so symmetry of the generator gives
$-\mathcal L(\boldsymbol n)^{\sf T}$ for the balance-to-auxiliary
coupling. Substitution of the principal constitutive response
$\mathbb R\mathcal L(\boldsymbol n)$ yields the    positive-semidefinite  dissipative
symbol above. The superscript ${\sf T}$ denotes transposition with respect
to the Euclidean constituent/vector and Frobenius tensor pairings. In particular,
\begin{equation}
  \ker\mathcal B(\boldsymbol n)
  =\left\{\boldsymbol r:\mathbb R^{1/2}\mathcal L(\boldsymbol n)\boldsymbol r=0\right\}.
\label{eq:dissipative-symbol-kernel}
\end{equation}
The thermo-diffusive $d_i^T\bu_i$ terms and the resistance, thermal-relaxation, and reaction sources are lower order for this second-order symbol.
After elimination, the momentum closure may contain first spatial
derivatives through $\boldsymbol\xi_i=\grad(1/T_i)$. Thus the Euler
flux Hessian is the reversible generator block. It need not be the
entire first-order coefficient matrix of the eliminated system. This frozen principal-symbol identity does not establish well-posedness or a complete symmetric hyperbolic--parabolic normal-form theorem of the type that additionally controls the kernel geometry of the second-order operator.
If the general nonlinear response
$\widehat{\wprime}(\vp,\boldsymbol Y)$ of
\Cref{rem:mixed-degenerate-lift} is differentiable in $\boldsymbol Y$,
the corresponding principal
constitutive coefficient is instead
$D_{\boldsymbol Y}\widehat{\wprime}$ at the frozen state.
The nonnegative entropy-source pairing alone does not imply that this
tangent coefficient is symmetric or positive semidefinite.
\end{remark}

\subsection{Principal subsystems of symmetric generators}
\label{subsec:common-temperature-principal}

The algebraic construction uses only the generator identities of the
Introduction. It first restricts the potentials to a constant affine
section of the complete main field, then retains the equations conjugate
to the free coordinates. The definition below includes singular temporal
matrices. The classical hyperbolic theorem is its positive-definite
specialization.

To interpret a linear recombination unambiguously, let $L$ be a constant
invertible matrix and put $\bU'=L\widetilde{\bU}'$.
Then $\widetilde h'^\alpha(\widetilde{\bU}')=
h'^\alpha(L\widetilde{\bU}')$ has flux and source vectors
$L^{\sf T}\bF^\alpha$ and $L^{\sf T}\boldsymbol J$, and Hessian
$L^{\sf T}(D_{\bU'}^2h'^\alpha)L$.
Thus symmetry and semidefiniteness are preserved by congruence.
For this abstract restriction only, write
$\bU'=(\Uret,\Ufix)$: the first block is retained and
the second is frozen at $\Ufix=\Ufixstar$. Both are
subblocks of the \emph{same} $\bU'$ (after a possible constant
linear recombination), not new main fields. This generic splitting
need not agree with the physically distinguished
$\bU'=(\vp,\wprime)$ used throughout the mixture model.

\begin{definition}[Degenerate principal subsystem]
\label[definition]{def:degenerate-principal-subsystem}
Let
\begin{equation}
  \partial_\alpha\frac{\partial h'^{\alpha}}{\partial \bU'}=\boldsymbol{J}(\bU'),
  \qquad
  h'^{\alpha}\in C^2,
  \qquad
  D_{\bU'}^2h'^{0}\succeq0,
\label{eq:abstract-degenerate-generator}
\end{equation}
be a symmetric generator system whose temporal Hessian may be singular. Here $\bU'$ is the full main field in the multiplier sense,
and $\boldsymbol{J}(\bU')$ is the source vector, including algebraic constitutive
terms when present. After a constant invertible change of main-field coordinates, let
$\bU'=(\Uret,\Ufix)$ be a fixed linear splitting, write $\boldsymbol{J}=(\boldsymbol J_{\mathrm{ret}},\boldsymbol J_{\mathrm{fix}})$ for the corresponding source blocks (not derivatives of a scalar), and let $\Ufixstar$ be a prescribed constant. The degenerate principal subsystem associated with $(\Uret,\Ufixstar)$ is obtained by setting $\Ufix=\Ufixstar$ and deleting the equations conjugate to $\Ufix$:
\begin{equation}
  \partial_\alpha\frac{\partial \overline h'^{\alpha}}{\partial \Uret}
  =\boldsymbol J_{\mathrm{ret}}(\Uret,\Ufixstar),
  \qquad
  \overline h'^{\alpha}(\Uret):=h'^{\alpha}(\Uret,\Ufixstar).
\label{eq:restricted-generator}
\end{equation}
The affine section must meet the main-field domain, and all statements
are local on an open set of retained coordinates in that section.
No hyperbolicity or characteristic statement is included in this
definition. In particular, Boillat--Ruggeri characteristic interlacing
applies under the hypothesis of a positive-definite temporal Hessian in
\Cref{thm:boillat-ruggeri-principal}.
\end{definition}

\paragraph{Affine restrictions and Galilean covariance.}
The definition permits any constant affine section meeting the admissible
main-field domain. In a hierarchy obtained by adjoining or deleting
balances, Galilean covariance of the restricted system requires checking
the transformation laws of the section and of the retained equations.
For the general transformation theorem and the block-triangular
hierarchy condition, see~\cite{Ruggeri1989Galilean} and
\cite[Sections~2.7 and~2.7.4]{RuggeriSugiyama2021}. For example, fixing
the energy main-field component at a constant value fixes the
temperature. Deleting the energy equation gives the isothermal Euler
principal subsystem \cite[Section~2.4.1]{RuggeriSugiyama2021}. For the
present mixture we consider thermal grouping and restriction to zero
constitutive multipliers. Their commutation is proved below.

\begin{proposition}[ Extension of the subentropy law to degenerate principal subsystems]
\label[proposition]{prop:degenerate-main-field-restriction}
For any $C^2$ generator, restriction to a constant affine section
preserves symmetry. Under the semidefiniteness hypothesis of
\Cref{def:degenerate-principal-subsystem}, every such subsystem has symmetric coefficient matrices and a positive-semidefinite temporal Hessian $D_{\Uret}^2\overline h'^{0}$. Define
\[
  h^\alpha:=\bU'\mathbin{\boldsymbol\cdot}\frac{\partial h'^{\alpha}}{\partial \bU'}-h'^{\alpha},
  \qquad
  g(\bU'):=\bU'\mathbin{\boldsymbol\cdot}\boldsymbol{J}(\bU').
\]
Then the reduced system satisfies $\partial_\alpha\overline h^\alpha=\overline g$, where
\begin{align}
  \overline h^\alpha
  &=h^\alpha\big|_{\Ufixstar}
    -\Ufixstar\mathbin{\boldsymbol\cdot}
     \left.\frac{\partial h'^{\alpha}}{\partial \Ufix}\right|_{\Ufixstar},
\label{eq:degenerate-subentropy-formula}
\\[0.6em]
  \overline g
  &=g\big|_{\Ufixstar}
    -\Ufixstar\mathbin{\boldsymbol\cdot}\boldsymbol J_{\mathrm{fix}}(\Uret,\Ufixstar).
\label{eq:degenerate-subentropy-production}
\end{align}
If $h^\alpha$ is minus the physical entropy four-vector and $\zeta=-g$ is the physical entropy production, then
\begin{equation}
  \overline\zeta
  =\zeta\big|_{\Ufixstar}
   +\Ufixstar\mathbin{\boldsymbol\cdot}\boldsymbol J_{\mathrm{fix}}(\Uret,\Ufixstar).
\label{eq:degenerate-physical-production}
\end{equation}
The    reduced mathematical entropy density $\overline h^0$ is called the \emph{subentropy}, and $\overline h^k$ are its fluxes. Its source is $\overline g$, while $\overline\zeta=-\overline g$ denotes the corresponding physical entropy production. In the degenerate case these quantities define a supplementary law. At a state with a singular temporal Hessian, the temporal Legendre map fails to be a local $C^1$ diffeomorphism. For a nonzero frozen value the sign of
$\overline\zeta$ must be checked separately.
For $\Ufixstar=0$, both the entropy four-vector and the physical entropy production are literal restrictions of the parent quantities. Successive zero restrictions along a fixed linear splitting commute.
\end{proposition}

\begin{proof}
Expanding the retained divergences in \eqref{eq:restricted-generator} gives the quasilinear coefficient matrices. The reduced Hessians are principal blocks of the parent Hessians and hence remain symmetric. The temporal block remains positive semidefinite. The same Legendre identity gives \eqref{eq:degenerate-subentropy-formula}. Contracting the retained equations with $\Uret$ gives
\[
  \overline g=\Uret\mathbin{\boldsymbol\cdot}\boldsymbol J_{\mathrm{ret}}
  =(\Uret,\Ufixstar)\mathbin{\boldsymbol\cdot}\boldsymbol{J}-\Ufixstar\mathbin{\boldsymbol\cdot}\boldsymbol J_{\mathrm{fix}},
\]
which proves \eqref{eq:degenerate-subentropy-production} and~\eqref{eq:degenerate-physical-production}. No positive-definiteness or characteristic argument is used.
\end{proof}

The familiar isothermal Euler model illustrates why the associated subentropy
must be constructed rather than simply obtained by restricting the entropy
\cite[Section~4.1]{BoillatRuggeri1997}. For a simple compressible fluid
satisfying the Gibbs relation on a stable domain, with $\rho>0$, $T>0$,
$c_v>0$, and $(\partial p/\partial\rho)_T>0$, fix the energy
component of the main field at $\Lambda^E=-1/T_*$, with constant
$T_*>0$, and retain the mass and momentum equations with
$p=p(\rho,T_*)$. The energy equation conjugate to the frozen
component is omitted. Because the frozen multiplier is nonzero, the
corresponding subentropy is
\[
  \overline h^0
  =\left(-\rho s+\frac{E}{T_*}\right)_{T=T_*},
  \qquad E=\rho e+\tfrac12\rho|\bv|^2,
\]
not merely the restriction of $-\rho s$. The grouped-temperature
principal subsystems constructed below instead freeze differences of
energy conjugates at zero, which simplifies the subentropy formula.

For the classical theorem, let $\bU'=(\Uret,\Ufix)$, with
$\Uret\in\mathbb R^\ell$ and $\Ufix\in\mathbb R^{m-\ell}$,
denote an arbitrary fixed linear splitting of a regular hyperbolic main
field.    \mbox{
\Cref{prop:degenerate-main-field-restriction} }
extends the algebraic subentropy identity to singular temporal Hessians. The following theorem recalls the classical positive-definite case and its characteristic inequalities.

\begin{definition}[Classical principal subsystem]
\label[definition]{def:classical-principal-subsystem}
Let
\begin{equation}
  \partial_\alpha\frac{\partial h'^{\alpha}}{\partial \bU'}=\boldsymbol J(\bU'),
  \qquad
  D_{\bU'}^2 h'^{0}\succ0,
\label{eq:abstract-classical-generator}
\end{equation}
be a symmetric-hyperbolic system of balance laws in main-field variables. Here $\partial h'^0/\partial \bU'$ gives the balance densities, $\partial h'^k/\partial \bU'$ gives their spatial fluxes, and $\boldsymbol J(\bU')$ is the source vector. Let $\bU'=(\Uret,\Ufix)$ be a fixed linear splitting, write $\boldsymbol J=(\boldsymbol J_{\mathrm{ret}},\boldsymbol J_{\mathrm{fix}})$ for the corresponding source blocks, and let $\Ufixstar$ be a prescribed constant. The principal subsystem associated with $(\Uret,\Ufixstar)$ is obtained by setting $\Ufix=\Ufixstar$ in the retained equations and deleting the equations conjugate to $\Ufix$:
\begin{equation}
  \partial_\alpha\frac{\partial \overline h'^{\alpha}}{\partial \Uret}
  =\boldsymbol J_{\mathrm{ret}}(\Uret,\Ufixstar),
  \qquad
  \overline h'^{\alpha}(\Uret):=h'^{\alpha}(\Uret,\Ufixstar).
\label{eq:classical-principal-restriction}
\end{equation}
\end{definition}

\begin{theorem}[ Classical Boillat--Ruggeri theorem: positive-definite case]
\label[theorem]{thm:boillat-ruggeri-principal}
Under the hypotheses of \Cref{def:classical-principal-subsystem}, the restricted system is again symmetric hyperbolic because $D_{\Uret}^2\overline h'^{0}$ is a positive-definite principal block of $D_{\bU'}^2h'^{0}$. Its convex mathematical entropy, called the subentropy, satisfies an entropy balance derived from the retained field equations. If
\[
  h^\alpha=\bU'\mathbin{\boldsymbol\cdot}\frac{\partial h'^{\alpha}}{\partial \bU'}-h'^{\alpha},
  \qquad
  g=\bU'\mathbin{\boldsymbol\cdot}\boldsymbol J,
\]
then
\begin{align}
  \overline h^\alpha
  &=h^\alpha\big|_{\Ufixstar}
    -\Ufixstar\mathbin{\boldsymbol\cdot}
     \left.\frac{\partial h'^{\alpha}}{\partial \Ufix}\right|_{\Ufixstar},
\label{eq:classical-subentropy-restriction}
\\[0.6em]
  \overline g
  &=g\big|_{\Ufixstar}
    -\Ufixstar\mathbin{\boldsymbol\cdot}\boldsymbol J_{\mathrm{fix}}(\Uret,\Ufixstar).
\label{eq:classical-production-restriction}
\end{align}
For the homogeneous principal part, the characteristic velocities obey the Boillat--Ruggeri interlacing/subcharacteristic inequalities at each state and each spatial direction~\cite{BoillatRuggeri1997}. In particular, for $\Ufixstar=0$ the entropy density and fluxes $\overline h^\alpha$, and their production $\overline g$, are obtained simply by evaluating the parent quantities at $\Ufix=0$.
\end{theorem}

\begin{proof}
The retained coefficient matrices are Hessians of the restricted potentials. Their symmetry and the entropy-balance identities follow by differentiation and the Legendre identity. More explicitly, if the retained conservative
density is $\boldsymbol D_{\mathrm{ret}}=\partial_{\Uret}\overline h'^0$, local Legendre duality gives
\[
  \overline h^0=\Uret\mathbin{\boldsymbol\cdot}\boldsymbol D_{\mathrm{ret}}-\overline h'^0,
  \qquad
  D_{\boldsymbol D_{\mathrm{ret}}}^2\overline h^0
  =\bigl(D_{\Uret}^2\overline h'^0\bigr)^{-1}\succ0.
\]
Thus the subentropy is convex in the retained conservative variables,
not merely in the dual coordinates. For the speed bounds, let the parent have dimension $m$ and the retained subspace dimension $\ell$, and order their characteristic values as $\lambda_1\le\cdots\le\lambda_m$ and $\overline\lambda_1\le\cdots\le\overline\lambda_\ell$. At a fixed state and spatial direction, apply the Rayleigh--Ritz principle to the symmetric spatial form with the positive temporal form as inner product. Restriction to the retained $\ell$-dimensional subspace gives
\[
 \lambda_j\le\overline\lambda_j\le\lambda_{j+m-\ell},
 \qquad j=1,\ldots,\ell.
\]
This argument also covers repeated characteristic values and does not require strict hyperbolicity in the distinct-eigenvalue sense.
\end{proof}

\begin{remark}[Source production at a nonzero frozen value]
In addition to the local domain condition in \Cref{def:degenerate-principal-subsystem}, no condition of the form $\Ufixstar\mathbin{\boldsymbol\cdot}\boldsymbol J_{\mathrm{fix}}\equiv0$ is part of \Cref{def:classical-principal-subsystem}. For a nonzero frozen value, however, the sign of the reduced physical entropy production must be checked from \eqref{eq:classical-production-restriction}. The condition $\Ufixstar\mathbin{\boldsymbol\cdot}\boldsymbol J_{\mathrm{fix}}\equiv0$ is sufficient for literal inheritance of the parent production. In the multi-temperature balance-law setting of Ruggeri--Simi\'c~\cite{RuggeriSimic2007}, the residual entropy inequality yields the corresponding condition. The characteristic interlacing statement itself concerns the principal matrices and is independent of this source admissibility issue.
\end{remark}

\subsection{Thermal and zero-storage restrictions of the mixture}
\label{subsec:degenerate-subentropy}

We now apply the general restriction to the complete mixture main field.
Both operations below retain the sources obtained by evaluating the specified enlarged source vector. They    introduce no independent constitutive closure for the reduced system.

\begin{corollary}[Euler-flux subsystem as the zero-storage degenerate principal subsystem]
\label[corollary]{cor:zero-storage-euler-restriction}
For the enlarged generator \eqref{eq:degenerate-generator-system}, set
\[
  \bU'=(\vp,\wprime),\qquad \wprime=(\boldsymbol\Sigma,\bQ).
\]
The zero restriction $\wprime=0$, together with deletion of the stress and heat-flux rows, gives, for the particular potentials \eqref{eq:extended-time-generator}--\eqref{eq:extended-spatial-generator},
\[
  \overline h'^{0}=h_E'^{0}(\vp),
  \qquad
  \overline h'^{k}=h_E'^{k}(\vp),
\]
and therefore exactly the physical Class-III Euler mass, momentum, and total-energy fluxes of \Cref{subsec:euler-rigidity}. The retained local source is $\boldsymbol J_{\vp}(\vp,0)$, with both auxiliary blocks set to zero. The resulting homogeneous principal part is symmetric hyperbolic, and its entropy density, fluxes, and production are inherited by restriction.
\end{corollary}

\begin{proof}
The flux assertion follows by evaluating the potentials at $\wprime=0$. By \Cref{prop:general-conservative-convexity} and~\eqref{eq:degenerate-temporal-hessian},
\[
  D_{\vp}^2h_E'^{0}
  =(D_{\bV}^2h^0)^{-1}\succ0,
\]
which gives symmetric hyperbolicity. The entropy and production identities follow from \Cref{prop:degenerate-main-field-restriction} with the frozen auxiliary block equal to zero.
\end{proof}

The retained sources need not be those of an independently chosen inviscid model. In particular, the restricted production is
\begin{equation}
 \left.\zeta_{\mathrm{lift}}\right|_{\wprime=0}
 =\sum_i\frac{(d_i^T)^2|\bu_i|^2}{\kappa_i}
  +\sum_i\bu_i\cdot(\bM\bu)_i
  +\zeta_{\mathrm{relax}}-\sum_aR_a\Acal_a\ge0.
\label{eq:euler-restricted-lifted-production}
\end{equation}
It need not coincide with the production of an Euler model    with an independently specified constitutive closure  with $d_i^T=0$. The deleted Fourier equations no longer impose $\grad(1/T_i)=-d_i^T\bu_i/\kappa_i$ on this subsystem.

\begin{remark}[Thermo-diffusive term in the Euler restriction]
\label[remark]{rem:euler-thermodiffusive-source}
The term $(d_i^T)^2|\bu_i|^2/\kappa_i$ in
\eqref{eq:euler-restricted-lifted-production} is a scalar contribution
to the entropy production from the negative contraction of the retained
source with the main field. The independently closed nonreacting Euler-mixture
models of \cite{RuggeriSimic2007} have no such contribution.
For $d_i^T=0$ for every species, it vanishes identically and the
restricted entropy production contains only diffusion resistance,
thermal relaxation and chemical contributions. If $d_i^T$ is fixed
and nonzero, this restricted expression may diverge as
$\kappa_i\downarrow0$ at a state with $\bu_i\ne0$.
The construction of \Cref{prop:degenerate-acta-generator} itself
assumes $\kappa_i>0$. It does not establish a finite zero-conductivity
limit. A limit with $d_i^T=O(\sqrt{\kappa_i})$ keeps this particular
source contribution bounded, but requires a separate consistent
constitutive scaling.
\end{remark}

\begin{corollary}[Thermal dissipative subsystems and commuting restrictions]
\label[corollary]{cor:commuting-zero-restrictions}
Assume the hypotheses of \Cref{prop:degenerate-acta-generator}.
Choose a partition into $K$ nonempty thermal groups and make the constant
dual recombination specified in
\eqref{eq:group-relative-energy-main-fields}--\eqref{eq:explicit-group-energy-duality}.
Write the resulting coordinates as
\[
 \widetilde{\bU}'=(\boldsymbol\Lambda_{\mathcal G},
                    \boldsymbol\Theta,\wprime),
 \qquad \wprime=(\boldsymbol\Sigma,\bQ),
\]
where $\boldsymbol\Lambda_{\mathcal G}$ consists of all mass and
momentum multipliers and the $K$ mean energy multipliers, and
$\boldsymbol\Theta$ consists of the $n-K$ intragroup differences.
On every nonempty admissible local section $\boldsymbol\Theta=0$:
\begin{enumerate}[label=\textnormal{(\roman*)}]
\item Deleting only the conjugate relative-energy equations retains
$4n+K$ stored fields and all $9n$ constitutive fields. The temporal
Hessian has a positive-definite stored block and precisely $9n$
zero-storage directions. The Newton and thermo-Fourier equations remain
in the subsystem, so elimination of their unknowns retains the
positive-semidefinite second-order dissipative symbol. For $K=1$ this
is a multi-velocity, single-temperature dissipative mixture.
\item The additional zero restriction $\wprime=0$, with deletion of
its constitutive equations, commutes with $\boldsymbol\Theta=0$.
Both orders give the same grouped-temperature Euler potentials and
restricted sources. Entropy and entropy production are inherited in
both steps.    The classical subcharacteristic inequalities apply between nested Euler subsystems with positive-definite temporal Hessians. They do not apply between  the singular dissipative    system and  its Euler restriction.
\end{enumerate}
\end{corollary}
\begin{proof}
The group recombination acts only on the conservative main field. Let
$C_{\mathcal G}$ be its constant full-column-rank inclusion on the
section $\boldsymbol\Theta=0$, so that
$\vp=C_{\mathcal G}\boldsymbol\Lambda_{\mathcal G}$.
The stored temporal Hessian is
$C_{\mathcal G}^{\sf T}P C_{\mathcal G}\succ0$, where
$P=D_{\vp}^2h_E'^0$. The constitutive temporal block remains zero.
All derivatives with respect to $\boldsymbol\Sigma_i$ and $\bQ_i$
are retained, giving the same Newton and Fourier rows evaluated at
the group temperatures. For a spatial covector $\boldsymbol n$, the
eliminated principal dissipative symbol is
\[
 \mathcal B_{\mathcal G}(\boldsymbol n)
 =C_{\mathcal G}^{\sf T}\mathcal B(\boldsymbol n)C_{\mathcal G}
 =(\mathcal L(\boldsymbol n)C_{\mathcal G})^{\sf T}
   \mathbb R(\mathcal L(\boldsymbol n)C_{\mathcal G})\succeq0.
\]
For one group the Fourier rows give
$\bq_i=d_i^T\bu_i+\kappa_i\grad(1/T)$. Thus the conductive part
of the summed heat flux has coefficient $\sum_i\kappa_i>0$.
The    separate viscous constitutive equations for all constituents are retained. Evaluating a potential and
source at $(\boldsymbol\Theta,\wprime)=(0,0)$ is independent of the
order of evaluation. The supplementary-law assertions follow from
\Cref{prop:degenerate-main-field-restriction}. No discarded equation
is imposed as an additional constraint.
\end{proof}

These algebraic restrictions do not imply dynamical invariance of the grouped-temperature manifold (see \Cref{sec:thermal-reduction}).

\paragraph{Common-temperature principal subsystem.}
Rewrite the $n$ total-energy equations as one equation for their sum and $n-1$ independent energy-difference equations. Main fields conjugate to the latter can be chosen as
\begin{equation}
  \Lambda_i^E-\Lambda_n^E
  =-\left(\frac{1}{T_i}-\frac{1}{T_n}\right),
  \qquad i=1,\ldots,n-1.
\label{eq:relative-energy-main-fields}
\end{equation}
Where the resulting section meets the admissible state domain, setting
these fields to zero gives
\begin{equation}
  \frac{1}{T_1}=\cdots=\frac{1}{T_n}=\frac{1}T,
  \qquad
  T_1=\cdots=T_n=T.
\label{eq:common-temperature}
\end{equation}

The fields in \eqref{eq:relative-energy-main-fields} are constant linear
combinations of the parent main field. Setting them to zero and deleting
their conjugate energy-difference equations therefore yields the common-temperature principal subsystem. Applied
to the Euler generator, it yields the Euler member of this hierarchy. By
\Cref{thm:boillat-ruggeri-principal}, the Euler member inherits a convex subentropy and
characteristic interlacing. This latter conclusion concerns the regular Euler equations:
unequal heating by reactions, exchange or transport may drive solutions of
the full Class-III system away from the common-temperature manifold.
Section~\ref{sec:thermal-reduction} gives the complementary
entropy-maximizing reduction and discusses the additional dynamical
assumptions needed to enforce thermal equality.

\subsection{Grouped temperatures and nested principal subsystems}
\label{subsec:grouped-principal-subsystems}

Under the stated convex-extension hypotheses, the same construction allows    constituents to be assigned a common temperature within each group  without imposing a single common temperature. Let
\begin{equation}
  \{1,\ldots,n\}
  =\bigcup_{a=1}^{K}\mathcal G_a,
  \qquad 1\le K\le n,
\label{eq:thermal-group-partition}
\end{equation}
be a partition into nonempty, pairwise disjoint thermal groups. For each group choose one reference index $i_a\in\mathcal G_a$ and introduce the group total-energy and thermal-energy densities
\begin{equation}
  \mathcal U_a^E
  :=\sum_{i\in\mathcal G_a}E_i
  =\mathcal E_a+\half\sum_{i\in\mathcal G_a}\rho_i|\bv_i|^2,
  \qquad
  \mathcal E_a:=\sum_{i\in\mathcal G_a}\rho_i e_i.
\label{eq:group-thermal-energy}
\end{equation}
Within each group, sum the constituent total-energy equations and retain $|\mathcal G_a|-1$ independent equations for differences of constituent energies. This is an invertible linear change with coefficients independent of state. It replaces the $n$ original energy equations by $K$ equations for the group sums $\mathcal U_a^E$ and $n-K$ equations for the relative energies. At fixed partial densities and velocities, fixing $\mathcal U_a^E$ is equivalent to fixing $\mathcal E_a$. Main fields conjugate to the independent intragroup relative-energy coordinates can be chosen as
\begin{equation}
  \Theta_i^{(a)}
  :=\Lambda_i^E-\Lambda_{i_a}^E
  =-\left(\frac{1}{T_i}-\frac{1}{T_{i_a}}\right),
  \qquad
  i\in\mathcal G_a\setminus\{i_a\}.
\label{eq:group-relative-energy-main-fields}
\end{equation}

An explicit dual coordinate choice removes any normalization ambiguity. With $n_a:=|\mathcal G_a|$, put
\[
 D_i^E:=E_i-\frac{\mathcal U_a^E}{n_a}
 \quad(i\in\mathcal G_a\setminus\{i_a\}),
 \qquad
 \overline\Lambda_a^E:=\frac{1}{n_a}\sum_{i\in\mathcal G_a}\Lambda_i^E.
\]
Then
\begin{equation}
 \sum_{i\in\mathcal G_a}\Lambda_i^E E_i
 =\overline\Lambda_a^E\mathcal U_a^E
   +\sum_{i\in\mathcal G_a\setminus\{i_a\}}
       \Theta_i^{(a)}D_i^E.
\label{eq:explicit-group-energy-duality}
\end{equation}
Thus the conjugate retained balance is exactly the physical group-energy sum. For one group this also justifies \eqref{eq:relative-energy-main-fields}.

Setting all fields in \eqref{eq:group-relative-energy-main-fields} to zero gives one temperature per group,
\begin{equation}
  T_i=T^{(a)}
  \qquad (i\in\mathcal G_a),
  \qquad a=1,\ldots,K,
\label{eq:grouped-temperature-manifold}
\end{equation}
while the $K$ temperatures $T^{(a)}$ remain independent. The resulting hierarchy is summarized in the following proposition.

\begin{proposition}[Nested grouped-temperature principal subsystems]
\label[proposition]{prop:nested-grouped-principal}
Assume the convex-extension and flux-potential hypotheses for the
$5n$-field Euler system in \Cref{subsec:euler-rigidity},
and work locally where the chosen grouped-temperature section meets the
admissible main-field domain. For every such partition \eqref{eq:thermal-group-partition}, the intragroup fields $\Theta_i^{(a)}$ in \eqref{eq:group-relative-energy-main-fields} are constant linear combinations of the parent main field. Their zero restriction, together with deletion of the conjugate intragroup energy-difference equations, therefore defines a principal subsystem in the sense of \Cref{def:classical-principal-subsystem}, with one independent temperature per group. It inherits the entropy restricted to the grouped-temperature manifold and the classical Boillat--Ruggeri interlacing/subcharacteristic inequalities. If a partition is coarsened by merging thermal groups, one sets the differences between the retained inverse group temperatures to zero. These differences are linear combinations of the finer subsystem's main-field components with state-independent coefficients. Successive zero restrictions therefore commute, and the family is nested in the ordinary Boillat--Ruggeri sense between the full Class-III Euler system and the one-temperature principal subsystem.
Partitions are partially ordered by refinement: each chain of successive
coarsenings gives a nested sequence, whereas incomparable partitions
need not be nested.
\end{proposition}

Table~\ref{tab:thermal-resolution-hierarchy} records what is retained
at each level. Thermal grouping removes $n-K$ scalar energy variables,
not species or momentum equations. Thus a common-temperature mixture
can still react, diffuse, and carry relative kinetic energy.

\begin{table}[htbp]

\centering
\caption{Resolution of the mixture hierarchy in three spatial dimensions.
Counts refer to independent scalar evolution fields, not auxiliary
stress or heat-flux variables.}
\label{tab:thermal-resolution-hierarchy}
\small
\renewcommand{\arraystretch}{1.16}
\begin{tabularx}{\linewidth}{@{}p{2.4cm}Xcc@{}}
\toprule
Description & Retained balances & Temperatures & Fields \\
\midrule
Class III & $n$ masses, $n$ momenta, $n$ energies & $n$ & $5n$ \\
Grouped theory & $n$ masses, $n$ momenta, $K$ group energies & $K$ & $4n+K$ \\
Class II & $n$ masses, $n$ momenta, one mixture energy & $1$ & $4n+1$ \\
Class I & $n$ masses, barycentric momentum, one mixture energy & $1$ & $n+4$ \\
\bottomrule
\end{tabularx}
\par\smallskip
\begin{minipage}{\linewidth}\footnotesize
The first three rows form the thermal Euler principal hierarchy under
the stated RET hypotheses. The Class-I row is a prospective mechanical
reduction, not a completed principal-subsystem or relaxation result
here. Its diffusion fluxes need not vanish.
\end{minipage}

\end{table}

The objective internal-exchange part of the retained group-energy balances contains the group sums
\begin{equation}
  \varepsilon^{(a)}:=\sum_{i\in\mathcal G_a}\varepsilon_i,
  \qquad
  \sum_{a=1}^{K}\varepsilon^{(a)}=0.
\label{eq:group-energy-production}
\end{equation}
The group sums retain the intergroup exchange terms of the selected closure, including any resolved reaction-energy transfers. The remaining stress, momentum-transfer, and mass-production terms are inherited by summing the constituent balances consistently. The state-resolved ionization construction in   \cite{BotheRuggeriPaperI2026} provides a concrete two-group instance with the partition
\begin{equation}
 \mathcal G_{\mathrm h}
 =\{\text{all heavy internal-state species and heavy colliders}\},
 \qquad
 \mathcal G_{\mathrm e}=\{\mathrm e^-\}.
\label{eq:plasma-thermal-groups}
\end{equation}
The level populations and their mass balances remain resolved, and specified event energy increments whose constituent sum is zero can transfer energy between the two thermal groups. The principal-subsystem argument uses only the grouping and the parent generator. The particular reaction and energy-exchange laws are still needed to specify the reduced sources.

In that example, identifying heavy-particle temperatures is not
the same as equilibrating the populations of electronic levels.
As shown in \cite[Section~9 and Appendix~B]{BotheRuggeriPaperI2026}, stationary populations
can coexist with energy transfer and positive entropy production
between thermal groups. The grouping considered here retains the
resolved reaction paths and their group-energy transfers. Eliminating
the level populations would require a separate kinetic justification.
The two-group example illustrates thermal resolution, not a complete
plasma model with electromagnetic or radiation balances.

For fully specified RET-compatible parent potentials and sources, this restriction defines an autonomous grouped-temperature Euler subsystem, but it does not make its manifold invariant under the unreduced dynamics. The principal-subsystem theorem does not determine transport coefficients or reaction and exchange rates. These must be supplied by the constitutive model. Its local temporal thermodynamics can equivalently be reconstructed by fixing all group energies \eqref{eq:group-thermal-energy} and maximizing the parent entropy subject to the $K$ constraints,
\begin{equation}
  \bigl((\rho_i e_i)^{*,\mathcal G}\bigr)_{i=1}^{n}
  :=\operatorname*{arg\,max}_{
  \sum_{i\in\mathcal G_a}\widetilde z_i=\mathcal E_a,\ a=1,\ldots,K}
  \eta(\widetilde z_1,\ldots,\widetilde z_n,\rho_1,\ldots,\rho_n).
\label{eq:grouped-entropy-projection}
\end{equation}
If the feasible group-energy slice lies in a convex part of the admissible thermodynamic domain, the negative-definite parent Hessian implies strict concavity on that slice and hence uniqueness whenever the constrained maximum is attained at an interior point of the admissible domain with positive temperatures. Its $K$ Lagrange multipliers yield \eqref{eq:grouped-temperature-manifold}. The corresponding value function defines the grouped entropy, while the remaining constitutive functions must be restricted and reconstructed consistently. \Cref{sec:thermal-reduction} below develops this thermodynamic completion in detail for the one-group reduction to class II. The same constrained-maximum construction supplies intermediate $K$-temperature reductions under the same attainment and domain assumptions.

\par\begingroup
\begin{proposition}[Grouped susceptibility and composition of thermal projections]
\label[proposition]{prop:grouped-projection-composition}
Fix the partial densities and use one convex admissible thermal-energy
domain with $-D_{\boldsymbol z}^2\eta\succ0$. Assume that every
constrained maximum occurring below is attained in its interior.
Let $\mathsf A_{\mathcal G}\in\R^{K\times n}$ be the group-incidence
matrix, with entry $1$ for $i\in\mathcal G_a$ and $0$ otherwise,
and put $\boldsymbol{\mathcal E}=\mathsf A_{\mathcal G}\boldsymbol z$.
Write $\boldsymbol z_{\mathcal G}^*$ for the maximizing partition and
$\eta_{\mathcal G}(\boldsymbol{\mathcal E},\boldsymbol\rho)$ for its entropy.
At this partition set
\[
 \mathsf H_{\mathcal G}:=-D_{\boldsymbol z}^2\eta|_{*,\mathcal G},
 \qquad
 \mathsf C_{\mathcal G}
 :=\mathsf A_{\mathcal G}\mathsf H_{\mathcal G}^{-1}
                    \mathsf A_{\mathcal G}^{\sf T}\succ0.
\]
Then, at fixed partial densities,
\begin{equation}
 D_{\boldsymbol{\mathcal E}}^2\eta_{\mathcal G}
       =-\mathsf C_{\mathcal G}^{-1},\qquad
 D_{\boldsymbol{\mathcal E}}\boldsymbol z_{\mathcal G}^*
       =\mathsf H_{\mathcal G}^{-1}\mathsf A_{\mathcal G}^{\sf T}
                         \mathsf C_{\mathcal G}^{-1}.
\label{eq:grouped-susceptibility}
\end{equation}
Thus the group energies and inverse group temperatures are equivalent
local thermal coordinates. Define
$\mathcal P_{\mathcal G}\boldsymbol z
:=\boldsymbol z_{\mathcal G}^*(\mathsf A_{\mathcal G}\boldsymbol z,\boldsymbol\rho)$,
leaving densities and velocities unchanged. If $\mathcal H$ coarsens
$\mathcal G$, then
\begin{align}
 \mathcal P_{\mathcal H}\mathcal P_{\mathcal G}
 &=\mathcal P_{\mathcal G}\mathcal P_{\mathcal H}
 =\mathcal P_{\mathcal H},\label{eq:grouped-projection-tower}\\
 \eta(\boldsymbol z)&\le\eta(\mathcal P_{\mathcal G}\boldsymbol z)
 \le\eta(\mathcal P_{\mathcal H}\boldsymbol z),\nonumber
\end{align}
where the fixed densities are suppressed. No additivity is needed.
\end{proposition}
\begin{proof}
Stationarity reads
$D_{\boldsymbol z}\eta
=\mathsf A_{\mathcal G}^{\sf T}\boldsymbol\vartheta_{\mathcal G}$,
where $(\boldsymbol\vartheta_{\mathcal G})_a=1/T^{(a)}$.
The bordered stationarity system is nonsingular because
$\mathsf H_{\mathcal G}\succ0$ and $\mathsf A_{\mathcal G}$ has
full row rank, so the maximizing branch is locally $C^1$.
Differentiating gives
$\dd\boldsymbol z^*
=-\mathsf H_{\mathcal G}^{-1}\mathsf A_{\mathcal G}^{\sf T}
 \dd\boldsymbol\vartheta_{\mathcal G}$ and hence
$\dd\boldsymbol{\mathcal E}
=-\mathsf C_{\mathcal G}\dd\boldsymbol\vartheta_{\mathcal G}$.
Full row rank proves positivity; the envelope identity
$D_{\boldsymbol{\mathcal E}}\eta_{\mathcal G}
=\boldsymbol\vartheta_{\mathcal G}$ proves
\eqref{eq:grouped-susceptibility}.
For composition, $\mathcal P_{\mathcal G}$ preserves every coarser
group total, so subsequent $\mathcal H$-maximization uses the same
feasible slice. Conversely, the $\mathcal H$-maximizer also maximizes
entropy on the smaller $\mathcal G$-slice through itself; uniqueness
makes the subsequent $\mathcal G$-projection leave it fixed.
The feasible-slice inclusions give both entropy inequalities.
\end{proof}
For $K=1$, $\mathsf C_{\mathcal G}$ reduces to the scalar $a_*$
in \eqref{eq:thermal-energy-susceptibility}. These are identities of
thermodynamic reconstruction, not assertions of dynamical invariance.
\par\endgroup

\begin{proposition}[Thermal equality and principal restrictions imposing equal velocities]
\label[proposition]{prop:thermal-before-mechanical}
For the Euler main-field variables \eqref{eq:complete-main-field-components}, equality of the velocities of constituents $i$ and $j$ is equivalent to the bilinear main-field condition
\begin{equation}
  \Lambda_j^E\boldsymbol\Lambda_i^{\bv}
  -\Lambda_i^E\boldsymbol\Lambda_j^{\bv}=\boldsymbol0.
\label{eq:bilinear-velocity-equality}
\end{equation}
Near a state with $\bv_i=\bv_j$, on a local dual thermodynamic chart on which $T_i$ and $T_j$ can vary independently, the velocity-equality manifold is not contained in any nontrivial affine level set of a fixed linear functional of the Euler main field $\vp$. Hence velocity equality alone cannot be obtained by fixing $\boldsymbol a\cdot \vp=c$ with state-independent coefficients $\boldsymbol a\ne0$ and constant $c$ and does not define a Boillat--Ruggeri principal subsystem of the unequal-temperature Euler system. After thermal merging, $T_i=T_j$, velocity equality is equivalent to
\begin{equation}
  \boldsymbol\Lambda_i^{\bv}-\boldsymbol\Lambda_j^{\bv}=\boldsymbol0,
\label{eq:linear-velocity-equality-after-thermal}
\end{equation}
which is a linear main-field restriction with constant coefficients. Thermal equality therefore allows velocity equality to be imposed by a classical principal restriction. This does not exclude smaller linear sections that impose velocity equality together with a fixed unequal temperature ratio.
\end{proposition}

\begin{proof}
Since $\boldsymbol\Lambda_i^{\bv}=\bv_i/T_i$ and $\Lambda_i^E=-1/T_i$, one has $\bv_i=-\boldsymbol\Lambda_i^{\bv}/\Lambda_i^E$, and cross multiplication gives \eqref{eq:bilinear-velocity-equality}. To address all components of the Euler main field, first vary the independent dual variables $\chi_k$ while holding all reciprocal temperatures and velocities fixed. Since $\Lambda_k^\rho=-\chi_k-\half\vartheta_k|\bv_k|^2$, any linear functional that is constant on the velocity-equality manifold must have zero coefficients for all mass-main-field components. Independent variation of the remaining constituents similarly removes their momentum and energy coefficients. It remains to consider constituents $i,j$. Suppose the remaining fixed scalar linear functional takes one constant value $c$ on every state with $\bv_i=\bv_j=\boldsymbol c$. Writing $\vartheta_k=1/T_k$, its restriction has the form
\[
  \vartheta_i(\boldsymbol a\cdot\boldsymbol c-a_0)
  +\vartheta_j(\boldsymbol b\cdot\boldsymbol c-b_0)=c
\]
on an open product neighbourhood of $\boldsymbol c,\vartheta_i,\vartheta_j$. Differentiation with respect to $\boldsymbol c$ gives $\vartheta_i\boldsymbol a+\vartheta_j\boldsymbol b=\boldsymbol0$. Independent variation of the reciprocal temperatures forces $\boldsymbol a=\boldsymbol b=\boldsymbol0$. The remaining affine identity then gives $a_0=b_0=c=0$. No assumption that the velocity neighbourhood contains the rest state is needed. Thus no nontrivial fixed linear functional is constant on the whole unequal-temperature velocity-equality manifold. If $T_i=T_j=T$, then $\boldsymbol\Lambda_i^{\bv}-\boldsymbol\Lambda_j^{\bv}=(\bv_i-\bv_j)/T$. Since $T>0$, its vanishing is equivalent to $\bv_i=\bv_j$, which proves \eqref{eq:linear-velocity-equality-after-thermal}.
\end{proof}

\begin{remark}

For any fixed $c>0$, $c\ne1$, the conditions
\[
 \Lambda_i^E=c\Lambda_j^E,\qquad
 \boldsymbol\Lambda_i^{\bv}=c\boldsymbol\Lambda_j^{\bv}
\]
are linear in the main-field components with constant coefficients and imply $T_i=T_j/c$ and $\bv_i=\bv_j$. They can be completed to a constant invertible main-field coordinate change and hence define a classical principal restriction wherever the parent convex-extension hypotheses hold. The retained energy and momentum combinations are weighted, for example $cE_i+E_j$, rather than the physical unweighted group totals. The preceding proposition excludes the whole velocity-equality manifold with independently variable temperatures, not these smaller sections. For equal temperatures, the retained group energies and momenta are the unweighted sums of the physical constituent quantities. None of these principal restrictions is the Class-II-to-Class-I reduction of \Cref{subsec:classII-classI-reduction} below, which retains diffusion fluxes and requires its own balance-compatible mechanical scaling.

\end{remark}

\subsection{Singular pencil and its relation to the Euler spectrum}
\label{subsec:singular-pencil}

The preceding principal restrictions do not require a characteristic
theory for the singular parent. The following separate calculation
records what its first-order pencil does and does not imply.

\begin{proposition}[Singular pencil and unbounded Rayleigh quotient at $\wprime=0$]
\label[proposition]{prop:singular-pencil-structure}
Fix an interior state with $\wprime=(\boldsymbol\Sigma,\bQ)=0$ and a unit spatial    vector $\boldsymbol n\in\R^3$. With the splitting $\bU'=(\vp,\wprime)$, let
\begin{equation}
  A^0:=D_{\bU'}^2h'^{0}
  =\begin{pmatrix}P&0\\0&0\end{pmatrix},
  \qquad P:=D_{\vp}^2h_E'^{0}\succ0,
\label{eq:singular-pencil-time}
\end{equation}
and
\begin{equation}
  A(\boldsymbol n):=\sum_{k=1}^3 n_kD_{\bU'}^2h'^{k}
  =\begin{pmatrix}
      a_E(\boldsymbol n)&b(\boldsymbol n)\\
      b(\boldsymbol n)^{\sf T}&0
    \end{pmatrix},
\label{eq:singular-pencil-space}
\end{equation}
where $a_E(\boldsymbol n)=\sum_kn_kD_{\vp}^2h_E'^{k}$ is the Euler spatial Hessian. The matrix pencil is the family $A(\boldsymbol n)-\lambda A^0$, parametrized by $\lambda$. For a perturbation $\boldsymbol r=(\boldsymbol x,\boldsymbol y)$, $\boldsymbol x$ changes the balance variables $\vp$, whereas $\boldsymbol y$ changes the auxiliary fields $\wprime$. Then:
\begin{enumerate}[label=\textnormal{(\roman*)}]
\item $b(\boldsymbol n)\ne0$ for every unit $\boldsymbol n$.
\item On the domain $\{(\boldsymbol x,\boldsymbol y):\boldsymbol x\ne0\}$, the extended generalized Rayleigh quotient
\begin{equation}
  \mathcal Q_{\boldsymbol n}(\boldsymbol x,\boldsymbol y)
  :=\frac{\boldsymbol r^{\sf T}A(\boldsymbol n)\boldsymbol r}{\boldsymbol r^{\sf T}A^0\boldsymbol r}
  =\frac{\boldsymbol x^{\sf T}a_E(\boldsymbol n)\boldsymbol x
    +2\boldsymbol x^{\sf T}b(\boldsymbol n)\boldsymbol y}{\boldsymbol x^{\sf T}P\boldsymbol x},
  \qquad \boldsymbol r=(\boldsymbol x,\boldsymbol y),
\label{eq:extended-singular-rayleigh}
\end{equation}
satisfies $\sup\mathcal Q_{\boldsymbol n}=+\infty$ and $\inf\mathcal Q_{\boldsymbol n}=-\infty$.
\item Let $K(\boldsymbol n):=\ker b(\boldsymbol n)^{\sf T}$. The set of finite generalized eigenvalues $\lambda$ admitting an eigenvector with nonzero balance-variable component $\boldsymbol x$ in
\begin{equation}
  A(\boldsymbol n)\boldsymbol r=\lambda A^0\boldsymbol r,
  \qquad A^0\boldsymbol r\ne0,
\label{eq:singular-generalized-eigenproblem}
\end{equation}
coincides with the spectrum of the Euler pencil compressed to $K(\boldsymbol n)$, namely the values $\lambda$ for which there is an $\boldsymbol x\in K(\boldsymbol n)\setminus\{0\}$ satisfying
\begin{equation}
  \boldsymbol z^{\sf T}a_E(\boldsymbol n)\boldsymbol x
  =\lambda \boldsymbol z^{\sf T}P\boldsymbol x
  \qquad\text{for every }\boldsymbol z\in K(\boldsymbol n).
\label{eq:compressed-euler-pencil}
\end{equation}
Consequently, if $K(\boldsymbol n)\ne\{0\}$, every finite generalized eigenvalue lies in the Euler characteristic interval,
\begin{equation}
  \lambda_{\min}^{E}(\boldsymbol n)
  \le \lambda^{\mathrm{fin}}
  \le \lambda_{\max}^{E}(\boldsymbol n),
\label{eq:finite-spectrum-euler-bounds}
\end{equation}
where the extremal Euler values are the minimum and maximum of $\boldsymbol x^{\sf T}a_E(\boldsymbol n)\boldsymbol x/(\boldsymbol x^{\sf T}P\boldsymbol x)$ over $\boldsymbol x\ne0$.
\end{enumerate}
The Rayleigh quotient of the singular pencil is unbounded above and
below. Its finite generalized eigenvalues admitting an eigenvector
with a nonzero stored component are the eigenvalues of the Euler pencil
compressed to $K(\boldsymbol n)$.
  \end{proposition}

\begin{proof}
The spatial generator \eqref{eq:extended-spatial-generator} is affine in $\wprime$, so the $\wprime$--$\wprime$ block in \eqref{eq:singular-pencil-space} is zero. On $\wprime=0$ its $\vp$--$\vp$ block is precisely the Euler Hessian $a_E(\boldsymbol n)$. Moreover the term $Q_{i,k}\Lambda_i^E$ gives
\[
  \frac{\partial^2h'^{k}}{\partial Q_{i,l}\,\partial\Lambda_i^E}=\delta_{kl}.
\]
After contraction with $n_k$, the $\Lambda_i^E$--$\bQ_i$ part of $b(\boldsymbol n)$ contains the components $n_l$. Since $\boldsymbol n$ is a unit covector, at least one $n_l$ is nonzero. Hence $b(\boldsymbol n)\ne0$.

Choose $\boldsymbol x$ with $b(\boldsymbol n)^{\sf T}\boldsymbol x\ne0$, which exists because $b(\boldsymbol n)\ne0$, and put $\boldsymbol y=t\,b(\boldsymbol n)^{\sf T}\boldsymbol x$. Then
\[
  \mathcal Q_{\boldsymbol n}(\boldsymbol x,tb^{\sf T}\boldsymbol x)
  =\frac{\boldsymbol x^{\sf T}a_E \boldsymbol x+2t|b^{\sf T}\boldsymbol x|^2}{\boldsymbol x^{\sf T}P\boldsymbol x},
\]
where the argument $\boldsymbol n$ is suppressed. The positive
denominator is independent of $t$, so the quotient is unbounded above
and below.

For a finite generalized eigenpair with $A^0\boldsymbol r\ne0$, one has $\boldsymbol x\ne0$ and \eqref{eq:singular-generalized-eigenproblem} is equivalent to
\begin{equation}
  a_E \boldsymbol x+b \boldsymbol y=\lambda P\boldsymbol x,
  \qquad
  b^{\sf T}\boldsymbol x=0.
\label{eq:singular-eigenpair-blocks}
\end{equation}
Hence $\boldsymbol x\in K:=\ker b^{\sf T}$, and testing the first equation against every $\boldsymbol z\in K$ removes $b\boldsymbol y$ and yields \eqref{eq:compressed-euler-pencil}. Conversely, if $\boldsymbol x\in K\setminus\{0\}$ satisfies \eqref{eq:compressed-euler-pencil}, then $(a_E-\lambda P)\boldsymbol x$ is orthogonal to $K$. In finite dimensions,
\[
  K^\perp=(\ker b^{\sf T})^\perp=\operatorname{Ran}b,
\]
so there is a $\boldsymbol y$ with $b\boldsymbol y=-(a_E-\lambda P)\boldsymbol x$, and \eqref{eq:singular-eigenpair-blocks} follows. This proves the exact characterization of the finite generalized eigenvalues. Finally, the Rayleigh--Ritz min--max principle on the subspace $K$ gives \eqref{eq:finite-spectrum-euler-bounds}. If $K=\{0\}$ there are no finite generalized eigenpairs with $A^0\boldsymbol r\ne0$ and the assertion is vacuous.
\end{proof}

\begin{corollary}[Explicit constrained subspace and finite stored spectrum]
\label[corollary]{cor:explicit-stored-spectrum}
For the full symmetric-stress and heat-flux generator \eqref{eq:extended-spatial-generator}, and every unit    vector $\boldsymbol n\in\R^3$,
\begin{equation}
 K(\boldsymbol n)
 =\{\boldsymbol x:\boldsymbol x_i^{\bv}=\boldsymbol0,\ x_i^E=0\text{ for all }i\},
 \qquad \dim K=n,\qquad\operatorname{rank}b(\boldsymbol n)=4n.
\label{eq:explicit-stored-kernel}
\end{equation}
In particular, this stored subspace is independent of direction. The finite generalized eigenvalues $\lambda$ admitting an eigenvector with nonzero stored component are exactly
\begin{equation}
 \lambda_i^{\mathrm{fin}}(\boldsymbol n)=\bv_i\cdot\boldsymbol n,
 \qquad i=1,\ldots,n,
\label{eq:explicit-finite-stored-speeds}
\end{equation}
with repetitions counted in the compressed pencil.
\end{corollary}
\begin{proof}
For an increment $\boldsymbol x$ of $\vp$, differentiation of $\bv_i=-\boldsymbol\Lambda_i^{\bv}/\Lambda_i^E$ gives $\delta\bv_i=T_i(\boldsymbol x_i^{\bv}+\bv_i x_i^E)$. The stress and heat components of the mixed block satisfy
\[
 (b^{\sf T}\boldsymbol x)_{\boldsymbol\Sigma_i}
 =-\operatorname{sym}(\delta\bv_i\otimes\boldsymbol n),
 \qquad
 (b^{\sf T}\boldsymbol x)_{\bQ_i}=\boldsymbol n x_i^E.
\]
Here $\operatorname{sym}A=(A+A^{\sf T})/2$. Since
\[
 |\operatorname{sym}(\boldsymbol a\otimes\boldsymbol n)|^2
 =\tfrac12\bigl(|\boldsymbol a|^2|\boldsymbol n|^2
                  +(\boldsymbol a\cdot\boldsymbol n)^2\bigr),
\]
these components vanish precisely when $x_i^E=0$ and $\boldsymbol x_i^{\bv}=\boldsymbol0$. There are $5n$ stored variables, proving the dimension and rank assertions. Constituent additivity makes the compressed temporal matrix diagonal with entries $P_i^\rho:=\partial\rho_i/\partial\Lambda_i^\rho>0$. Since $\bv_i$ is independent of $\Lambda_i^\rho$, differentiation of the generated mass flux $\rho_i\bv_i\cdot\boldsymbol n$ gives the compressed spatial entries $(\bv_i\cdot\boldsymbol n)P_i^\rho$. This proves \eqref{eq:explicit-finite-stored-speeds}.
\end{proof}

The spectral convention matters because the full pencil is singular for every $\lambda$. Its common null vectors, annihilated by both $A^0$ and $A(\boldsymbol n)$, are $(0,\boldsymbol y)$ with $\boldsymbol y\in\ker b(\boldsymbol n)$. Explicitly they satisfy $\delta\boldsymbol\Sigma_i\boldsymbol n=\boldsymbol0$ and $\delta\bQ_i\cdot\boldsymbol n=0$, and form a $5n$-dimensional space. It is this auxiliary kernel that depends on direction. Such vectors solve the pencil equation for every $\lambda$ but are excluded by $A^0\boldsymbol r\ne0$ in \eqref{eq:singular-generalized-eigenproblem}. No ordinary determinant-based spectrum of the unreduced singular pencil is being asserted.

\begin{remark}[Relation to the classical subcharacteristic hierarchy]
\label[remark]{rem:singular-versus-classical-subcharacteristic}

The speeds $\bv_i\cdot\boldsymbol n$ in
\eqref{eq:explicit-finite-stored-speeds} characterize the eigenpairs of
the singular first-order pencil with a nonzero stored component. The
Euler acoustic speeds are obtained from the full Euler pencil. The
frequency-dependent dispersion relation of the viscous mixture
requires linearizing the evolution after constitutive elimination.

The quotient \eqref{eq:extended-singular-rayleigh} has supremum and
infimum $+\infty$ and $-\infty$, respectively. For a fixed $\boldsymbol x$
with $b^{\sf T}\boldsymbol x\ne0$, its numerator varies linearly with
$\boldsymbol y$ and its positive denominator is fixed. Eliminating the zero-storage constitutive variables through their equations produces the viscous and heat-conducting second-order terms. Setting $\boldsymbol y=0$ gives the Euler Rayleigh quotient,
whose extrema are finite. The Boillat--Ruggeri theorem applies to the
Euler subsystems along each chain of coarsening thermal partitions by
\Cref{thm:boillat-ruggeri-principal,prop:nested-grouped-principal}.
Including a further mechanical or chemical restriction in that hierarchy
requires an admissible affine main-field section with state-independent
coefficients and retention of the equations conjugate to the free
coordinates.
\Cref{prop:thermal-before-mechanical} gives the corresponding
velocity-equality conditions.
\end{remark}

\begin{remark}[Qualitative PDE scope and the Shizuta--Kawashima question]
\label[remark]{rem:degenerate-qualitative-scope}
The Euler principal subsystems have positive-definite temporal Hessians. Applying a local symmetric-hyperbolic Cauchy theorem additionally requires its coefficient, source, and initial-data regularity hypotheses and uniform positivity on the chosen state neighbourhood. The $C^2$ thermodynamic assumption alone is not a complete well-posedness hypothesis. For the enlarged first-order system of \Cref{subsec:complete-field-and-generator}, what is proved here is symmetry of the generator matrices, positive semidefiniteness of the temporal block, the entropy identities, and the singular-pencil structure of \Cref{prop:singular-pencil-structure}. These facts alone do not provide a Cauchy theory for the differential--algebraic system. After elimination of the zero-storage constitutive variables, however, \Cref{rem:factorized-dissipative-symbol} gives the frozen highest-order dissipative symbol
\[
  \mathcal B(\boldsymbol n)
  =\mathcal L(\boldsymbol n)^{\sf T}\mathbb R\mathcal L(\boldsymbol n)\succeq0.
\]

\par\begingroup
For the strictly positive transport coefficients of
\Cref{prop:degenerate-acta-generator}, a sharper algebraic statement holds:
with $P$ and $a_E(\boldsymbol n)$ as in
\Cref{prop:singular-pencil-structure}, for every unit $\boldsymbol n$,
\[
 \ker\mathcal B(\boldsymbol n)=K(\boldsymbol n),\qquad
 \ker\mathcal B(\boldsymbol n)\cap
 \ker\bigl(a_E(\boldsymbol n)-\lambda P\bigr)=\{0\}
 \quad(\lambda\in\R).
\]
Indeed, \Cref{cor:explicit-stored-spectrum} and
$\mathbb R\succ0$ give $\delta\bv_i=0$ and $\delta T_i=0$ for any
$\boldsymbol x\in\ker\mathcal B(\boldsymbol n)$.
If $\boldsymbol x$ were also an Euler eigenvector, its mass and momentum
equations would imply
\[
 (\bv_i\cdot\boldsymbol n-\lambda)\delta\rho_i=0,\qquad
 \bv_i(\bv_i\cdot\boldsymbol n-\lambda)\delta\rho_i
       +\boldsymbol n\,\delta p_i=0 .
\]
Hence $\delta p_i=0$. At fixed $T_i$ and $\bv_i$, the constituent
Gibbs--Duhem relation gives
$\delta p_i=\rho_iT_i\,x_i^\rho$, so $\boldsymbol x=0$.
Thus no Euler characteristic eigenvector lies in the dissipative kernel.
The eigenvectors of the compressed pencil need not be eigenvectors of the
full Euler pencil, so this does not contradict
\eqref{eq:explicit-finite-stored-speeds}.
A related single-fluid kernel argument for a distinct hyperbolic
relaxation model is given in \cite{Freistuehler2021}.
\par\endgroup
\par\begingroup
\paragraph{Sharp transport condition in three dimensions.}
The bulk viscosity need not be positive for this coupling property.
Retain the nonnegative diagonal transport laws and use the mixed
formulation when a coefficient vanishes. For a unit $\boldsymbol n$ set
$\delta\bv_{i,\parallel}=(\delta\bv_i\cdot\boldsymbol n)\boldsymbol n$
and $\delta\bv_{i,\perp}=\delta\bv_i-\delta\bv_{i,\parallel}$.
The factorization \eqref{eq:factorized-dissipative-symbol} gives
\begin{equation}
 \boldsymbol x^{\sf T}\mathcal B(\boldsymbol n)\boldsymbol x
 =\sum_i\left[
 \frac{\eta_i^{\mathrm v}}{T_i}|\delta\bv_{i,\perp}|^2
 +\frac{4\eta_i^{\mathrm v}/3+\lambda_i^{\mathrm v}}{T_i}
       |\delta\bv_{i,\parallel}|^2
 +\kappa_i|\delta\vartheta_i|^2\right].
 \label{eq:sharp-transport-symbol}
\end{equation}
Indeed,
$|\operatorname{sym}(\boldsymbol a\otimes\boldsymbol n)^\circ|^2
=\tfrac12|\boldsymbol a|^2+\tfrac16|\boldsymbol a\cdot\boldsymbol n|^2$.
Consequently, the condition
\[
 \ker\mathcal B(\boldsymbol n)\cap
 \ker(a_E(\boldsymbol n)-\lambda P)=\{0\}
 \quad\text{for every unit }\boldsymbol n\text{ and }\lambda\in\R
\]
holds if and only if $\eta_i^{\mathrm v}>0$ and $\kappa_i>0$
for every constituent; the bulk viscosities may be zero.
Sufficiency follows from \eqref{eq:sharp-transport-symbol} and the
preceding Gibbs--Duhem argument. For necessity, if
$\eta_i^{\mathrm v}=0$, a perturbation of constituent $i$ alone with
$\delta\rho_i=\delta z_i=0$ and nonzero
$\delta\bv_i\perp\boldsymbol n$ is an Euler eigenvector at speed
$\bv_i\cdot\boldsymbol n$ in the dissipative kernel.
If $\kappa_i=0$, choose a nonzero thermodynamic perturbation of constituent $i$ alone with
$\delta\bv_i=0$ and $\delta p_i=0$. Such a perturbation exists:
in local coordinates $(\vartheta_i,\chi_i)$, the Gibbs--Duhem identity
gives $\delta\chi_i=-(z_i+p_i)\delta\vartheta_i/\rho_i$ at fixed
pressure, and one may take $\delta\vartheta_i\ne0$.
This is an isobaric Euler contact eigenvector at the same speed and
again belongs to the dissipative kernel. These necessity statements
concern the Euler and second-order matrices only; reaction and
exchange sources and additional first-order terms may affect the
complete linearized dynamics.
\par\endgroup

The Shizuta--Kawashima K-condition is a genuine-coupling condition
for systems with a regular hyperbolic evolution part (and suitable
source or dissipative structure). See~\cite{KawashimaShizuta1988,ShizutaKawashima1985}.
Its application to hyperbolic Euler-mixture systems, including a binary
example and large-time estimates, was already studied in
\cite{RuggeriSimic2007}. See also the treatment in
   \mbox{
\cite[Chapters~2 and~28]{RuggeriSugiyama2021}}.
The unreduced generator here has a singular temporal block, so that
earlier hyperbolic result cannot simply be applied to this first-order
differential--algebraic    formulation. 

The kernel statement above concerns the Euler spatial matrix and the
highest-order dissipative symbol. A local Cauchy theorem requires
regularity and estimates for the full eliminated operator, including
its first-order terms. Stability also requires analysis of the
linearized reaction and relaxation sources.
\end{remark}

\section{Entropy-production-invariant Class-III-to-Class-II reduction}
\label{sec:thermal-reduction}

The grouped main-field construction of \Cref{subsec:grouped-principal-subsystems} defines autonomous Euler restrictions for fully specified RET-compatible parent generators. Constrained entropy maximization gives the same local reduced entropy
and temporal Legendre potential as these restrictions. The construction
applies also to parents without RET compatibility and is used below
to reduce the specified dissipative constitutive laws. Dynamical
invariance of the reduced manifold is a separate question. Unless
explicitly stated otherwise, the analysis uses the general Class-III
parent satisfying \eqref{eq:negative-definite-entropy-hessian} and~\eqref{eq:positive-temperature-domain} and the pressure    and entropy-flux  representation of \Cref{subsec:entropy-flux-pressure}. The RET additivity condition of \Cref{cor:ret-compatible-classIII} is not assumed. The one-group reduction is developed in detail, and the same constrained-maximum construction applies to intermediate group-energy constraints.

Let $\mathcal P_T$ denote the thermal reduction map: it preserves the partial densities, constituent velocities, and total thermal-energy density, eliminates the $n-1$ independent relative-energy variables, and reconstructs the constituent thermal energies on the common-temperature manifold. We call this constitutive reduction \emph{entropy-production invariant} (``entropy-invariant'' in the terminology of the parent reduction principle) if
\begin{equation}
  \zeta^{\mathrm{II}}
  \equiv \mathcal P_T\!\left(\zeta^{\mathrm{III}}\right)
\label{eq:entropy-invariant-reduction-principle}
\end{equation}
as an identity of entropy-production expressions before substitution
of the constitutive laws. The right-hand side denotes evaluation of
$\zeta^{\mathrm{III}}$ on reconstructed fields, with their spatial
derivatives obtained by the chain rule. It does not mean that the entropy of an arbitrary off-manifold Class-III state equals the entropy of its reduced image, nor that every solution of the reduced equations is a solution of the parent system. A contribution conjugate to an eliminated variable may vanish identically on the reduced manifold. If its elimination leaves a nonzero finite remainder, that remainder must be retained as a reduced constitutive mechanism or neglected only under an explicit approximation.

\subsection{Thermodynamic projection and definition of the common temperature}
\label{subsec:thermal-projection}

A model retaining only the total thermal energy does not resolve its
distribution among constituents. At fixed
partial densities and velocities, many energy partitions represent
the same retained state. Entropy maximization selects the thermally
equilibrated partition compatible with those data.    This  local state reconstruction    preserves the chemical composition and does not assert convergence of the full dynamics to the maximizing  partition. In particular, the chemical mass balances and the
resolved mechanical energy are not removed by this thermal operation.

Fix the partial densities and the total thermal-energy density
\begin{equation}
  \rho e=\sum_i\rho_i e_i.
\label{eq:fixed-total-thermal-energy}
\end{equation}
The thermodynamic projection onto the Class-II state space is defined by constrained entropy maximization,
\begin{equation}
  \bigl((\rho_1e_1)^{*},\ldots,(\rho_ne_n)^{*}\bigr)
  :=\operatorname*{arg\,max}_{\sum_i\widetilde z_i=\rho e}
  \eta(\widetilde z_1,\ldots,\widetilde z_n,\rho_1,\ldots,\rho_n).
\label{eq:entropy-maximization}
\end{equation}
Assume that the feasible energy slice lies in a convex part of the admissible thermodynamic domain, intersects the interior of the positive-temperature domain specified by \eqref{eq:entropy-state} and~\eqref{eq:positive-temperature-domain}, and that the maximum is attained there. The negative-definite Hessian \eqref{eq:negative-definite-entropy-hessian} implies strict concavity on this slice and hence uniqueness. With $1/T$ denoting the Lagrange multiplier of the energy constraint, stationarity gives
\begin{equation}
  \left.\frac{1}{T_1}\right|_{*}
  =\cdots=
  \left.\frac{1}{T_n}\right|_{*}
  =\frac{1}T.
\label{eq:entropy-max-temp}
\end{equation}
The domain assumption \eqref{eq:positive-temperature-domain} implies $1/T>0$ at the maximizing state and hence $T>0$. We denote the attained maximum by $\rho s^{\mathrm{II}}(\rho e,\boldsymbol\rho)$, the reduced entropy defined explicitly in \eqref{eq:reduced-classII-entropy} below.

\begin{definition}[Entropy-maximizing energy-equivalent common temperature]
The common temperature associated with a Class-III state is the temperature $T$ of the entropy-maximizing state \eqref{eq:entropy-maximization} with the same partial densities and the same total thermal energy. For the general nonadditive parent this is the definition. Its reduction to the earlier average-temperature formula for additive mixtures is shown below.
\label[definition]{def:average-temperature}
\end{definition}

Equivalently, let $\rho e^{\mathrm{II}}(T,\rho_1,\ldots,\rho_n)$ denote the total thermal energy evaluated on the common-temperature manifold. Then $T$ is determined locally by
\begin{equation}
  \rho e^{\mathrm{II}}(T,\rho_1,\ldots,\rho_n)
  :=\sum_i(\rho_i e_i)^{*}(T,\rho_1,\ldots,\rho_n)
  =\sum_i\rho_i e_i.
\label{eq:average-temp-energy-equation}
\end{equation}

The required positive reduced heat capacity follows from the parent entropy Hessian. At fixed partial densities, use $\boldsymbol z=(z_1,\ldots,z_n)^{\sf T}$ from
\eqref{eq:thermal-density-notation}, let $\boldsymbol z^*(\rho e)$ be its maximizing partition, and set
\begin{equation}
  \mathsf H_*:=-D_{\boldsymbol z}^2\eta\big|_*\succ0,
  \qquad
  a_*:=\sum_{i,j=1}^n(\mathsf H_*^{-1})_{ij}>0.
\label{eq:thermal-energy-susceptibility}
\end{equation}
The quantity $a_*$ is positive because $\mathsf H_*^{-1}$ is
positive definite. Differentiating the stationarity equations
$\partial_{z_i}\eta|_*=1/T$ ($i=1,\ldots,n$) and the energy constraint
gives, componentwise,
\begin{equation}
  \partial_{\rho e}z_i^*
  =\frac{\sum_{j=1}^n(\mathsf H_*^{-1})_{ij}}{a_*}
  \quad(i=1,\ldots,n),
  \qquad
  \sum_{i=1}^n\partial_{\rho e}z_i^*=1.
\label{eq:thermal-optimizer-tangent}
\end{equation}
The envelope identity and reduced curvature are therefore
\begin{equation}
  \frac{\partial\rho s^{\mathrm{II}}}{\partial(\rho e)}=\frac{1}T,
  \qquad
  \frac{\partial^2\rho s^{\mathrm{II}}}{\partial(\rho e)^2}
  =-\frac{1}{a_*}<0.
\label{eq:reduced-entropy-curvature}
\end{equation}
Consequently
\begin{equation}
  \frac{\partial T}{\partial(\rho e)}=\frac{T^2}{a_*}>0,
  \qquad
  \frac{\partial\rho e^{\mathrm{II}}}{\partial T}=\frac{a_*}{T^2}>0.
\label{eq:positive-total-heat-capacity}
\end{equation}
The linearization of the stationarity equations together with the energy constraint is invertible. The implicit-function theorem therefore gives a local $C^1$ maximizing partition under the stated $C^2$ entropy hypothesis. Higher branch regularity requires corresponding higher regularity of the entropy. These statements require neither constituent additivity nor ideal-mixture equations of state. Positivity holds for the total reduced heat capacity. For a nonadditive entropy it does not force every component of $\partial_T\boldsymbol z^*$ to be positive. Global invertibility on a large state domain remains a separate question.

For an additive simple mixture with $\rho_i e_i=\rho_i e_i(\rho_i,T_i)$, \eqref{eq:average-temp-energy-equation} is the energy definition used in the multi-temperature literature \cite{GouinRuggeri2008}. Let $T_0$ be a reference common temperature and set $c_{V,i}(T_0):=(\partial e_i/\partial T_i)_{\rho_i,T_0}$. Assuming bounded second temperature derivatives of the caloric functions near $T_0$, for $T_i-T_0$ small energy equivalence gives
\begin{equation}
  T-T_0
  =\frac{\sum_i\rho_i c_{V,i}(T_0)(T_i-T_0)}{\sum_i\rho_i c_{V,i}(T_0)}
  +O\!\left(\max_i|T_i-T_0|^2\right).
\label{eq:heat-capacity-average}
\end{equation}

For calorically perfect ideal gases with
$e_i=e_i^0+c_{V,i}T_i$, this relation is exact:
\[
 T=\frac{\sum_i\rho_i c_{V,i}T_i}{\sum_i\rho_i c_{V,i}}.
\]
The weights are volumetric heat capacities, not equal weights and not
mass fractions alone. At fixed partial densities the energy offsets
$e_i^0$ cancel. For a general nonadditive entropy, this formula is
replaced by the constrained maximization and energy relation
\eqref{eq:entropy-maximization} and~\eqref{eq:average-temp-energy-equation}, rather
than postulated as an averaging rule.

For additive nonreacting mixtures this is exactly the energy-equivalent average temperature of the earlier multi-temperature theory~\cite{GouinRuggeri2008,RuggeriSimic2009}. Its local entropy maximum is the additive specialization of \eqref{eq:entropy-maximization}. The present construction extends the definition to one possibly nonadditive Class-III entropy potential.

\par\begingroup
\paragraph{Range of the reconstructed temperature.}
For nonadditive entropy, positivity of the reduced heat capacity does
not imply $\min_iT_i\le T\le\max_iT_i$.
For a local example in dimensionless variables, take two constituents,
$\boldsymbol z_0=\boldsymbol\rho_0=(1,1)^{\sf T}$, and
\[
 \eta=4+\boldsymbol1^{\sf T}(\boldsymbol z-\boldsymbol z_0)
       -\frac12(\boldsymbol z-\boldsymbol z_0)^{\sf T}
          \begin{pmatrix}5&2\\2&1\end{pmatrix}
          (\boldsymbol z-\boldsymbol z_0)
       -\frac12|\boldsymbol\rho-\boldsymbol\rho_0|^2 .
\]
The full entropy Hessian is negative definite. At
$\boldsymbol\rho=\boldsymbol\rho_0$ and
$\boldsymbol z-\boldsymbol z_0=(-1/10,1/5)^{\sf T}$,
the temperatures are $T_1=10/11$ and $T_2=1$.
At the same total thermal energy the maximizing partition is
$\boldsymbol z^*-\boldsymbol z_0=(-1/20,3/20)^{\sf T}$, giving
$T=20/19>\max(T_1,T_2)$ and entropy increase $1/400$.
Both states lie in a convex positive-energy, positive-temperature
neighbourhood. The common temperature is therefore an
energy-equivalent entropy-maximizing reconstruction, not in general
an average with nonnegative constituent weights.
\par\endgroup

The projection is entropy increasing unless the original state already has a common temperature:
\begin{equation}
  \rho s^{\mathrm{II}}(\rho e,\rho_1,\ldots,\rho_n)
  -\eta(\rho_1e_1,\ldots,\rho_ne_n,\rho_1,\ldots,\rho_n)
  \ge0.
\label{eq:thermal-projection-entropy-deficit-exact}
\end{equation}
With
\begin{equation}
  \delta z_i:=\rho_i e_i-(\rho_i e_i)^{*},
  \qquad
  \delta\boldsymbol z:=(\delta z_1,\ldots,\delta z_n)^{\sf T},
  \qquad
  \sum_i \delta z_i=0,
\label{eq:relative-energy-coordinates}
\end{equation}
choose a constant full-column-rank matrix
$\mathsf S_{\mathrm{rel}}\in\mathbb R^{n\times(n-1)}$ whose columns
have zero sum, $\sum_{i=1}^n(\mathsf S_{\mathrm{rel}})_{ij}=0$
for $j=1,\ldots,n-1$, and write
$\delta\boldsymbol z=\mathsf S_{\mathrm{rel}}\boldsymbol z_{\mathrm{rel}}$.
The vector $\boldsymbol z_{\mathrm{rel}}\in\mathbb R^{n-1}$ contains
independent relative-energy coordinates. The near-manifold deficit is
\begin{equation}
  \eta(\boldsymbol z,\boldsymbol\rho)
  =\rho s^{\mathrm{II}}
  -\half\boldsymbol z_{\mathrm{rel}}^{\sf T}
        \boldsymbol C_{\mathrm{rel}}^{-1}\boldsymbol z_{\mathrm{rel}}
  +o(|\boldsymbol z_{\mathrm{rel}}|^2),
\label{eq:entropy-deficit}
\end{equation}
where
\begin{equation}
  \boldsymbol C_{\mathrm{rel}}^{-1}
  :=\mathsf S_{\mathrm{rel}}^{\sf T}\mathsf H_*\mathsf S_{\mathrm{rel}}
  =-\mathsf S_{\mathrm{rel}}^{\sf T}
     \left.D_{\boldsymbol z}^2\eta\right|_*
     \mathsf S_{\mathrm{rel}}\succ0.
\label{eq:relative-energy-hessian}
\end{equation}
The negative-definite parent Hessian and the full column rank of
$\mathsf S_{\mathrm{rel}}$ prove positivity. This explicitly distinguishes
the $n$-component increment $\delta\boldsymbol z$ from its $n-1$
independent coordinates. $\boldsymbol C_{\mathrm{rel}}$ acts on the latter. The entropy increase associated with approach to the manifold belongs to the eliminated thermal-equilibration dynamics. It is not a production term of the strict Class-II slow model.

The positive quadratic deficit therefore measures the local
thermodynamic departure associated with the discarded relative-energy
variables. It can indicate whether a state is close to thermal
equilibrium, but by itself it does not bound the error between
solutions of the full and reduced evolution equations.

\begin{remark}[Local primal--dual form of the thermal reduction]
\label[remark]{rem:primal-dual-thermal-reduction}
After a constant invertible recombination of the constituent total energies into retained group energies and unshifted relative-energy coordinates, write the mathematical entropy as $h^{\mathrm{III}}(\boldsymbol X,\boldsymbol Y)=-\rho s$, where $\boldsymbol X$ collects the retained conservative variables and $\boldsymbol Y$ collects the eliminated relative-energy variables. On a convex local coordinate neighbourhood containing the minimizing branch, the thermodynamic reduction is
\begin{equation}
  h^{\mathrm{red}}(\boldsymbol X)=\inf_{\boldsymbol Y} h^{\mathrm{III}}(\boldsymbol X,\boldsymbol Y).
\label{eq:primal-infimal-thermal-reduction}
\end{equation}
Here subscripts $X,Y$ denote derivatives with respect to the vector
blocks $\boldsymbol X,\boldsymbol Y$. At the minimizing branch $h_Y^{\mathrm{III}}=0$, which is exactly equality of the reciprocal temperatures within the corresponding groups. Since the parent conservative Hessian is positive definite by \Cref{prop:general-conservative-convexity}, the $YY$ block is invertible there and the reduced Hessian is the positive Schur complement
\begin{equation}
  D_{\boldsymbol X}^2 h^{\mathrm{red}}
  =h_{XX}^{\mathrm{III}}
   -h_{XY}^{\mathrm{III}}(h_{YY}^{\mathrm{III}})^{-1}h_{YX}^{\mathrm{III}}
  \succ0.
\label{eq:thermal-reduction-hessian-schur}
\end{equation}
On the corresponding neighbourhood where the local Legendre maps are diffeomorphisms, the Legendre potentials satisfy
\begin{equation}
  (h^{\mathrm{red}})^*(\boldsymbol\Lambda_X)
  =(h^{\mathrm{III}})^*(\boldsymbol\Lambda_X,0).
\label{eq:thermal-reduction-dual-restriction}
\end{equation}

   Minimization over the eliminated energy variables imposes the stationarity condition $h_Y^{\mathrm{III}}=0$. The Schur complement    is the Hessian of the resulting reduced entropy.
This preserves local thermodynamic stability, but does not supply
missing spatial flux potentials for a nonadditive parent.

Thus entropy maximization in the energy variables and zero restriction of the conjugate relative-temperature variables give the same local reduced entropy density and Legendre potential. This is not a global Fenchel-duality assertion on an unspecified nonconvex state domain. The two operations also remain distinct dynamically: \eqref{eq:thermal-reduction-dual-restriction} does not make the reduced manifold invariant and does not replace the constitutive reconstruction required for an autonomous lower-class model.
\end{remark}

\par\begingroup
\begin{proposition}[Exact relative-entropy decomposition under thermal projection]
\label[proposition]{prop:thermal-relative-entropy-decomposition}
Let $\Omega$ be an open convex conservative-coordinate neighbourhood
on which $h^0(\bV)=-\rho s$ belongs to $C^2(\Omega)$ and has
positive-definite Hessian. Let $\Pi$ be the constant full-row-rank
map retaining the partial densities, partial momenta and the selected
sums of constituent total energies. Let $\mathcal X\subset\Pi\Omega$
be open and convex, and assume that, for every
$\boldsymbol X\in\mathcal X$, the minimum of $h^0$ over
$\{\bV\in\Omega:\Pi\bV=\boldsymbol X\}$ is attained in $\Omega$.
Denote its unique minimizer by $\mathcal M(\boldsymbol X)$ and put
$h^{\mathrm{red}}=h^0\circ\mathcal M$.
Then $\mathcal M\in C^1(\mathcal X)$ and
$h^{\mathrm{red}}\in C^2(\mathcal X)$, with positive-definite
reduced Hessian.
For a differentiable function $f$, write
\[
 f(\boldsymbol a\mid\boldsymbol b)
 :=f(\boldsymbol a)-f(\boldsymbol b)
   -Df(\boldsymbol b)\cdot(\boldsymbol a-\boldsymbol b).
\]
For every $\bV\in\Omega$ with $\Pi\bV\in\mathcal X$ and every
$\overline{\boldsymbol X}\in\mathcal X$, one has
\begin{equation}
 h^0\bigl(\bV\mid\mathcal M(\overline{\boldsymbol X})\bigr)
 =h^0(\bV)-h^{\mathrm{red}}(\Pi\bV)
  +h^{\mathrm{red}}\bigl(\Pi\bV\mid\overline{\boldsymbol X}\bigr).
\label{eq:thermal-relative-entropy-decomposition}
\end{equation}
The first term on the right is the exact thermal entropy deficit
on the retained fibre; it is nonnegative and vanishes precisely
when $\bV=\mathcal M(\Pi\bV)$. The second is nonnegative and
vanishes precisely when $\Pi\bV=\overline{\boldsymbol X}$.
More explicitly,
\begin{align}
 h^0(\bV)-h^{\mathrm{red}}(\Pi\bV)
 &=h^0\bigl(\bV\mid\mathcal M(\Pi\bV)\bigr),\nonumber\\
 h^{\mathrm{red}}\bigl(\Pi\bV\mid\overline{\boldsymbol X}\bigr)
 &=h^0\bigl(\mathcal M(\Pi\bV)\mid
                         \mathcal M(\overline{\boldsymbol X})\bigr).
\label{eq:thermal-relative-entropy-identifications}
\end{align}
Thus \eqref{eq:thermal-relative-entropy-decomposition} is also an
exact decomposition into two parent relative entropies. These
identities hold for nonadditive parent entropies as well.
\end{proposition}
\begin{proof}
Strict convexity on each feasible slice gives uniqueness. At its
interior minimum there is a unique multiplier $\boldsymbol\lambda$
such that
\[
 Dh^0\bigl(\mathcal M(\boldsymbol X)\bigr)
   =\Pi^{\sf T}\boldsymbol\lambda(\boldsymbol X),
 \qquad \Pi\mathcal M(\boldsymbol X)=\boldsymbol X.
\]
Writing $\mathsf H_c=D_{\bV}^2h^0(\mathcal M(\boldsymbol X))$,
the Jacobian of this bordered stationarity system with respect to
the minimizing state and multiplier is
\[
 \begin{pmatrix}\mathsf H_c&-\Pi^{\sf T}\\\Pi&0\end{pmatrix}.
\]
It is nonsingular: a null vector $(\boldsymbol a,\boldsymbol b)$
satisfies $\Pi\boldsymbol a=0$ and
$\mathsf H_c\boldsymbol a=\Pi^{\sf T}\boldsymbol b$, hence
$\boldsymbol a^{\sf T}\mathsf H_c\boldsymbol a=0$; positivity
and full row rank then give $\boldsymbol a=\boldsymbol b=0$.
The implicit-function theorem and uniqueness therefore give
$\mathcal M,\boldsymbol\lambda\in C^1(\mathcal X)$.
The chain rule gives $Dh^{\mathrm{red}}=\boldsymbol\lambda$,
so $h^{\mathrm{red}}\in C^2$, although only $C^1$ regularity
of $\mathcal M$ is needed. Differentiating the stationarity system
yields the coordinate-independent form of
\eqref{eq:thermal-reduction-hessian-schur}:
\begin{equation}
 D^2h^{\mathrm{red}}
   =(\Pi\mathsf H_c^{-1}\Pi^{\sf T})^{-1}\succ0,
 \qquad
 D\mathcal M=\mathsf H_c^{-1}\Pi^{\sf T}D^2h^{\mathrm{red}}.
\label{eq:thermal-relative-entropy-reconstruction-derivatives}
\end{equation}

The gradient identity
$Dh^0(\mathcal M(\boldsymbol X))
=\Pi^{\sf T}Dh^{\mathrm{red}}(\boldsymbol X)$ now follows.
Insert it into the left-hand side of
\eqref{eq:thermal-relative-entropy-decomposition} and add and
subtract $h^{\mathrm{red}}(\Pi\bV)$.
The same identity annihilates the fibre increment
$\bV-\mathcal M(\Pi\bV)$ and identifies the relative entropy
between two reconstructed states, proving
\eqref{eq:thermal-relative-entropy-identifications}.
Strict convexity of $h^0$ on $\Omega$ and of
$h^{\mathrm{red}}$ on $\mathcal X$ proves nonnegativity
and both equality characterizations.
\end{proof}

In particular, for a reconstructed reference state,
\begin{equation}
 h^0\bigl(\mathcal M(\Pi\bV)\mid
                    \mathcal M(\overline{\boldsymbol X})\bigr)
 \le h^0\bigl(\bV\mid\mathcal M(\overline{\boldsymbol X})\bigr),
\label{eq:thermal-relative-entropy-projection-inequality}
\end{equation}
with equality if and only if $\bV=\mathcal M(\Pi\bV)$.
Moreover, for fixed $\bV$, the unique minimum of
$h^0(\bV\mid\mathcal M(\overline{\boldsymbol X}))$ over
$\overline{\boldsymbol X}\in\mathcal X$ occurs at
$\overline{\boldsymbol X}=\Pi\bV$. The reconstructed comparison
state occupies the second argument throughout these statements.

For two nested thermal groupings, with $\mathcal H$ coarser than
$\mathcal G$, impose the common-domain and interior-attainment
hypotheses of \Cref{prop:grouped-projection-composition}, with all
reconstructed states and retained comparison states in the domains
specified above.
Since partial densities and momenta are retained, fixing group
total energies is equivalent to fixing group thermal energies.
The same identity and \eqref{eq:grouped-projection-tower} give
\begin{equation}
 h^0(\bV\mid\mathcal P_{\mathcal H}\bV)
 =h^0(\bV\mid\mathcal P_{\mathcal G}\bV)
  +h^0(\mathcal P_{\mathcal G}\bV\mid\mathcal P_{\mathcal H}\bV).
\label{eq:thermal-relative-entropy-grouped-decomposition}
\end{equation}
Indeed, the tower property makes $\mathcal P_{\mathcal H}\bV$
a fixed point of $\mathcal P_{\mathcal G}$, so it is an admissible
reconstructed reference for the finer projection. The identity
separates the entropy gain of the finer equilibration from the
additional gain caused by coarsening.
\par\endgroup

\subsection{Reduced entropy, conjugate variables, and Euler relation}
\label{subsec:reduced-thermodynamics}

Define the reduced Class-II entropy by the value function
\begin{equation}
  \rho s^{\mathrm{II}}(\rho e,\rho_1,\ldots,\rho_n)
  :=\max_{\sum_i\widetilde z_i=\rho e}
  \eta(\widetilde z_1,\ldots,\widetilde z_n,\rho_1,\ldots,\rho_n).
\label{eq:reduced-classII-entropy}
\end{equation}
The envelope theorem and \eqref{eq:entropy-conjugates} and~\eqref{eq:entropy-max-temp} yield
\begin{equation}
  \dd(\rho s^{\mathrm{II}})
  =\frac{1}T\,\dd(\rho e)
  -\sum_i\frac{\mu_i^{\mathrm{II}}}{T}\,\dd\rho_i,
\label{eq:reduced-classII-gibbs-relation}
\end{equation}
where
\begin{equation}
  \mu_i^{\mathrm{II}}
  :=\left.\mu_i\right|_{*},
  \qquad
  p_i^{\mathrm{II}}:=\left.p_i\right|_{*},
  \qquad
  p^{\mathrm{II}}:=\sum_i p_i^{\mathrm{II}}.
\label{eq:reduced-classII-conjugates}
\end{equation}
In particular,
\begin{equation}
  \frac{\partial\rho s^{\mathrm{II}}}{\partial(\rho e)}=\frac{1}T,
  \qquad
  \frac{\partial\rho s^{\mathrm{II}}}{\partial\rho_i}
  =-\frac{\mu_i^{\mathrm{II}}}{T}.
\label{eq:reduced-classII-derivatives}
\end{equation}
The    quantities $1/T$ and $-\mu_i^{\mathrm{II}}/T$  are the thermodynamic    variables conjugate to  the retained Class-II energy and mass    densities in the entropy representation. They determine the reduced temperature and chemical potentials without a separate averaging assumption.

Restriction of the local Class-III Euler relation gives
\begin{equation}
  \rho e-T\rho s^{\mathrm{II}}
  +p^{\mathrm{II}}-\sum_i\rho_i\mu_i^{\mathrm{II}}=0.
\label{eq:euler-recovered}
\end{equation}
Differentiating this Euler relation and using the Gibbs relation gives the Class-II Gibbs--Duhem relation
\begin{equation}
  \rho s^{\mathrm{II}}\,\dd T-\dd p^{\mathrm{II}}
  +\sum_i\rho_i\,\dd\mu_i^{\mathrm{II}}=0.
\label{eq:gibbs-duhem-recovered}
\end{equation}
Equivalently,
\begin{equation}
  \grad p^{\mathrm{II}}
  =\rho s^{\mathrm{II}}\grad T
  +\sum_i\rho_i\grad\mu_i^{\mathrm{II}}.
\label{eq:classII-gibbs-duhem-spatial}
\end{equation}
For the strict constitutive restriction of the specified constitutive model used below, retained state-dependent equations of state and phenomenological coefficients are obtained by evaluating the corresponding Class-III functions at the maximizing partition \eqref{eq:entropy-maximization}. In particular, the reduced partial pressures $p_i^{\mathrm{II}}$ are the restrictions of the prescribed Class-III partial thermodynamic pressure functions. The thermodynamic part of the reduced state may subsequently be parametrized by $(T,\rho_1,\ldots,\rho_n)$ through \eqref{eq:average-temp-energy-equation}. The reduced constitutive laws retain any dependence on the relative velocities. This evaluation rule is not a general statement about singular elimination or about coupled dissipative mechanisms: eliminating a fast conjugate variable can instead generate effective Schur-complement coefficients, as discussed in \eqref{eq:thermal-schur-reduced} below.

For gradient-dependent constitutive laws, substitute the $C^1$ reconstructed field $z_i=z_i^*(\rho e,\boldsymbol\rho)$ and compute its spatial derivatives by the chain rule. In particular, $T_i|_*=T$ implies $\grad T_i|_*=\grad T$. The eliminated temperature gradients are not independent data on the reduced manifold.

\begin{remark}[Thermodynamic information inherited at class II]
\label[remark]{rem:classII-inherited-thermodynamics}
The value function $\rho s^{\mathrm{II}}$ determines the common temperature, the chemical potentials, and the mixture pressure through \eqref{eq:reduced-classII-gibbs-relation} and~\eqref{eq:euler-recovered}. It does not determine an individual pressure partition $p_i^{\mathrm{II}}$ by itself. In the present reduction, reconstructing the constituent stresses and
energy fluxes uses both partitions. Entropy maximization gives
$(\rho_i e_i)^*$, and restriction of the Class-III pressure partition
gives $p_i^{\mathrm{II}}$. A directly formulated Class-II model that
reconstructs these same constituent quantities must supply the
corresponding constitutive information.
\end{remark}

\subsection{The resulting reacting Class-II model}
\label{subsec:reduced-classII-model}

The retained unknowns are
\begin{equation}
  \rho_1,\ldots,\rho_n,
  \qquad
  \bv_1,\ldots,\bv_n,
  \qquad
  T
  \quad\text{or, equivalently, }\rho e.
\label{eq:classII-unknowns}
\end{equation}
The $n$ partial mass balances and $n$ partial momentum balances are retained, while the $n$ partial thermal-energy balances are replaced by their sum:
\begin{align}
  \partial_t\rho_i+\Div(\rho_i\bv_i)&=r_i^{\mathrm{II}},
\label{eq:classII-mass-balance}
\\
  \partial_t(\rho_i\bv_i)
  +\Div(\rho_i\bv_i\otimes\bv_i-\bS_i^{\mathrm{II}})
  &=\rho_i\bb_i+\bmom_i^{\mathrm{II}}+r_i^{\mathrm{II}}\bv,
\label{eq:classII-momentum-balance}
\\
  \partial_t(\rho e)
  +\Div\bigl(\rho e\bv+\bq^{\mathrm{II}}\bigr)
  &={}
  \sum_i\left(
    (\bS_i^{\mathrm{II}})^{\circ}:\bD_i^{\circ}
    -\Pi_i^{\mathrm{II}}\Div\bv_i
  \right)
  -p^{\mathrm{II}}\Div\bv
\nonumber\\
  &\quad
  -\sum_i\bu_i\cdot\left(
    \bmom_i^{\mathrm{II}}
    -\half r_i^{\mathrm{II}}\bu_i
    -\grad p_i^{\mathrm{II}}
  \right),
\label{eq:classII-thermal-energy-balance}
\end{align}
where
\begin{equation}
  \bq^{\mathrm{II}}
  :=\sum_i\left[
    \bq_i^{\mathrm{II}}
    +\bigl((\rho_i e_i)^{*}+p_i^{\mathrm{II}}\bigr)\bu_i
  \right].
\label{eq:classII-total-heat-flux-definition}
\end{equation}
The flux $\bq^{\mathrm{II}}$ is the restriction of $\bq^{\mathrm h}$ in \eqref{eq:thermal-flux-and-irreversible-stress}, not of the physical $\bq^{\mathrm{tot}}$. Their difference remains \eqref{eq:physical-versus-thermal-energy-flux} evaluated on the common-temperature manifold. Equation~\eqref{eq:classII-thermal-energy-balance} follows directly from \eqref{eq:thermal-energy-balance} by the pressure shift used in the entropy derivation. It is one evolution equation for the common thermal state. The zero-sum energy exchanges $\varepsilon_i$ do not appear because their sum vanishes. The    constituent energy balance  equations are no longer part of the reduced model.

The selected entropy-flux representation reduces to
\begin{equation}
  \bPhi^{\mathrm{II}}
  =\frac{\bq^{\mathrm{II}}}{T}
  -\sum_i\frac{\rho_i\mu_i^{\mathrm{II}}}{T}\bu_i.
\label{eq:common-T-entropy-flux}
\end{equation}
Introduce the common-temperature molar affinity
\begin{equation}
  \Acal_a^{\mathrm{II}}
  :=\sum_iM_i\nu_i^a\mu_i^{\mathrm{II}}.
\label{eq:classII-affinity}
\end{equation}
Then the Class-III affinity satisfies
\begin{equation}
  \left.\Acal_a\right|_{*}
  =\frac{\Acal_a^{\mathrm{II}}}{T}.
\label{eq:classIII-to-classII-affinity}
\end{equation}
Restriction of \eqref{eq:entropy-production} and use of $\sum_i\varepsilon_i=0$ gives
\begin{align}
  \zeta^{\mathrm{II}}={}&
  \frac{1}T\sum_i\left(
    (\bS_i^{\mathrm{II}})^{\circ}:\bD_i^{\circ}
    -\Pi_i^{\mathrm{II}}\Div\bv_i
  \right)
  +\bq^{\mathrm{II}}\cdot\grad\frac{1}T
\label{eq:classII-entropy-production}
\\
  &-\sum_i\bu_i\cdot\left[
    \rho_i\grad\frac{\mu_i^{\mathrm{II}}}{T}
    +\frac{1}T\left(
      \bmom_i^{\mathrm{II}}
      -\half r_i^{\mathrm{II}}\bu_i
      -\grad p_i^{\mathrm{II}}
    \right)
  \right]
  -\frac{1}T\sum_aR_a^{\mathrm{II}}\Acal_a^{\mathrm{II}}.
\nonumber
\end{align}
This is the reacting Class-II entropy-production structure. In particular,
\begin{equation}
  \sum_i\frac{\varepsilon_i}{T}=0
\label{eq:thermal-exchange-vanishes-reduced}
\end{equation}
identically, so thermal relaxation is absent as an explicit Class-II mechanism.

\begin{theorem}[Entropy-production-invariant thermal reduction]
\label[theorem]{thm:classIII-classII-reduction}
Assume the thermodynamic hypotheses of Section~\ref{subsec:thermal-projection},
including an attained interior maximum on the stated convex energy slice,
and restrict to the resulting local $C^1$ reconstruction branch.
The displayed balance and supplementary-law identities are evaluated on fields for which all flux divergences exist classically.
Let the reduced constitutive functions of the strict thermal restriction be obtained by restriction to the entropy-maximizing partition \eqref{eq:entropy-maximization}, and let the reduced balance system be \eqref{eq:classII-mass-balance}, \eqref{eq:classII-momentum-balance}, and~\eqref{eq:classII-thermal-energy-balance}. Then the entropy-production invariance identity
\eqref{eq:entropy-invariant-reduction-principle} holds, with
$\zeta^{\mathrm{II}}$ given by \eqref{eq:classII-entropy-production}. Hence the strict thermal reduction of the present Class-III theory is a multi-velocity, single-temperature, chemically reacting Class-II model.
\end{theorem}

\begin{proof}
At the interior maximizing energy partition the $n-1$ relative-energy
entropy derivatives vanish. The envelope theorem thus gives
\eqref{eq:reduced-classII-gibbs-relation}: the single retained energy
conjugate is $1/T$, and the mass conjugates are
$-\mu_i^{\mathrm{II}}/T$. Restricting the weighted Euler relation and
differentiating it yields \eqref{eq:euler-recovered} and
\eqref{eq:gibbs-duhem-recovered}. In particular, the reduced entropy
is a function of the retained Class-II state alone.

The selected constituent entropy flux is evaluated at the maximizing
partition. Grouping each diffusive enthalpy contribution with the
intrinsic heat flux yields \eqref{eq:classII-total-heat-flux-definition}
and then the entropy flux \eqref{eq:common-T-entropy-flux}.
For the sources, the eliminated internal-energy exchanges cancel
because all reciprocal temperatures agree:
\[
 \left.\sum_i\frac{\varepsilon_i}{T_i}\right|_*
 =\frac{1}T\sum_i\varepsilon_i=0.
\]
Likewise the reaction mass sources satisfy
\[
 -\sum_i\frac{\mu_i^{\mathrm{II}}}{T}r_i^{\mathrm{II}}
 =-\frac{1}T\sum_a R_a^{\mathrm{II}}
       \sum_i M_i\nu_i^a\mu_i^{\mathrm{II}}
 =-\frac{1}T\sum_aR_a^{\mathrm{II}}\Acal_a^{\mathrm{II}}.
\]
Substituting the retained mass and common thermal-energy balances into
the Gibbs identity, and adding the divergence of the reduced entropy
flux, leaves the viscous, Fourier and diffusion terms displayed in
\eqref{eq:classII-entropy-production}. The two preceding identities
account for the eliminated exchange and the chemical contribution.
This proves \eqref{eq:entropy-invariant-reduction-principle} for the
strictly restricted constitutive functions. An independent direct
balance-law calculation is given immediately below.
\end{proof}

The identity uses the entropy-flux and pressure representation selected in \Cref{subsec:entropy-flux-pressure}. This strict restriction introduces no additional phenomenological closure. Eliminating fast variables in coupled dissipative mechanisms is a separate operation and may modify the effective coefficients.

The supplementary entropy law also follows directly from the retained balances, without imposing the discarded constituent-energy equations. Write $D_t:=\partial_t+\bv\cdot\grad$. The reduced Gibbs and Euler relations, together with $D_t\rho_i=r_i^{\mathrm{II}}-\rho_i\Div\bv-\Div\bj_i$, give
\begin{align}
 \partial_t(\rho s^{\mathrm{II}})+\Div(\rho s^{\mathrm{II}}\bv)
 ={}&\frac{1}T\bigl[\partial_t(\rho e)+\Div(\rho e\bv)
                         +p^{\mathrm{II}}\Div\bv\bigr]\nonumber\\
 &-\sum_i\frac{\mu_i^{\mathrm{II}}}{T}r_i^{\mathrm{II}}
   +\sum_i\frac{\mu_i^{\mathrm{II}}}{T}\Div\bj_i.
\label{eq:direct-reduced-entropy-identity}
\end{align}
Insert \eqref{eq:classII-thermal-energy-balance} and add the divergence of \eqref{eq:common-T-entropy-flux}. Product differentiation and $r_i^{\mathrm{II}}=M_i\sum_a\nu_i^aR_a^{\mathrm{II}}$ then yield
\begin{equation}
 \partial_t(\rho s^{\mathrm{II}})
 +\Div(\rho s^{\mathrm{II}}\bv+\bPhi^{\mathrm{II}})
 =\zeta^{\mathrm{II}}.
\label{eq:autonomous-classII-entropy-law}
\end{equation}
Thus the reduced entropy balance is a supplementary law of the
autonomous Class-II equations, with production $\zeta^{\mathrm{II}}$.

For the specified constitutive relations, the reduction is explicit. The viscous laws retain their diagonal form with coefficients evaluated on the common-temperature manifold. The thermo-diffusion closure gives
\begin{equation}
  \bq^{\mathrm{II}}
  =\left(\sum_i\kappa_i^{\mathrm{II}}\right)\grad\frac{1}T
  +\sum_i\bigl((\rho_i e_i)^{*}+p_i^{\mathrm{II}}+d_i^{T,\mathrm{II}}\bigr)\bu_i,
\label{eq:common-T-total-heat-flux}
\end{equation}
with $\sum_i d_i^{T,\mathrm{II}}=0$. The common-temperature thermodynamic diffusion co-factor is
\begin{equation}
  \bB_i^{\mathrm{II}}
  :=\rho_i\grad\frac{\mu_i^{\mathrm{II}}}{T}
  -\frac{1}T\grad p_i^{\mathrm{II}}
  -\bigl((\rho_i e_i)^{*}+p_i^{\mathrm{II}}\bigr)\grad\frac{1}T,
  \qquad
  \sum_i\bB_i^{\mathrm{II}}=\boldsymbol 0.
\label{eq:common-T-diffusion-cofactor}
\end{equation}
For the pairwise-resistance choice \eqref{eq:pairwise-resistance}, the restricted resistance retains the binary form
\begin{equation}
  (\bM^{\mathrm{II}}\bu)_i
  =\sum_{j\ne i}f_{ij}^{\mathrm{II}}\rho_i\rho_j(\bu_i-\bu_j).
\label{eq:common-T-BD-resistance}
\end{equation}
The physical momentum productions are reconstructed without an additional constitutive assumption. Define
\begin{equation}
  \blambda^{\mathrm{II}}
  :=-\frac{1}{2T}\sum_j r_j^{\mathrm{II}}\bu_j.
\label{eq:classII-momentum-multiplier}
\end{equation}
Then the common-temperature specialization of the projected law is
\begin{equation}
  \bmom_i^{\mathrm{II}}
  =T\left[
    -(\bM^{\mathrm{II}}\bu)_i
    +y_i\blambda^{\mathrm{II}}
    -\bB_i^{\mathrm{II}}
    +d_i^{T,\mathrm{II}}\grad\frac{1}T
  \right]
  +\half r_i^{\mathrm{II}}\bu_i.
\label{eq:classII-momentum-reconstruction}
\end{equation}
Indeed, \eqref{eq:classII-momentum-multiplier} follows from $\sum_i\bmom_i^{\mathrm{II}}=\boldsymbol0$, $\sum_i\bB_i^{\mathrm{II}}=\boldsymbol0$, $\sum_i d_i^{T,\mathrm{II}}=0$, and $\sum_i(\bM^{\mathrm{II}}\bu)_i=\boldsymbol0$. The kinetic mass-production contribution is therefore retained in the diffusion block exactly as in the parent Class-II construction.

Finally, the specified reaction law becomes
\begin{equation}
  \log\frac{R_a^{f,\mathrm{II}}}{R_a^{b,\mathrm{II}}}
  =-\frac{\gamma_a}{\mathcal R T}\Acal_a^{\mathrm{II}}.
\label{eq:classII-reaction-closure-recovered}
\end{equation}
For $\gamma_a=1$ this is the Marcelin--De Donder rate-ratio law of the parent common-temperature theory. In the thermochemical process law of \Cref{sec:entropy-principle}, zero-sum constituent energy increments have zero thermal entropy pairing at common temperature. The detailed-balance condition is $\Acal_a^{\mathrm{II}}=0$ for a process supported on path $a$ alone and $\sum_a\omega_a^\ell\Acal_a^{\mathrm{II}}=0$ for a process $\ell$ with general reaction increments. Mass-action and Saha--Eggert formulas require the corresponding ideal-mixture and internal-state constitutive assumptions. They do not follow from the general reduced entropy alone.

Substitution of the reduced constitutive relations into \eqref{eq:classII-entropy-production} gives
\begin{align}
  \zeta^{\mathrm{II}}={}&
  \sum_i\frac{2\eta_i^{\mathrm v,\mathrm{II}}}{T}|\bD_i^{\circ}|^2
  +\sum_i\frac{\lambda_i^{\mathrm v,\mathrm{II}}}{T}(\Div\bv_i)^2
  +\left(\sum_i\kappa_i^{\mathrm{II}}\right)
   \left|\grad\frac{1}T\right|^2
\label{eq:closed-classII-zeta}
\\
  &+\sum_i\bu_i\cdot(\bM^{\mathrm{II}}\bu)_i
  -\frac{1}T\sum_aR_a^{\mathrm{II}}\Acal_a^{\mathrm{II}}
  \ge0.
\nonumber
\end{align}
Thus the explicit thermal-relaxation square of the Class-III model is the only dissipation block of the specified model forced to vanish by temperature equality alone. Every retained block has the established Class-II form.

The fully coupled constitutive alternative developed in   \cite{BotheRuggeriPaperI2026} reduces to the same entropy-production structure \eqref{eq:classII-entropy-production}. Its restriction gives the separated Class-II closure when the
retained diagonal laws agree and the additional couplings have
vanishing net contributions to the retained laws. Nonzero retained
cross-contributions give additional Class-II cross-effects.

For linear reaction blocks, comparison with \eqref{eq:classII-reaction-closure-recovered} is restricted to the near-detailed-balance approximation. Exact nonlinear agreement also requires the same kinetic activities and logarithmic or resolved-process completion \cite[Appendices~A.5--A.6]{BotheRuggeriPaperI2026}.

\subsection{Principal subsystem, constrained dynamics, and singular relaxation}
\label{subsec:three-thermal-reductions}

The thermodynamic projection just constructed must be distinguished
from three operations on the equations: principal restriction,
constrained dynamics, and fast relaxation. Table~\ref{tab:reduction-operations}
collects all four viewpoints. Their common-temperature states can
coincide while their dynamical claims and constitutive consequences
differ.

\begin{table}[htbp]

\centering
\caption{Distinct meanings of a thermal reduction. The stated conclusions
are conditional on the hypotheses of the corresponding construction.}
\label{tab:reduction-operations}
\small
\renewcommand{\arraystretch}{1.18}
\begin{tabularx}{\linewidth}{@{}p{3.0cm}XX@{}}
\toprule
Operation &    Construction  &    Established conclusions  \\
\midrule
Principal restriction & Fix linear main-field components and delete their conjugate equations. & A reduced generator. Convex subentropy and characteristic interlacing for Euler systems with positive-definite temporal Hessians. \\
Entropy projection & Maximize entropy at fixed densities and retained thermal energies. & Reduced thermodynamics and an energy partition. No dynamical invariance or convergence assertion. \\
Constrained dynamics & Supply energy exchanges that maintain exact temperature equality in all partial-energy equations. & All partial-energy balances are satisfied with the required constraint exchanges. \\
Fast relaxation & Scale interconstituent thermal exchange and eliminate fast relative-temperature modes. & A formal slow-time approximation under additional assumptions. Effective coefficients may change. \\
\bottomrule
\end{tabularx}

\end{table}

\paragraph{Principal-subsystem restriction.}

Setting the $n-1$ relative-energy main fields in
\eqref{eq:relative-energy-main-fields} to zero and deleting their conjugate
equations gives the common-temperature Euler principal subsystem.
\Cref{thm:boillat-ruggeri-principal} supplies characteristic interlacing
for its positive-definite temporal matrix. For the viscous,
heat-conducting generator, only the restriction and entropy identities
extend to the semidefinite setting
(\Cref{def:degenerate-principal-subsystem,prop:degenerate-main-field-restriction}).
The connection with the Euler hierarchy is given by
\Cref{cor:commuting-zero-restrictions}.
The restricted temporal potential and entropy-maximizing value function
give the same local thermodynamics by
\eqref{eq:thermal-reduction-dual-restriction}. Entropy projection also
applies without spatial RET integrability. Neither construction alone
establishes dynamical invariance or convergence in a relaxation limit.

\paragraph{Constrained common-temperature dynamics.}
Let the partial thermal-energy balance be written schematically as
\begin{equation}
  \partial_t(\rho_i e_i)+\Div(\rho_i e_i\bv_i)
  =F_i+\varepsilon_i,
\label{eq:energy-channel-F}
\end{equation}
where $F_i$ collects the heat-flux divergence, stress power, mechanical exchange, and mass-production terms. This scalar $F_i$ is distinct from the spatial balance-flux vectors $\boldsymbol F^k$. On the common-temperature manifold,
\begin{equation}
  \rho_i e_i=(\rho_i e_i)^{*}(\rho e,\rho_1,\ldots,\rho_n),
\label{eq:common-temp-energy-partition}
\end{equation}
the energy exchanges required to maintain the constraint are
\begin{equation}
  \varepsilon_i^{\mathrm{con}}
  =\partial_t(\rho_i e_i)^{*}
   +\Div\bigl((\rho_i e_i)^{*}\bv_i\bigr)-F_i.
\label{eq:constraint-reaction-energy}
\end{equation}
The summed energy equation implies $\sum_i\varepsilon_i^{\mathrm{con}}=0$. Since every reciprocal temperature equals $1/T$,
\begin{equation}
  \sum_i\frac{\varepsilon_i^{\mathrm{con}}}{T}=0.
\label{eq:constraint-reaction-zero-entropy}
\end{equation}
Thus finite constituent energy exchanges can remain as entropy-neutral constraint reactions even though the finite-coefficient relaxation law vanishes at exact temperature equality. These reactions enforce the embedding of the reduced solution into the full set of partial energy equations. They are not additional variables of the autonomous Class-II model.

This neutrality holds on the constrained manifold. It does not
supply a freely selectable, everywhere entropy-neutral source on an
open Class-III state neighbourhood containing common-temperature
states. The argument in \cite[Section~6.4]{BotheRuggeriPaperI2026} shows why a continuous zero-sum source
that is independently neutral throughout such a neighbourhood must
vanish at full common temperature. The constraint exchanges here are
instead reactions of the imposed constraint, or possible limits of a
separately justified fast-exchange dynamics.

\paragraph{Fast thermal relaxation: asymptotic approximation.}
Unlike the entropy-maximizing projection, the following construction is a singular time-scale approximation of the dynamics and may    modify the effective  constitutive coefficients.
Let
\begin{equation}
  b_{ij}^{\delta}=\delta^{-1}\widehat b_{ij},
  \qquad \delta\downarrow0.
\label{eq:fast-relaxation-scaling}
\end{equation}
The relative-energy equations have the schematic form
\begin{equation}
  \boldsymbol C(\boldsymbol z,\boldsymbol\rho)\dot{\boldsymbol\theta}
  +\boldsymbol G_{\mathrm{rel}}
  =-\delta^{-1}\boldsymbol L_{\theta}\boldsymbol\theta.
\label{eq:relative-temp-system}
\end{equation}
Here $\boldsymbol\theta=\mathsf S_{\mathrm{rel}}^{\sf T}\boldsymbol\vartheta$
is the vector of reciprocal-temperature combinations dual to
$\boldsymbol z_{\mathrm{rel}}$ (differences for the standard difference basis), a dot denotes the time derivative in this schematic equation, and $\boldsymbol G_{\mathrm{rel}}$ collects the    remaining terms in the relative-energy balances. With $\mathsf S=\mathsf S_{\mathrm{rel}}$,
$\mathsf S^+=(\mathsf S^{\sf T}\mathsf S)^{-1}\mathsf S^{\sf T}$, and
$\mathsf H=-D_{\boldsymbol z}^2\eta$, the matrices are
\[
 \boldsymbol C(\boldsymbol z,\boldsymbol\rho)
   =(\mathsf S^{\sf T}\mathsf H\mathsf S)^{-1},
 \qquad
 \boldsymbol L_\theta
   =\mathsf S^+\widehat{\boldsymbol L}(\mathsf S^+)^{\sf T},
\]
where $\widehat{\boldsymbol L}$ is the Laplacian of $\widehat b_{ij}$.
At fixed retained variables
$D_{\boldsymbol z_{\mathrm{rel}}}\boldsymbol\theta=-\boldsymbol C^{-1}$.
Projecting the thermal-energy balances by $\mathsf S^+$ and reversing
the sign gives \eqref{eq:relative-temp-system}, with the time derivatives
of the retained variables included in $\boldsymbol G_{\mathrm{rel}}$.
In particular $\boldsymbol C|_*=\boldsymbol C_{\mathrm{rel}}$.
We assume a uniformly coercive $\boldsymbol L_\theta$ along the chosen
branch. For a disconnected graph with a fixed partition into connected
components, the relative-energy variations have zero sum within each
component. The corresponding reduction requires uniform coercivity
of the exchange Laplacian on that subspace. The following Maxwellian
iteration approximates the reciprocal-temperature combinations
$\boldsymbol\theta$ by balancing their relaxation against the remaining
forcing. It is a formal expansion for sufficiently smooth fields on
the slow time scale after the initial transient. The retained fields
and $\boldsymbol G_{\mathrm{rel}}$ vary on an $O(1)$ time scale, and we
assume both $\boldsymbol\theta=O(\delta)$ and
$\dot{\boldsymbol\theta}=O(\delta)$ along the branch.
The expansion is taken near a compact subset of the stable
common-temperature branch, with $\boldsymbol C(\boldsymbol z,\boldsymbol\rho)$ and
$\boldsymbol L_\theta^{-1}$ uniformly bounded as $\delta\downarrow0$.
The term
$\delta\boldsymbol L_\theta^{-1}\boldsymbol C\dot{\boldsymbol\theta}$
is therefore $O(\delta^2)$.
\begin{equation}
  \boldsymbol\theta
  =-\delta\boldsymbol L_{\theta}^{-1}\boldsymbol G_{\mathrm{rel}}+O(\delta^2).
\label{eq:maxwellian-thermal-iterate}
\end{equation}

Assume the thermal-exchange graph is connected at the
leading common-temperature state. The slow-branch calculation below
uses an $O(\delta)$ state error and convergence of the full-state
derivatives entering the partial-energy balances.

\par\begingroup
The constraint exchanges have a direct interpretation on this smooth
slow branch for the separated baseline law.
Let $\boldsymbol L_*$ be the graph Laplacian of
$\widehat b_{ij}$ evaluated at the leading common-temperature state,
with $\ker\boldsymbol L_*=\operatorname{span}\{\boldsymbol1\}$.
For a smooth expansion
$\boldsymbol\vartheta^\delta=T^{-1}\boldsymbol1
+\delta\boldsymbol\beta_1+O(\delta^2)$, and smooth coefficients,
the scaled law gives
\[
 \boldsymbol\varepsilon^{T,\delta}
       =\boldsymbol L_*\boldsymbol\beta_1+O(\delta).
\]
The leading partial-energy balances therefore require
$\boldsymbol L_*\boldsymbol\beta_1=\boldsymbol\varepsilon^{\mathrm{con}}$,
which is solvable because the constraint exchanges sum to zero.
With $\boldsymbol L_*^\dagger$ the inverse on $\boldsymbol1^\perp$,
extended by zero on $\operatorname{span}\{\boldsymbol1\}$,
\[
 \zeta_{\mathrm{relax}}^\delta
   =\delta\,\boldsymbol\beta_1^{\sf T}\boldsymbol L_*
                    \boldsymbol\beta_1+O(\delta^2)
   =\delta\,(\boldsymbol\varepsilon^{\mathrm{con}})^{\sf T}
          \boldsymbol L_*^\dagger
          \boldsymbol\varepsilon^{\mathrm{con}}+O(\delta^2).
\]
Finite limiting exchanges can thus coexist with vanishing entropy
production on the slow branch. This expansion does not account for
entropy production in the initial transient.
\par\endgroup
\par\begingroup
For comparison, isolated spatially homogeneous thermal exchange at fixed
partial densities obeys $\dot{\boldsymbol z}
=\boldsymbol L(\boldsymbol z)\nabla_{\boldsymbol z}\eta$.
For a connected graph with smooth coefficients, linearization at the
constrained entropy maximum gives
$\dot{\delta\boldsymbol z}=-\boldsymbol L_*\mathsf H_*
\delta\boldsymbol z$.
This matrix is similar to
$-\mathsf H_*^{1/2}\boldsymbol L_*\mathsf H_*^{1/2}$.
The nullspace of $\boldsymbol L_*\mathsf H_*$ is spanned by $\mathsf H_*^{-1}\boldsymbol1$,
which is not tangent to the fixed-energy slice because $a_*>0$.
All eigenvalues on that slice are therefore negative, giving local
asymptotic stability of the entropy maximum for this isolated
finite-dimensional exchange system.
\par\endgroup
\par\begingroup
\begin{proposition}[Quantitative isolated thermal equilibration]
\label[proposition]{prop:quantitative-thermal-equilibration}
Fix the partial densities and total thermal energy, and let
$\boldsymbol z^*$ be the constrained entropy maximum. For $n\ge2$,
on a convex admissible neighbourhood of $\boldsymbol z^*$ assume
\[
 m\mathsf I_n\preceq
 \mathsf H(\boldsymbol z):=-D_{\boldsymbol z}^2\eta
 \preceq M\mathsf I_n,\qquad
 \boldsymbol L(\boldsymbol z)\succeq\ell\mathsf P_\perp,\qquad
 \mathsf P_\perp:=\mathsf I_n-\boldsymbol1\boldsymbol1^{\sf T}/n,
\]
where $m,M,\ell>0$, $\boldsymbol L$ is a symmetric graph Laplacian
with $\boldsymbol L\boldsymbol1=0$, and its coefficients are locally
Lipschitz. For the isolated exchange system
$\dot{\boldsymbol z}
=\delta^{-1}\boldsymbol L(\boldsymbol z)\nabla_{\boldsymbol z}\eta$,
$\delta>0$, define
$\mathcal E_T(t):=\eta(\boldsymbol z^*)-\eta(\boldsymbol z(t))$.
Every solution on the fixed-energy slice remaining in the neighbourhood
satisfies
\begin{align}
 \tfrac m2|\boldsymbol z-\boldsymbol z^*|^2
 &\le\mathcal E_T\le\tfrac M2|\boldsymbol z-\boldsymbol z^*|^2,
 \label{eq:thermal-deficit-bounds}\\
 \mathcal E_T(t)&\le\mathcal E_T(0)e^{-2m\ell t/\delta},
 \label{eq:thermal-exponential-decay}\\
 \int_0^t\zeta_{\mathrm{relax}}^\delta(s)\,\dd s
 &=\mathcal E_T(0)-\mathcal E_T(t).
 \label{eq:thermal-integrated-production}
\end{align}
These estimates hold for all positive times for initial data in
sufficiently small sublevel sets of the entropy deficit compactly
contained in the neighbourhood.
\end{proposition}
\begin{proof}
The first variation at $\boldsymbol z^*$ vanishes on the energy slice.
Writing $\boldsymbol d=\boldsymbol z-\boldsymbol z^*$, Taylor's
integral formula gives
\[
 \mathcal E_T=\int_0^1(1-s)\,
 \boldsymbol d^{\sf T}
 \mathsf H(\boldsymbol z^*+s\boldsymbol d)\boldsymbol d\,\dd s,
\]
proving \eqref{eq:thermal-deficit-bounds}.
Strong concavity on the slice gives
$\mathcal E_T\le
|\mathsf P_\perp\nabla_{\boldsymbol z}\eta|^2/(2m)$:
apply the strong-concavity inequality at $\boldsymbol z$ to
$\boldsymbol z^*$ and maximize the resulting linear-minus-quadratic
bound over tangent displacements. The exchange law therefore yields
\[
 \dot{\mathcal E}_T=-\zeta_{\mathrm{relax}}^\delta
 \le-\frac{\ell}{\delta}
       |\mathsf P_\perp\nabla_{\boldsymbol z}\eta|^2
 \le-\frac{2m\ell}{\delta}\mathcal E_T.
\]
Integration proves the remaining identities and estimate.
Small closed deficit sublevel sets are compactly contained in the
neighbourhood and positively invariant. Local Lipschitz continuity
and continuation then give solutions for all positive times there.
\end{proof}
The uniform bounds hold locally at a stable common-temperature state
when the exchange graph is connected there and the exchange weights
are continuous. Local Lipschitz continuity of the weights gives
uniqueness of the exchange dynamics. For fixed initial data
in one of these invariant deficit sublevel sets, the entropy produced over $[0,t]$ for any
fixed $t>0$ tends to the initial projection deficit as $\delta\downarrow0$.
This quantifies the finite initial-transient contribution absent from
the zero thermal-exchange production of the strict Class-II slow model.
It is a result for isolated homogeneous exchange, not a convergence
theorem for the coupled balance-law system.
\par\endgroup

For the specified uncoupled thermal-exchange law,
with the remaining constitutive laws held fixed, the formal leading slow system is the strict Class-II model of Theorem~\ref{thm:classIII-classII-reduction}. First-order elimination may, however,    modify  the retained constitutive laws. Let $\boldsymbol X$ collect the retained scalar entropy co-factors. For a quadratic near-equilibrium contribution written schematically as
\begin{equation}
  \zeta_{\mathrm{sc}}
  =\boldsymbol X\cdot\mathbb L_{XX}\boldsymbol X
  +2\boldsymbol X\cdot\mathbb L_{X\theta}\boldsymbol\theta
  +\boldsymbol\theta\cdot\mathbb L_{\theta\theta}\boldsymbol\theta,
\label{eq:thermal-schur-parent}
\end{equation}
assume that the complete reciprocal parent block is symmetric positive semidefinite on the retained/relative scalar space,
\begin{equation}
  \begin{pmatrix}
    \mathbb L_{XX} & \mathbb L_{X\theta}\\
    \mathbb L_{\theta X} & \mathbb L_{\theta\theta}
  \end{pmatrix}\succeq0,
  \qquad
  \mathbb L_{\theta X}=\mathbb L_{X\theta}^{\sf T},
  \qquad
  \mathbb L_{\theta\theta}\succ0
\label{eq:thermal-schur-hypotheses}
\end{equation}
on the relative-energy subspace. Assume that the leading
quasistationary fast balance gives
$\mathbb L_{\theta X}\boldsymbol X+
\mathbb L_{\theta\theta}\boldsymbol\theta=0$.
Eliminating $\boldsymbol\theta$ using this relation yields the quadratic
form of the positive-semidefinite Schur complement
\begin{equation}
  \zeta_{\mathrm{sc}}^{\mathrm{eff}}
  =\boldsymbol X\cdot\left(
    \mathbb L_{XX}
    -\mathbb L_{X\theta}\mathbb L_{\theta\theta}^{-1}
     \mathbb L_{\theta X}
  \right)\boldsymbol X,
\label{eq:thermal-schur-reduced}
\end{equation}
with positive semidefiniteness following from the full block condition
in \eqref{eq:thermal-schur-hypotheses}. Applying this constitutive
reduction to \eqref{eq:maxwellian-thermal-iterate} requires verifying
the stated quasistationary relation for the particular forcing
$\boldsymbol G_{\mathrm{rel}}$.

Thus simply evaluating a constitutive coefficient at equal
temperatures and eliminating a rapidly adjusting thermal mode are
not interchangeable procedures. The former restricts a given law.
The latter    may modify the constitutive relations for  the retained mechanisms. The displayed Schur complement    gives this modification only under the stated positivity assumption on the complete block and the matching quasistationary relation. No convergence estimate for the nonlinear parent PDE is
established by this algebraic calculation.

\par\begingroup
The scaling of the cross-block determines the order of this correction.
If $\mathbb L_{\theta\theta}=\delta^{-1}\mathbb K$ and
$\mathbb L_{X\theta}=\delta^{-\alpha}\mathbb C$, with fixed
$\mathbb K\succ0$ and $\mathbb C\ne0$, then
\[
 \mathbb L_{X\theta}\mathbb L_{\theta\theta}^{-1}\mathbb L_{\theta X}
 =\delta^{\,1-2\alpha}\mathbb C\mathbb K^{-1}\mathbb C^{\sf T}.
\]
For bounded $\mathbb L_{XX}$, positivity of the complete block requires
$\alpha\le\tfrac12$. An order-one cross-block therefore gives an
$O(\delta)$ correction, whereas an admissible $O(\delta^{-1/2})$
cross-block may alter the leading reduced coefficients.
\par\endgroup

In the nonreacting near-equilibrium simple-mixture reference regime, when the first Maxwellian iteration is evaluated at a common-velocity zeroth-order state and other    forcing of the relative-energy equations  is absent, the compression forcing is proportional to $\Div\bv$. With the corresponding thermodynamically compatible relaxation closure, its first-order pressure contribution has the form~\cite{GouinRuggeri2008,RuggeriSimic2009}
\begin{equation}
  \pi_{\theta}=-L_{\pi}\Div\bv,
  \qquad L_{\pi}\ge0.
\label{eq:dynamical-pressure}
\end{equation}

Under the scaling \eqref{eq:fast-relaxation-scaling}, this relaxation-induced contribution is of order $\delta$ on the stated slow branch. Neither proportionality to $\Div\bv$ nor the sign of an arbitrary pressure response follows solely from the schematic forcing in \eqref{eq:relative-temp-system}.

For comparison, the nonreacting simple-mixture analyses of Ruggeri and Simi\'c~\cite{RuggeriSimic2009,RuggeriSimic2017} provide reference cases for diffusion temperatures and Maxwellian iteration. Temperature differences must be distinguished from their contribution to the summed thermodynamic pressure. For calorically perfect ideal gases, let $e_i=e_i^0+c_{V,i}T_i$ and $p_i=\rho_i R_iT_i$. If every $R_i=(\gamma-1)c_{V,i}$ with the same constant $\gamma$, then
\[
 \sum_i p_i=(\gamma-1)\left(\sum_i\rho_i e_i-\sum_i\rho_i e_i^0\right).
\]
At fixed partial densities and total thermal energy, this pressure is independent of the temperature partition exactly. Consequently its difference from the pressure at the energy-equivalent common temperature vanishes at every order.   Any nonzero effective pressure in that case must arise from separately identified mechanical, constitutive, or nonideal contributions. Unequal-$\gamma$ temperature-induced pressure coefficients remain useful comparison quantities after matching the model assumptions and relaxation scaling.

In the general viscous, heat-conducting, and reacting theory, $\boldsymbol G_{\mathrm{rel}}$ also contains differential stress powers, heat-flux divergences, reaction heating, and diffusion work. Hence \eqref{eq:dynamical-pressure} is a special effective reduction, not a    general  Class-III    constitutive law. A direct constitutive coupling between dilatation and thermal exchange, as allowed by the fully coupled scalar constitutive block of   \cite{BotheRuggeriPaperI2026}, may contribute to or    modify  the reduced pressure correction, but is not required for its appearance.

\subsection{Structural constraints on a further Class-II-to-Class-I reduction}
\label{subsec:classII-classI-reduction}

A further mechanical reduction would replace the independent constituent momentum equations by the barycentric momentum equation and constitutive relations for diffusion. The present paper records the balance and entropy constraints on such a reacting Class-I reduction, rather than deriving a complete closure. Unlike the equal-velocity principal restrictions in \Cref{prop:thermal-before-mechanical}, this reduction retains relative motion through constitutive diffusion fluxes. The diffusion mass fluxes
\begin{equation}
  \bj_i=\rho_i\bu_i,
  \qquad
  \sum_i\bj_i=\boldsymbol 0,
\label{eq:classI-diffusion-flux}
\end{equation}
remain constitutive quantities. The diffusion kinetic stress and energy in \eqref{eq:mixture-stress} and~\eqref{eq:energy-decomposition} must be retained, absorbed consistently into reduced constitutive quantities, or neglected only under an additional small-relative-inertia approximation.

\paragraph{Reactive mechanical reduction versus the nonreactive diffusional approximation.}
In the nonreactive reduction from class II to Maxwell--Stefan diffusion, the constituent relative-inertia terms may be small in a suitable low-diffusion-Mach-number regime~\cite{BotheDreyer2015}. In the balance decomposition used for the Class-II reduction, write
\begin{equation}
  \boldsymbol{\mathcal I}_i
  :=\rho_i(\partial_t+\bv\cdot\grad)\bu_i
    +\rho_i(\bu_i\cdot\grad)\bv_i.
\label{eq:relative-inertia-reactive-reduction}
\end{equation}
The barycentric constraint $\sum_i\rho_i\bu_i=\boldsymbol0$ and the partial mass balances give the exact compatibility relation
\begin{equation}
  \sum_i\boldsymbol{\mathcal I}_i
  =
  \Div\!\left(\sum_i\rho_i\bu_i\otimes\bu_i\right)
  -\sum_i r_i\bu_i.
\label{eq:reactive-relative-inertia-sum}
\end{equation}
The first term is the divergence of the diffusion-kinetic stress contribution. The commonly used small-relative-inertia reduction therefore contains two logically separate approximations: neglect of the divergence of the quadratic diffusion-kinetic
stress under a suitable spatial scaling and quasistationary elimination of the $\boldsymbol{\mathcal I}_i$. After the first approximation,
\[
  \sum_i\boldsymbol{\mathcal I}_i\approx-\sum_i r_i\bu_i.
\]
Thus setting every $\boldsymbol{\mathcal I}_i$ to zero by applying the nonreactive scaling argument also requires $\sum_i r_i\bu_i\approx0$, which is not an identity of a reacting mixture. In a spatially homogeneous reacting state the exact relation already reduces to $\sum_i\boldsymbol{\mathcal I}_i=-\sum_i r_i\bu_i$, which determines the sign of the reaction term. The chemical contribution may be asymptotically small in a specified regime, but that must be justified jointly with the chemical and diffusion time scales. Any entropy-production-invariant Class-II-to-Class-I reduction must therefore proceed through balance-compatible identities before any further small-relative-inertia approximation is made.

\par\begingroup
An exact projection makes the compatibility explicit. Define
$\widetilde{\boldsymbol{\mathcal I}}_i
 :=\boldsymbol{\mathcal I}_i-y_i\sum_j\boldsymbol{\mathcal I}_j$
and $\bb:=\sum_i y_i\bb_i$. Subtracting $y_i$ times the summed
momentum balance from the $i$th constituent momentum balance, and using
\eqref{eq:reactive-relative-inertia-sum}, gives
\[
 \bmom_i=
 y_i\Div\sum_j\bS_j-\Div\bS_i+\rho_i(\bb-\bb_i)
 +r_i\bu_i-y_i\sum_jr_j\bu_j
 +\widetilde{\boldsymbol{\mathcal I}}_i .
\]
Here $\sum_i\widetilde{\boldsymbol{\mathcal I}}_i=\boldsymbol0$.
The diffusion-kinetic stress cancels from this projected identity
without approximation. If a specified scaling permits
$\widetilde{\boldsymbol{\mathcal I}}_i$ to be neglected, the remaining
relation still satisfies $\sum_i\bmom_i=\boldsymbol0$ and retains the
reaction-dependent correction. Neglect of diffusion-kinetic stress in
the separate barycentric equation requires its own scaling assumption.
\par\endgroup

\paragraph{Constitutive information inherited from the resolved momentum level.}
The exact mixture stress of the parent model contains $-\sum_i\rho_i\bu_i\otimes\bu_i$ by \eqref{eq:mixture-stress}. A Class-I stress defined by the sum of the partial stresses omits
this term, which is quadratic in the diffusion velocities. Neglecting
its divergence in the momentum balance requires spatial derivative
control, as discussed above. In addition, the balance-compatible flux identities inherited from the resolved Class-II level show how partial-stress information can enter a reduced diffusion and transport representation. On the common-temperature manifold define the partial enthalpy density and irreversible stress by
\begin{equation}
  h_i^{\mathrm{II}}:=(\rho_i e_i)^*+p_i^{\mathrm{II}},
  \qquad
  \boldsymbol\tau_i^{\mathrm{II}}
  :=-\Pi_i^{\mathrm{II}}\bI+(\bS_i^{\mathrm{II}})^\circ.
\label{eq:classII-partial-enthalpy-irreversible-stress}
\end{equation}
Separate the part of the inherited physical energy flux that excludes the cubic diffusion-kinetic term:
\begin{equation}
  \bq^{\mathrm I}_{\mathrm{inh}}
  =\sum_i\left[
     \bq_i^{\mathrm{II}}
     +h_i^{\mathrm{II}}\bu_i
     -\boldsymbol\tau_i^{\mathrm{II}}\bu_i
   \right],
\label{eq:inherited-classI-energy-flux}
\end{equation}

The exact relation to the parent physical flux is
\begin{equation}
  (\bq^{\mathrm{tot}})|_*
  =\bq^{\mathrm I}_{\mathrm{inh}}
   +\half\sum_i\rho_i|\bu_i|^2\bu_i.
\label{eq:inherited-classI-physical-flux}
\end{equation}
Using only $\bq^{\mathrm I}_{\mathrm{inh}}$ in a physical total-energy balance requires an explicit truncation consistent with the diffusion-kinetic storage and stress. Common temperature alone does not justify it. With this separation of the cubic term, the selected entropy flux is exactly

\begin{equation}
  \bPhi^{\mathrm I}_{\mathrm{inh}}
  =\frac{\bq^{\mathrm I}_{\mathrm{inh}}}{T}
   -\sum_i\frac{\mu_i^{\mathrm{II}}\bj_i}{T}
   +\frac{1}T\sum_i\boldsymbol\tau_i^{\mathrm{II}}\bu_i.
\label{eq:inherited-classI-entropy-flux}
\end{equation}
When the entropy flux is instead expressed through $(\bq^{\mathrm{tot}})|_*$, the cubic term in \eqref{eq:inherited-classI-physical-flux}, divided by $T$, must be subtracted, as in \eqref{eq:physical-shifted-entropy-flux}. The projected momentum balance above gives the partial irreversible
stress divergences in the relative momentum equations from which a
reduced diffusion relation would be derived. A Class-I theory formulated directly from the barycentric balances may omit this contribution. It is then a different constitutive model, not merely another notation for \eqref{eq:inherited-classI-energy-flux} and~\eqref{eq:inherited-classI-entropy-flux}.

\paragraph{Reactive resistance after elimination.}
When the chemically inherited momentum source of \eqref{eq:BD-chemical-force} and~\eqref{eq:chemical-matrix} is resolved before mechanical reduction, it must enter any reduced relative-momentum/resistance relation together with the kinetic mass-production terms required by the compatibility identity \eqref{eq:reactive-relative-inertia-sum}. The chemical momentum-transfer matrix $C^{\mathrm{chem}}_{ij}$ is generally nonsymmetric away from pathwise detailed chemical equilibrium, but this fact alone neither proves nor obstructs entropy admissibility of a reduced operator. An entropy-production-invariant mechanical reduction must reproduce the parent nonnegative entropy pairing on the reduced constrained space. The symmetric and skew-adjoint parts can be assessed only after the complete conjugate reduced operator has been derived. In particular, nonsymmetry of $C^{\mathrm{chem}}_{ij}$ does not by itself identify an independently selectable entropy-neutral mechanism.

No closed reacting Class-I resistance operator is derived here. Deriving such an operator with \eqref{eq:reactive-relative-inertia-sum}, a stated mechanical scaling, and the chosen chemical momentum model is a separate mechanical-reduction problem. The conclusion established here is that a Class-I model closed directly from barycentric balances need not coincide with a Class-I model inherited from a resolved momentum theory. The inherited model may contain stress-driven and chemical-momentum-dependent transport together with compensating energy/entropy-flux terms. Equivalence of the two constructions requires additional constitutive or asymptotic simplifications to be stated explicitly.

The constructed thermal reduction and the prospective mechanical
reduction can be summarized as follows.
\begin{equation}
  \text{class III}
  \xrightarrow{\ \mathcal P_T\ }
  \text{class II}
  \xrightarrow{\ \mathcal P_{\bv}\ }
  \text{class I},
\label{eq:reduction-hierarchy}
\end{equation}
where the first map is constructed explicitly in this paper and the second denotes a prospective entropy-production-invariant mechanical reduction constrained as above. A complete reacting realization of $\mathcal P_{\bv}$ is not derived here. More generally, an intermediate grouped projection $\mathcal P_{\mathcal G}$ may be inserted before $\mathcal P_T$. It eliminates only intragroup relative energies, retains one thermal-energy equation per group, and leaves all constituent mass and momentum variables intact.

\section{Conclusions}
\label{sec:conclusions}

The main result is a complete main-field generator for the dissipative
Class-III mixture in the constitutive subclass where the prescribed
constituent Euler fluxes admit spatial potentials. The main field is
identified through the full supplementary-entropy multiplier identities,
including the zero-storage Newton and Fourier equations. Its generating
potentials yield a symmetric first-order system whose temporal Hessian is
positive semidefinite and singular. After eliminating the auxiliary
constitutive variables, the highest-order dissipative symbol admits an
explicit nonnegative factorization.

For the separated Newton--Fourier laws with positive shear viscosities
and heat conductivities, the kernel of the dissipative principal
symbol contains no Euler characteristic eigenvector, including when
the bulk viscosities vanish. The mixed formulation also accommodates
local constitutive responses that admit the stated representation
and a nonnegative complete entropy-source pairing.

A second result is the extension of the algebraic Boillat--Ruggeri
principal-subsystem construction to such degenerate symmetric
generators. The thermal and zero-storage restrictions commute:
temperature grouping can retain a dissipative multi-velocity  mixture,
whereas setting the constitutive main-field components to zero and
deleting their equations recovers the constituent Euler fluxes. Further
grouping produces a nested hierarchy of Euler subsystems with one
temperature per group. Classical characteristic interlacing applies
within this Euler hierarchy because its temporal Hessians are positive
definite. The associated constraint manifolds need not be invariant
under the parent evolution.

The   separate
Gibbs--Duhem conditions $\omega_i=0$ are necessary and sufficient for the
spatial generator of the prescribed Euler fluxes under the hypotheses
of the integrability theorem, and they imply locally additive entropy
$\rho s=\sum_i\eta_i(\rho_i e_i,\rho_i)$ with compatible partial
pressures. They are stronger than the mixture-wide compatibility of
   \mbox{
\cite{BotheRuggeriPaperI2026} }
 and are not a consequence of entropy admissibility alone.

Independently of that integrability restriction, the thermodynamic
Class-III-to-Class-II reduction uses constrained entropy maximization
to reconstruct an energy-equivalent common temperature. It applies
locally also to nonadditive parent entropies, yields a positive reduced
heat capacity, and identifies the reduced entropy and its conjugate
variables. The exact relative-entropy decomposition
separates the thermal deficit from the retained-state relative entropy
and gives a projection inequality with a reconstructed reference state.
We formulate the entropy-production-invariance requirement
for the reduced constitutive laws and distinguish this projection
from constrained dynamics and formal fast-relaxation approximations.
A complete reactive Class-II-to-Class-I reduction requires further
mechanical and chemical assumptions.

Grouped entropy projections satisfy the composition identity under
coarsening and have an explicit positive-definite susceptibility matrix.
For isolated homogeneous thermal exchange, uniform strict concavity
on the fixed-energy slice and a spectral gap of the exchange Laplacian
give exponential equilibration and an exact identity for time-integrated
thermal entropy production.

These results    extend  the established symmetric-hyperbolic theory of Euler mixtures
\cite{MuellerRuggeri1998,RuggeriSimic2007,RuggeriSugiyama2021}.
The two papers therefore address complementary questions:    the theory in \mbox{
\cite{BotheRuggeriPaperI2026}
}
 establishes the general entropy-admissible constitutive framework,
whereas    the present paper  identifies the integrable fixed-flux subclass,
constructs its degenerate symmetric generator and principal hierarchy,
and develops the reduced thermodynamics and thermal-reduction principle
for the general parent entropy. The fixed-flux obstruction applies to
representations that recover the prescribed Euler fluxes on a constant
auxiliary section. Symmetric formulations outside that setting and a
full well-posedness theory for the singular dissipative system require
additional analysis.
\appendix
\section{Conservative Liu multipliers and the conservative block of the main field}
\label{app:liu-consistency}
This appendix checks the conservative part of the multiplier identities.
The complete main field is characterized by the supplementary law and
all its differential flux identities, including the spatial identities.
The storage-rate Liu calculation determines only the $5n$ conservative
multipliers, not the $9n$ auxiliary multipliers of the  enlarged  first-order dissipative    formulation. This is a consistency check on the already
specified Class-III state and balance    equations, not a derivation of    those assumptions. Liu's exploitation method also applies to more general,
including nonquasilinear, differential constraints. Its multipliers depend
on the form in which those constraints are stated. In particular, the
coincidence found below cannot be transferred without qualification to
multipliers for equations expressed using material derivatives. By
contrast, the main field is defined algebraically through its action on
the full quasilinear balance-law system and its supplementary entropy law.
The storage-rate calculation here checks only the conservative block.

For this check we apply the storage-rate part of the Liu multiplier procedure~\cite{Liu1972} to all three constituent balances in conservative variables. Let
\begin{equation}
  \bV_i=(\rho_i,\rho_i\bv_i,E_i),
  \qquad
  E_i:=z_i+\frac{|\rho_i\bv_i|^2}{2\rho_i},
\label{eq:liu-conservative-variables}
\end{equation}
with exactly the conservative variables of \eqref{eq:conservative-variables}. Since
\begin{equation}
  z_i=E_i-\frac{|\rho_i\bv_i|^2}{2\rho_i},
\end{equation}
the storage-rate Liu equations for the mass, momentum, and total-energy balances give
\begin{equation}
  \boldsymbol\lambda_i^{\mathrm L}
  =\frac{\partial\eta}{\partial\bV_i}
  =\left(
  \frac{\half|\bv_i|^2-\mu_i}{T_i},
  -\frac{\bv_i}{T_i},
  \frac{1}{T_i}
  \right)
  =-\vp_i,
\label{eq:liu-main-field-relation}
\end{equation}
where $\vp_i=\partial h^0/\partial\bV_i$ is the conservative constituent block of the main field, given in \eqref{eq:complete-main-field-components}, and $h^0=-\eta$. The equality is therefore a sign convention, not an additional constitutive assumption.

The same conservative calculation independently checks the source pairing. The conservative source vector in the first three rows of \eqref{eq:acta-FNS-first-order-form} gives
\begin{align}
&\lambda_i^{\rho,\mathrm L}r_i
+\boldsymbol\lambda_i^{m,\mathrm L}\cdot(\rho_i\bb_i+\bfv_i)
+\lambda_i^{E,\mathrm L}(\rho_i\bb_i\cdot\bv_i+k_i)
\nonumber\\
&\qquad=
\frac{1}{T_i}\left(k_i-\bv_i\cdot\bfv_i+\half r_i|\bv_i|^2-\mu_i r_i\right)
\nonumber\\
&\qquad=
\frac{1}{T_i}\left(\varepsilon_i-\bu_i\cdot\bmom_i
+\half r_i|\bu_i|^2-\mu_i r_i\right),
\label{eq:liu-source-pairing}
\end{align}
where the last equality uses \eqref{eq:objective-sources} and~\eqref{eq:barycentric-definitions}. Summing over the constituents gives \eqref{eq:ruggeri-source-pairing}.

Finally consider the conservative Euler principal part and suppose, in addition, that it possesses a supplementary entropy balance with physical entropy-flux components $j_s^k(\bV)$ (distinct from the generating potentials $h_E'^{k}$). The spatial Liu equations are then
\begin{equation}
  \frac{\partial j_s^k}{\partial\bV}
  =\boldsymbol\lambda^{\mathrm L}\cdot
  \frac{\partial\boldsymbol F_E^k}{\partial\bV}.
\label{eq:liu-spatial-compatibility}
\end{equation}
With $h^0=-\eta$, $h^k=-j_s^k$, and $\vp=-\boldsymbol\lambda^{\mathrm L}$ this is
\begin{equation}
  \dd h^k=\vp\mathbin{\boldsymbol\cdot}\dd\boldsymbol F_E^k.
\label{eq:liu-main-field-differential}
\end{equation}
Hence the Legendre-type flux potential
\begin{equation}
  h_E'^{k}:=\vp\mathbin{\boldsymbol\cdot}\boldsymbol F_E^k-h^k
\end{equation}
satisfies
\begin{equation}
  \dd h_E'^{k}=\boldsymbol F_E^k\mathbin{\boldsymbol\cdot}\dd \vp,
  \qquad
  \boldsymbol F_E^k=\frac{\partial h_E'^{k}}{\partial \vp},
\label{eq:liu-flux-potential}
\end{equation}

which is the RET flux-potential condition \eqref{eq:main-field-flux-potential-condition}. The main field is defined independently through the supplementary-law structure described in \Cref{sec:main-field}. Liu multipliers are, a priori, constraint multipliers and need not themselves provide independent coordinates or a symmetrizing field. Under the supplementary-law and spatial-integrability hypotheses used here, the conservative Liu multipliers reduce, up to the entropy-sign convention, to the conservative block of the independently defined main field. This calculation neither defines nor determines the auxiliary multipliers $\wprime$ of the enlarged system.

\begin{proposition}[Scope of the Liu consistency calculation]
\label[proposition]{prop:liu-consistency}
On the   specified Class-III state and balance    equations  summarized in \Cref{sec:balances,sec:model-class-specification}, the conservative Liu multiplier vector is $-\vp$ and its source contraction is \eqref{eq:ruggeri-source-pairing}. For an Euler subsystem with a supplementary entropy law, spatial Liu compatibility is equivalent to \eqref{eq:main-field-flux-potential-condition}. These are consistency statements on an already selected constitutive class. They neither select the thermodynamic state and mechanism list nor replace the balance-anchoring principle used in the   continuum-thermodynamic construction  of \mbox{
\cite{BotheRuggeriPaperI2026}}.
\end{proposition}

\end{document}